\documentclass[sigconf]{acmart}
\usepackage{mathtools}
\usepackage{breakurl}
\usepackage{makecell}
\usepackage{algorithm}
\usepackage{algpseudocode}
\usepackage{booktabs} 
\usepackage{enumitem}
\usepackage{letltxmacro}
\usepackage{multirow,array}
\usepackage{subfig}
\usepackage{scalerel}
\usepackage{listings}
\usepackage{multicol}

\definecolor{gre}{RGB}{15, 140, 0}
\definecolor{evalOrange}{RGB}{239, 155, 0}

\definecolor{page1color}{RGB}{34, 185, 4}
\definecolor{page2color}{RGB}{128, 255, 104}
\definecolor{page3color}{RGB}{230, 230, 0}
\definecolor{page4color}{RGB}{109, 109, 109}
\definecolor{page5color}{RGB}{251, 0, 6}

\newcommand{\maxColor}[1]{\textcolor[RGB]{15, 140, 0}{#1}}
\newcommand{\minColor}[1]{\textcolor[RGB]{251, 0, 6}{#1}}

\newcommand{\myfbox}[2]{{\color{#1}\fbox{\normalcolor#2}}}

\begin{document}
\settopmatter{printacmref=false} 
\renewcommand\footnotetextcopyrightpermission[1]{} 
\pagestyle{plain}

\title{Using Micro-collections in Social Media\\to Generate Seeds for Web Archive Collections}
\titlenote{This is an extended version of the ACM/IEEE Joint Conference on Digital Libraries (JCDL 2019) full paper. Some figures have been enlarged, and appendices of additional figures included.}

\author{Alexander C. Nwala}
\affiliation{%
  \institution{Old Dominion University}
  \city{Norfolk} 
  \state{Virginia} 
  \postcode{23529}
  \country{USA}
}
\email{anwala@cs.odu.edu}

\author{Michele C. Weigle}
\affiliation{%
  \institution{Old Dominion University}
  \city{Norfolk} 
  \state{Virginia} 
  \postcode{23529}
  \country{USA}
}
\email{mweigle@cs.odu.edu}

\author{Michael L. Nelson}
\affiliation{%
  \institution{Old Dominion University}
  \city{Norfolk} 
  \state{Virginia} 
  \postcode{23529}
  \country{USA}
}
\email{mln@cs.odu.edu}

\pagestyle{empty}

\begin{abstract}
In a Web plagued by disappearing resources, Web archive collections provide a valuable means of preserving Web resources important to the study of past events ranging from elections to disease outbreaks. These archived collections start with seed URIs (Uniform Resource Identifiers) hand-selected by curators. Curators produce high quality seeds by removing non-relevant URIs and adding URIs from credible and authoritative sources, but it is time consuming to collect these seeds. Two main strategies adopted by curators for discovering seeds include scraping Web (e.g., Google) Search Engine Result Pages (SERPs) and social media (e.g., Twitter) SERPs. In this work, we studied three social media platforms in order to provide insight on the characteristics of seeds generated from different sources. First, we developed a simple vocabulary for describing social media posts across different platforms. Second, we introduced a novel source for generating seeds from URIs in the threaded conversations of social media posts created by single or multiple users. Users on social media sites routinely create and share posts about news events consisting of hand-selected URIs of news stories, tweets, videos, etc. In this work, we call these posts \textit{micro-collections}, and we consider them as an important source for seeds because the effort taken to create micro-collections is an indication of editorial activity, and a demonstration of domain expertise. Third, we generated 23,112 seed collections with text and hashtag queries from 449,347 social media posts from Reddit, Twitter, and Scoop.it. We collected in total 120,444 URIs from the conventional scraped SERP posts and micro-collections. We characterized the resultant seed collections across multiple dimensions including the distribution of URIs, precision, ages, diversity of webpages, etc. We showed that seeds generated by scraping SERPs had a higher median probability (0.63) of producing relevant URIs than micro-collections (0.5). However, micro-collections were more likely to produce seeds with a higher precision than conventional SERP collections for Twitter collections generated with hashtags. Also, micro-collections were more likely to produce older webpages and more non-HTML documents.
\end{abstract}

\maketitle
\sloppy
\section{Introduction and Background}
\label{intro}
\begin{figure*}[h!]
    \centering
    \subfloat[Subset 1: tweet reply thread]{{ \fbox{\includegraphics[width=0.46\textwidth]{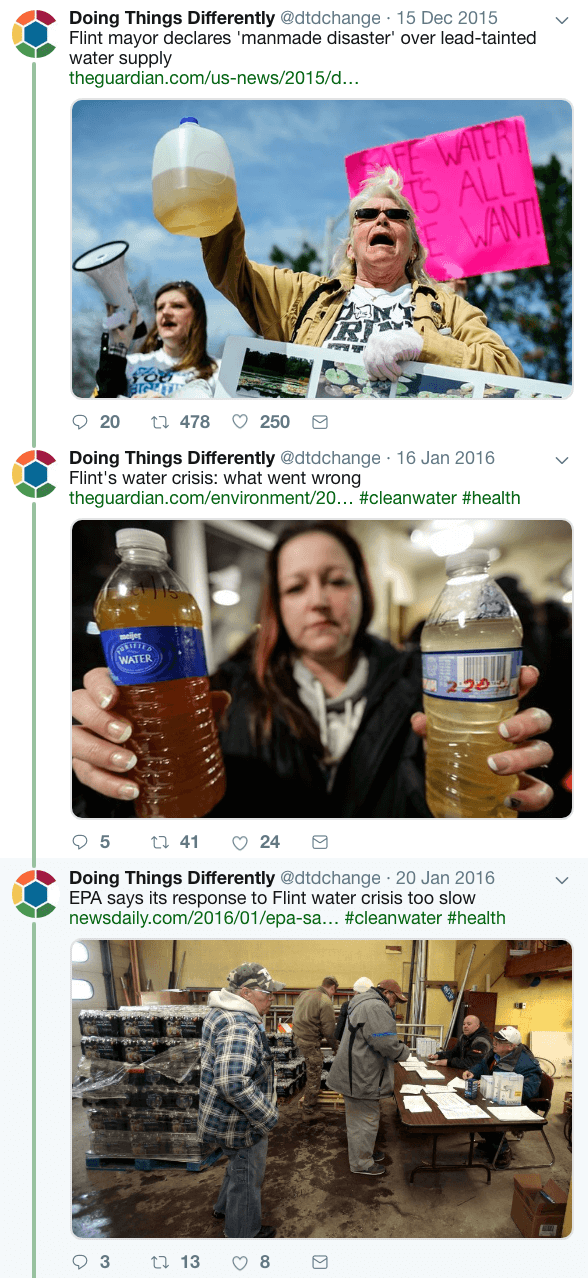}} }}%
    \,
    \subfloat[Subset 2: tweet reply thread]{{ \fbox{\includegraphics[width=0.46\textwidth]{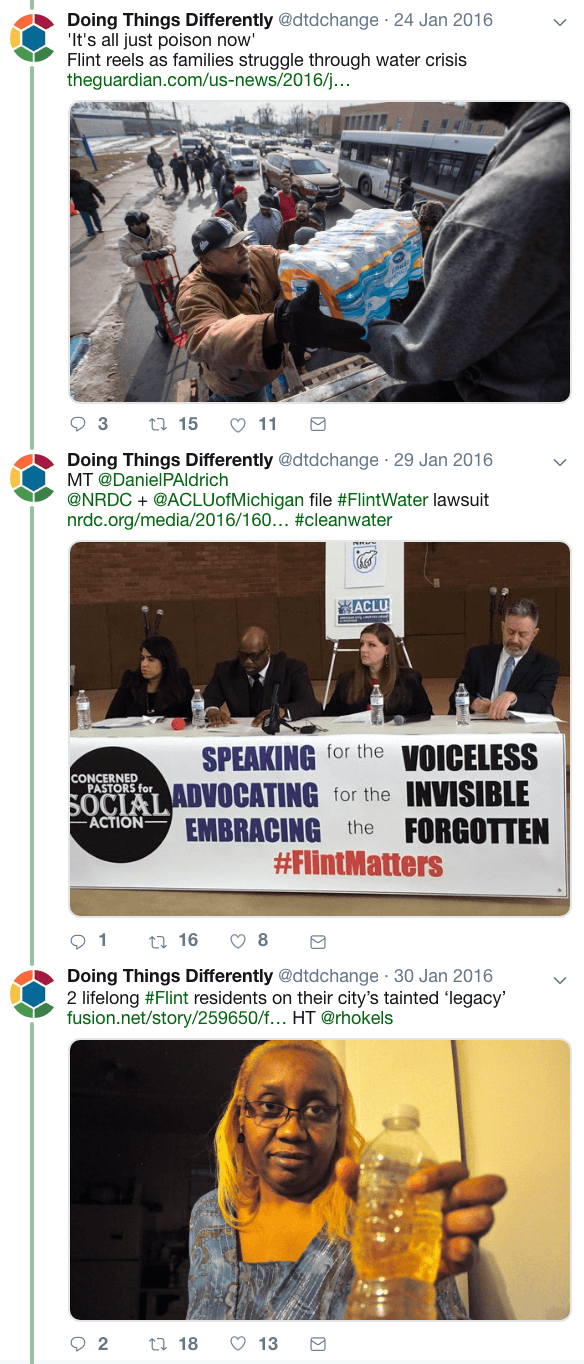}} }}
    \caption{Example of micro-collections from Twitter by a single author (\texttt{@dtdchange}) consisting of a pair of three tweets that are part of a reply thread \cite{dtdchangeBigThread} about the \textit{Flint water crisis}. This micro-collection is of post class \textbf{P$_n$A$_1$} since it consists of multiple Posts from a single Author.}%
    \label{fig:mcsTwitter}%
\end{figure*}

\begin{figure*}[h!]
    \centering
    \subfloat[Part 1: Reddit post]{{ \fbox{\includegraphics[width=0.46\textwidth]{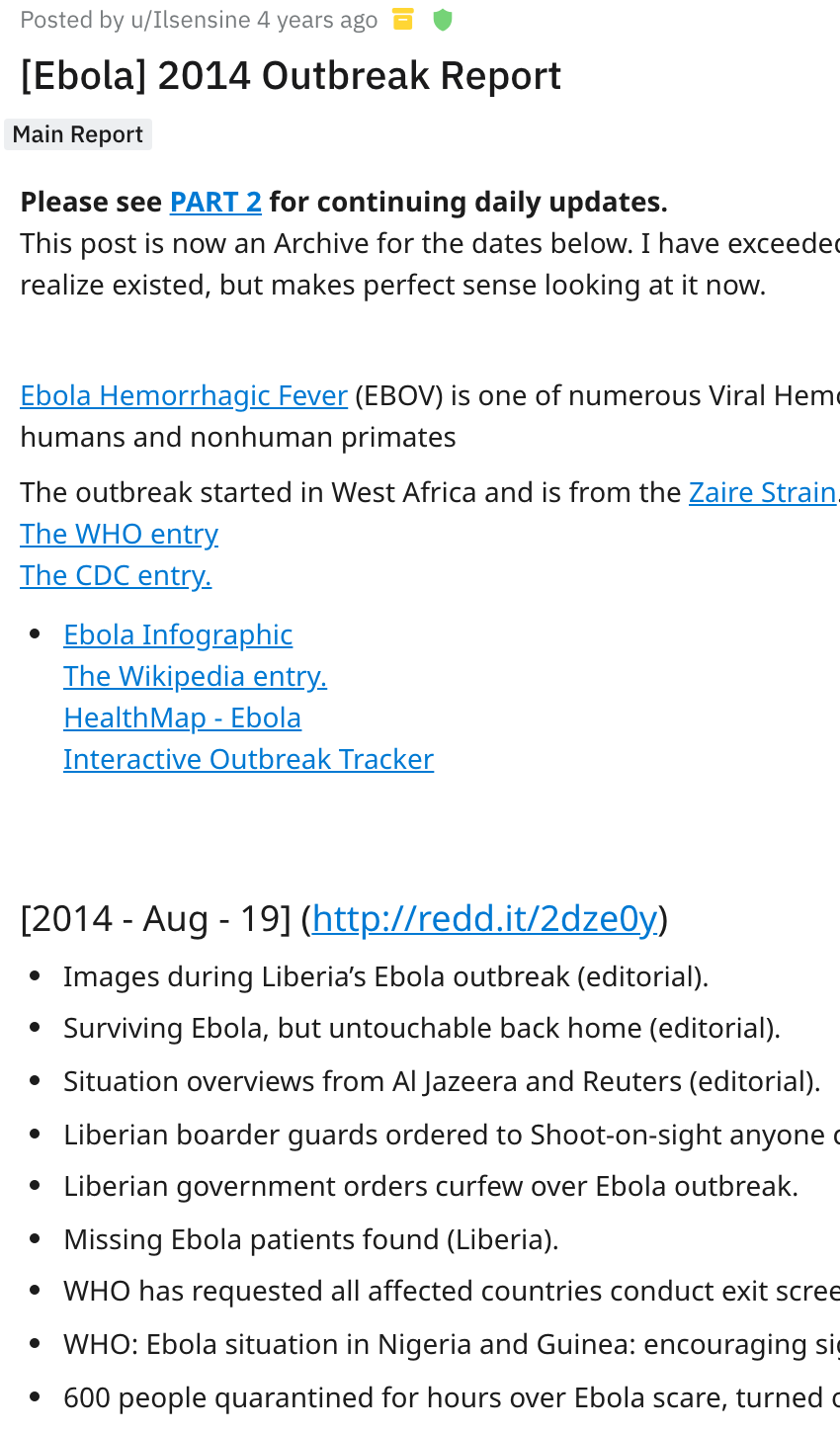}} }}
    \,
    \subfloat[Part 2: Reddit post]{{ \fbox{\includegraphics[width=0.46\textwidth]{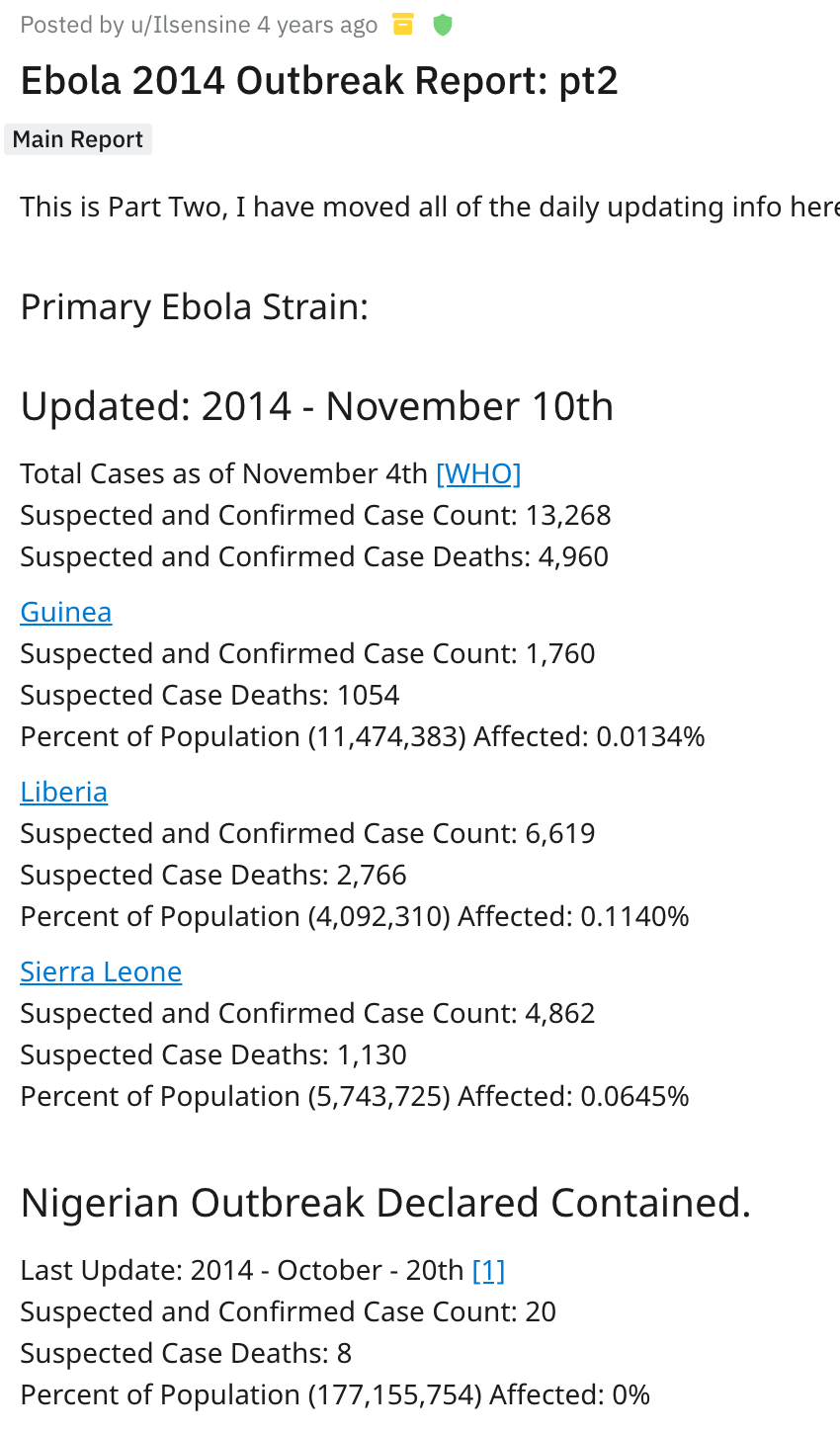}} }}
    \caption{Example of a pair of micro-collection Reddit posts \cite{IlsensinePost0, IlsensinePost1} consisting of 102 external references for the 2014 Ebola outbreak. Both micro-collections are of type \textbf{P$_1$A$_1$} (single Posts from a single Author).}%
    \label{fig:mcsReddit}%
\end{figure*}

\begin{figure}
    \centering
    \begin{tabular}{c}
      \fbox{\includegraphics[width=0.45\textwidth]{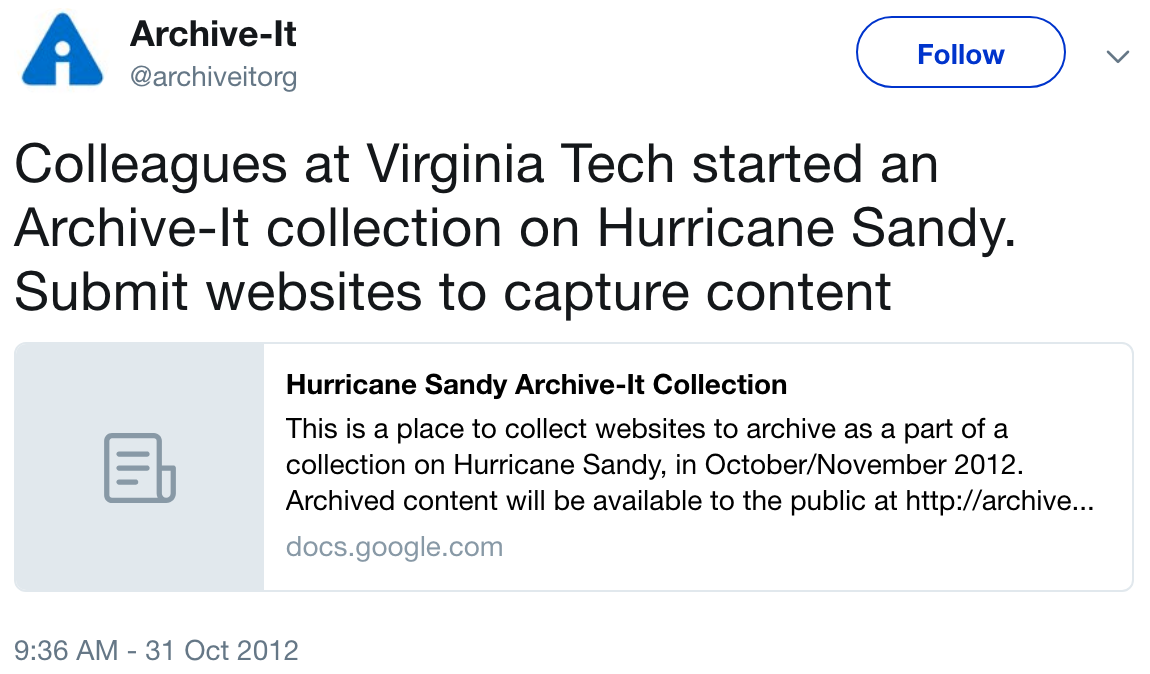}}
    \end{tabular}
    \caption{A tweet from Archive-It requesting seeds for the \textit{2012 Hurricane Sandy} collection \cite{twtReq1}.}%
    \label{fig:IAColTwtRequest}%
\end{figure}
In an attempt to save the digital history of unfolding world events before they are lost due to link rot \cite{KleinJCDL2018, zittrain2014perma, bar2004sic}, we often see the creation of Web archive collections following the occurrence of a major news event. For example, an Archive-It Ebola virus collection \cite{ChristieEbolaVirus} was created months after the 2014 Ebola outbreak. It consists of groups of webpages of government organizations and public health care workers associated with the Ebola outbreak event. However, some important events occur without the creation of Web archive collections. For example, on February 14, 2018, there was a shooting that claimed the lives of 17 people at the Marjory Stoneman Douglas (MSD) High School in Florida. In the aftermath of the tragic event, the teenage students boldly stepped into the highly politically divisive gun control debate demanding stricter gun control measures \cite{stonemanDemandGunControl0, stonemanDemandGunControl1}. Less than two weeks after the shooting, major retailers Walmart and Dick's Sporting Goods increased the minimum age required to purchase firearms and ammunition from 18 to 21 \cite{walmartDickIncreaseGunAge0}. Dick's additionally discontinued the sale of assault-style rifles, high capacity magazines, and bump stocks. On March 9, 2018, Governor Rick Scott of Florida signed the \textit{Marjory Stoneman Douglas High School Public Safety Act} bill into law. Among other gun control measures, it raised the minimum age for buying rifles to 21, banned bump stocks, and instituted background checks. The ripple effects of the activism of the MSD students is still being felt, and most would agree that this incident deserves highlight as part of the broader gun control discourse in the US, thus worthy of a Web archive collection. However, one year after the shooting there is still no corresponding Archive-It collection. Any subsequent collection would likely not be able to collect all the contemporary resources shared in social media or even reported in the news \cite{salaheldeen2012losing, NwalaJCDL2018}.

\begin{figure*}[h]
  \centering
  \fbox{\includegraphics[width=0.98\textwidth]{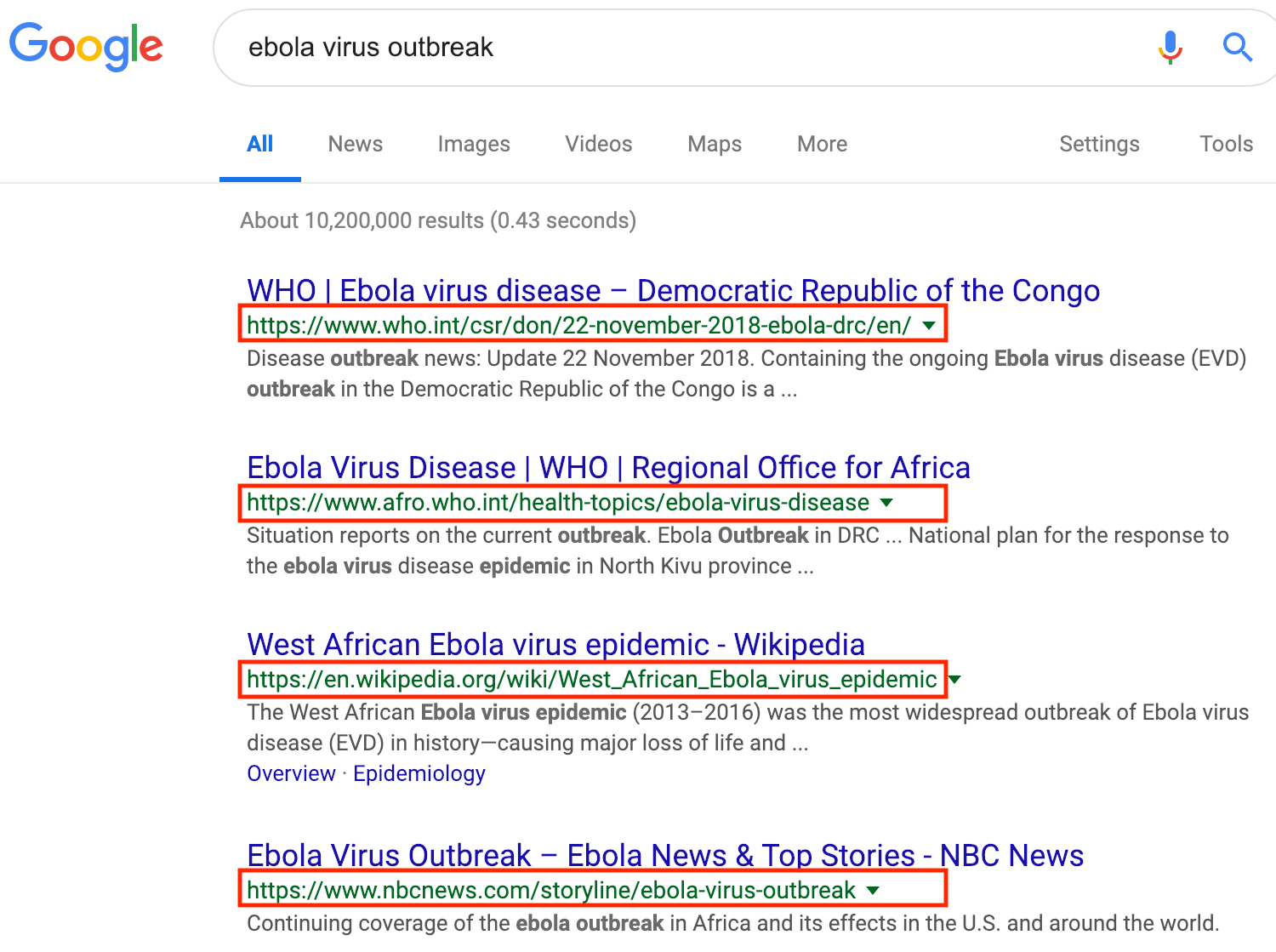}}

  \caption{Potential seeds (red annotation) from Google (All SERP) for query: ``ebola virus outbreak.'' Seeds can be extracted by issuing queries to SERPs and scraping the links returned. This post has been edited to show more details.}
  \label{fig:scrapingSERPsGoogle}
\end{figure*}
\begin{figure*}[h]
  \centering

  \fbox{\includegraphics[width=0.98\textwidth]{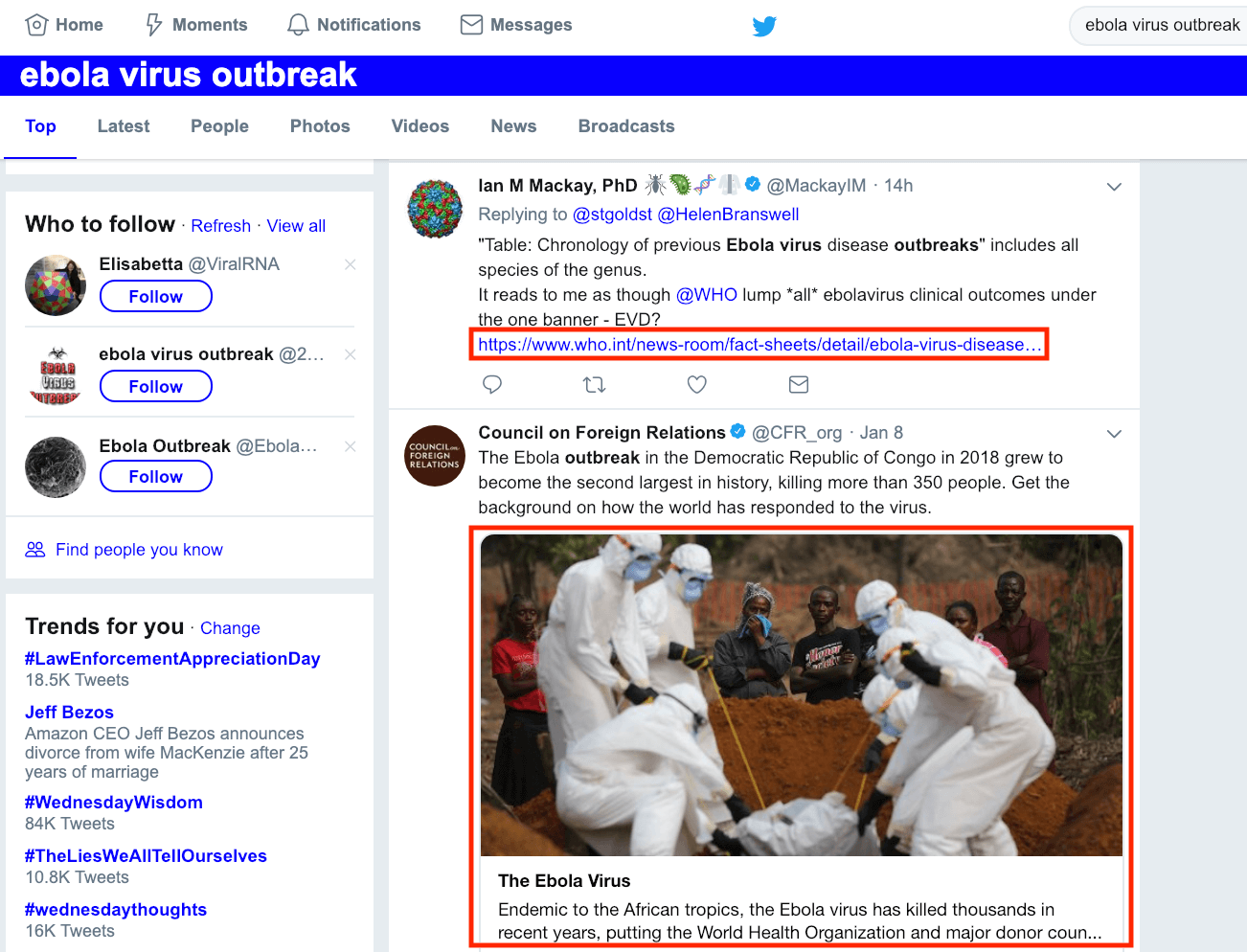}}
  \caption{Potential seeds (red annotation) from Twitter (Top SERP) for query: ``ebola virus outbreak.'' Seeds can be extracted by issuing queries to SERPs and scraping the links returned. This post has been edited to show more details.}
  \label{fig:scrapingSERPsTwitter}
\end{figure*}

The MSD shooting example illustrates gaps in Web archive collections for important events. One major reason for the lack of Web archive collections for important events is tied to how Web archive collections are created. Web archive collections begin with high quality seeds URIs (Uniform Resource Identifiers) selected by curators, a time consuming process often done manually. Amidst an abundance of important local and global events, various organizations such as the Internet Archive cope with the shortage of curators by routinely requesting (Fig. \ref{fig:IAColTwtRequest}) for users to contribute links (seeds) to Archive-It collections, e.g., the \textit{2012 Hurricane Sandy} \cite{twtReq1}, and the \textit{2013 Boston Marathon Bombing} \cite{twtReq0} collections. But this crowd-sourced approach to collection building, while useful, is not enough. In some other cases, archived collections are initiated months or years after the precipitating event. This could have serious consequences since Web archive collections that start late could omit webpages that address the early stages of events \cite{NwalaJCDL2018, kleinWebSci}. Consequently, it is important to start collecting seeds for Web archive collections early. This calls for a method for generating seeds automatically and on demand. Two prominent sources for automatically generating seeds have been adopted for extracting (scraping) links over the years: Web (e.g., Google, Fig. \ref{fig:scrapingSERPsGoogle}) and social media (e.g., Twitter, Fig. \ref{fig:scrapingSERPsTwitter}) Search Engine Result Pages (SERPs).

In this work, we explore a new source (we call \textit{micro-collections}) for generating seeds beyond URIs returned by SERPs. 
It is important to note that our proposed method is not just concerned about finding what a search engine such as Google may find. Even though search engines often produce quality seeds, our micro-collection method of generating seeds is more concerned about finding quality, ``hard-to-find,'' and heterogeneous seeds that may not be popular enough to be easily retrievable through a simple Web search. We define a micro-collection as a post or group of social media posts that exhibit some properties associated with collection building. Web archive curators spend time selecting and filtering seed URI candidates. Similarly, social media users often perform similar tasks when faced with the decision of choosing what URIs to include in a ``non-standard'' social media post. For example, the Twitter account \textit{Doing Things Differently} (\texttt{@dtdchange}) \cite{dtdchange} created a chain of tweets (Fig. \ref{fig:mcsTwitter} \cite{dtdchangeBigThread}) by replying to each subsequent tweet in order to chronicle the \textit{Flint water crisis} story. This reply thread spans almost 3 years and consists of 75 tweets (as of April 8, 2019) each containing a URI. These tweets exhibit curatorial discretion (selection and filtering), and thus we consider the thread a micro-collection for the \textit{Flint water crisis} story. Another example of micro-collections are Reddit posts (Fig. \ref{fig:mcsReddit} \cite{IlsensinePost0, IlsensinePost1}) created by the user \textit{Ilsensine} \cite{Ilsensine} for the 2014 \textit{Ebola virus outbreak} story. In total, the posts contain over 102 external references and were published less than two weeks after the World Health Organization (WHO) declared the 2014 Ebola outbreak a Public Health Emergency of International Concern \cite{whoDeclarationEbola}. We distinguish micro-collections from standard social media posts by showing that micro-collections can be identified by considering the properties of the posts. The rationale for considering micro-collections as a good source for seeds is that the effort taken to create micro-collections is an indication of editorial effort and a demonstration of domain expertise.

Conventional techniques use SERPs from search engines and social media to extract seeds. Since seeds highly influence the nature of collections generated after the seeds are crawled, we consider it pertinent to understand the nature of the seeds returned from the services often used to generate seeds. Accordingly, we conducted a study to investigate the nature of the seeds generated from different sources on popular social media sites (Reddit and Twitter) and a less popular social media site (Scoop.it). First, we created a classification called \textit{post class} from four pairs (\textbf{P$_1$A$_1$}, \textbf{P$_1$A$_n$}, \textbf{P$_n$A$_1$}, \textbf{P$_n$A$_n$} - Table \ref{tab:postClasses}) of acronyms for identifying social media posts regardless of platform. A post class is formed by combining two acronyms, \textbf{P} and \textbf{A}, with subscripts (1 - single or $n$ - multiple), both combined to represent the count of \textbf{P}osts and \textbf{A}uthors, respectively. Second, we generated 23,112 collections of seeds extracted from the various post classes by issuing five queries against the following social media sources: Reddit, Twitter (and Twitter Moments\footnote{A service launched by Twitter (in October 6, 2015) that enables users to collect and share tweets of noteworthy events as they unfold. A collection of tweets is called a moment.}), and Scoop.it. In total we collected 120,444 URIs from 449,347 social media posts. Third, for a combination of social media and post classes, we studied the resultant collections across the following dimensions: the distribution of links, the probability distribution of URI counts for various post classes, the precision of the seeds, ages of webpages, the diversity of seed hostnames, and overlap with the Google SERPs.

Our study resulted in the following contributions that collectively provide some insight on the nature of seeds generated from various social media post classes. First, the provision of a simple cross-platform vocabulary (post class) for describing social media posts facilitates comparing different social media posts across different platforms, ranging from tweets (\textbf{P$_1$A$_1$}) on Twitter to Reddit (\textbf{P$_1$A$_1$}) posts. 
\begin{table*}
   \setlength{\tabcolsep}{1pt}
   \centering
   \caption{Post Classes for Social Media Posts. All non-\textbf{P$_1$A$_1$} collections are combined to create Micro-Collections (MC).}
   \begin{tabular}{|c|c|c|c|}
          \hline
          \textbf{Acronym} & \textbf{Post Count} & \textbf{Author Count} & \textbf{Definition/Example}  \\ \hline
          \textbf{P$_1$A$_1$} & Single (1) & Single (1) & \makecell[l]{A Single \textbf{P}ost from a single \textbf{A}uthor, e.g., an isolated tweet or post on Reddit (Fig. \ref{fig:mcsReddit}) or\\Facebook. These posts are visible to seeds generators that scrape SERPs.} \\ \hline

          \textbf{P$_1$A$_n$} & Single (1) & Multiple (n) & \makecell[l]{Single \textbf{P}ost from multiple \textbf{A}uthors, e.g., the references contributed by multiple Wikipedia editors.} \\ \hline

          \textbf{P$_n$A$_1$} & Multiple (n) & Single (1) & \makecell[l]{Multiple \textbf{P}osts from a single \textbf{A}uthor, e.g., a thread of tweets (Fig. \ref{fig:mcsTwitter}) from a Twitter user.} \\ \hline

          \textbf{P$_n$A$_n$} & Multiple (n) & Multiple (n) & \makecell[l]{Multiple \textbf{P}osts from multiple \textbf{A}uthors, e.g., a tweet conversation consisting of multiple tweets\\or posts from different Twitter or Reddit (or Facebook) users.} \\ \hline
   \end{tabular}
   \label{tab:postClasses}
\end{table*}
\begin{table*}
\setlength{\tabcolsep}{1pt}
\centering
\caption{Temporal characteristics of dataset topics}
\begin{tabular}{|c|c|c|c|c|}
  \hline
  & & & \multicolumn{2}{c|}{ \textbf{Occurrence definition} }                                                                                                                                                                                \\ \hline
  \textbf{Topic [Wikipedia Page]} & 
  \multicolumn{1}{c|}{\textbf{ \makecell{Expectation\\ (Expected/Unexpected)} }} & 
  \multicolumn{1}{c|}{\textbf{ \makecell{Recurrence\\ (Recurring/Non-Recurring)} }} & 
  \multicolumn{1}{c|}{\textbf{ \makecell{Start definition\\ (Defined/Undefined)} }} & 
  \multicolumn{1}{c|}{\textbf{ \makecell{End definition\\ (Defined/Undefined)} }} \\ \hline
  
  Ebola Virus Outbreak \cite{wikiEbola} & 
  Unexpected & 
  Recurring (Irregular) & 
  December 2013 \cite{cdcEbolaChronology} &
  June 2016 \cite{cdcEbolaChronology} \\ \hline

  Flint Water Crisis \cite{wikiFlint} & 
  Unexpected & 
  Non-Recurring & 
  March 2014 \cite{DRobbins} &
  Undefined \\ \hline

  MSD Shooting \cite{wikiParkland} & 
  Unexpected & 
  Non-Recurring & 
  February 14, 2018 &
  February 14, 2018 \\ \hline

  2018 World Cup \cite{wikiWorldCup} & 
  Expected & 
  Recurring & 
  June 14, 2018 &
  July 15, 2018 \\ \hline

  2018 Midterm Elections \cite{wikiUSElections} & 
  Expected & 
  Recurring & 
  November 6, 2018 &
  November 6, 2018 \\ \hline

\end{tabular}
\label{tab:datasetQueries}
\end{table*}
Second, we introduced micro-collections (\textbf{MC}s) as social media posts that exhibit properties associated with collection building, and proposed generating seeds from them. \textbf{MC}s are formed by combining seeds from \textbf{P$_1$A$_n$}, \textbf{P$_n$A$_1$}, and \textbf{P$_n$A$_n$}. Seeds generated from scraping SERPs belong to the \textbf{P$_1$A$_1$} post class. We showed that seeds generated from social media sources are not easily discoverable from Google. Third, we provided a means of characterizing and comparing seeds generated from different post classes. Fourth, we showed that \textbf{MC}s produced more seeds than \textbf{P$_1$A$_1$}, but \textbf{P$_1$A$_1$} had a higher median probability (0.63) of producing relevant URIs than \textbf{MC}s (0.5) for all social media and SERP combination excluding seeds generated with hashtags. Finally, we showed that the ages of webpages extracted depends on a combination of features such as the topic and SERP vertical (e.g., Top vs. New). Similarly, we showed that the diversity of seed hostnames varied with post classes. These findings may provide useful information to curators using social media to generate seeds. For example, if a seed generation process prioritizes quantity of URIs (HTML and non-HTML), the curator may consult \textbf{MC}s first. However, if precision is the priority, then \textbf{P$_1$A$_1$}. Our research dataset comprising of 120,444 links extracted from 449,347 social media posts, as well as the source code for the application utilized to generate the seeds, are publicly available \cite{jcdl2019Repo}.
\section{Related work}
\label{sec:relatedwork}
The collection building process starts with seeds. The seeds are fed into a focused crawler's crawl frontier to start the process of discovering more Web resources related to the collection topic. Chakrabarti et al. \cite{chakrabarti1999focused} introduced the first focused crawler in the 1999s as a means to build collections for specific topics, as opposed to a general-purpose crawler which does not take the topics of the documents under consideration during the crawling process. Since the first focused crawlers, there have been many variants of focused crawlers. Bergmark \cite{bergmark2002collection} used a focused crawler to crawl and classify webpages into various topics in science, mathematics, engineering and technology, discarding off-topic pages. Farag et al. \cite{farag2017focused} introduced the Event Focused Crawler, a focused crawler for events that uses an event model to represent documents and a similarity measure to quantify the degree of relevance between a candidate URI and a collection. An event is represented as a triple - \textit{Topic}, \textit{Location}, and \textit{Date}. Similar to Farag et al., Risse et al. \cite{risse2014arcomem} introduced a new crawler architecture based on the ARCOMEM project. Instead of the conventional crawling of all webpages, ARCOMEM performs a semantic crawl of only webpages related to \textit{events} and \textit{entities} such as persons, locations, and organizations. Most focused crawling is performed on the live Web. Unfortunately, the live web is plagued by link rot and content drift, consequently, Klein et al. \cite{kleinWebSci} demonstrated that focused crawling on the archived Web results in more relevant collections than focused crawling on the live Web, for events that occurred in the distant past. Additionally, Klein et al. proposed extracting seeds from external references contained in the Wikipedia page of an event. We consider Wikipedia references examples of \textbf{P$_1$A$_n$} micro-collections. Our focus in this work is the seed generation process, therefore we did not utilize a focused crawler. Instead we explored the various sources for extracting seeds from social media posts.

Not all collection building uses focused crawling. Gossen et al. \cite{gossen2016analyzing} proposed a methodology for extracting sub-collections from Web archive collections focused on specific topics and events (called the \textit{topic and event focused sub-collection}). The \textit{topic and event focused sub-collection} is defined as a collection of documents in a Web archive collected using a \textit{sub-collection specification}. Our research differs from Gossen et al. in two major ways. First, Gossen proposes generating collections from within the Web archives, but we propose generating seeds from the live social Web. Second, Gossen proposed running an algorithm over a sub-collection specification on a Web archive to generate a sub-collection. This means the decision of whether a URI belongs in a sub-collection is encoded in the specification of an algorithm. However, in this work, we leverage the judgment of humans on social media.

In a similar work, Gossen et al. \cite{gossen2017extracting} adapted some portions of the \textit{topic and event focused sub-collection} in a method to extract event-centric documents from Web archives based on a specialized focused extraction algorithm. They defined two broad kinds of events based on time: \textit{planned} and \textit{unexpected}. The goal of the event-centric extraction process is, given an event input and a Web archive, generate an interlinked collection of documents relevant to the input event that meet the \textit{collection specification}. The differences of our research with Gossen's previous work \cite{gossen2016analyzing} transfer to this work. However, we adapted Gossen's categorization of events as either \textit{planned} or \textit{unexpected}, and we renamed \textit{planned} to \textit{expected} (Table \ref{tab:datasetQueries}). Similar to Gossen et al., Nanni et al. \cite{nanni2018toward} presented an approach for extracting event-centric sub-collection from Web Archives. Their method extracts documents not only related to the input event, but also documents describing related events (e.g., premises and consequences). Nanni et al.'s method utilized Wikipedia pages as inputs to generate event-centric collections. In this work, however, we used Wikipedia references to generate our gold standard dataset.

Selecting good seeds is challenging and has not been extensively studied. Collection building researches often acknowledge the importance of selecting good seeds, and admit its link to the performance of their systems, but often they pay more attention on the mechanisms of building the collection, and not seed selection. The challenge of selecting good seeds is embodied in the idea that it is difficult to define ``good.'' This challenge is captured by Bergmark's statement \cite{bergmark2002collection}: ``It is unclear what makes a good seed URL, but intuitively it should be rich in links, yet not too broad in scope.'' Zheng et al. \cite{zheng2009graph} argued that the seed selection problem for Web crawlers is not a trivial, and proposed different seed selection strategies based on PageRank, number of outlinks, and website importance. They also showed that different seeds may result in collections that are considered ``good'' or ``bad.'' While there have been efforts made to automatically generate seeds, many of these methods (e.g., Prasath and \"Ozt\"urk \cite{prasath2011finding}) target generating seeds for Web crawlers that build indexes for search engines, and not seeds for focused crawlers or Web archive collections.

Du et al. \cite{du2014approach} proposed a customized method of generating seeds for focused crawlers based on user past Web usage information that captures the interests of the user. Since this method depends on historical use information, its performance is tied to the availability of such historical data, which might be lacking due to the absence of domain knowledge or privacy concerns. As part of the Crisis, Tragedy, and Recovery Network project, Yang et al. \cite{yang2012study} proposed using URIs found in tweet collections (generated with hashtags and keywords) as seeds to bootstrap Web archiving tasks quickly for sudden emergencies and disasters. Similarly, we consider extracting seeds from tweets, but expand the areas for extracting seeds beyond scraping Twitter SERPs. Additionally, we identify post classes of tweets as part of an effort to characterize the nature of seeds generated from different post classes (Table \ref{tab:postClasses}). Priyatam et al. \cite{priyatam2014seed} proposed extracting diverse seeds from tweets in a Twitter URI graph for the Web crawlers of digital libraries such as CiteSeerX. Even though their work does not target the generation of seeds for collections of stories and events, which is a focus of our work, the notion of diversity of seeds is adopted in our work (Section \ref{subsec:uriDiv}).

In previous work \cite{NwalaHT2018}, we showed that collections generated from social media sources such as Reddit, Storify, Twitter, and Wikipedia are similar to Archive-It collections across multiple dimensions such as the distribution of sources and topics, content and URI diversity, etc. These findings suggest that curators may consider extracting URIs from these sources in order to begin or augment collections about various news topics. Here, we adopt a subset of the dimensions for comparing collections. Similarly, in another previous work \cite{NwalaJCDL2018} as part of an effort to understand the behavior of SERPs, a popular source for generating seeds, we investigated ``refinding'' news stories on the Google SERP by tracking the URIs returned from Google, everyday for over seven months. We discovered that the probability of finding the same URI of a news story diminished drastically after a week (0.01 -- 0.11). These findings suggest it becomes more difficult to find the same news story with the same query on the Google SERP. Therefore, collection building efforts that scrape SERPs are highly sensitive to the query issue dates.

\section{Research questions}
Before generating seeds from micro-collections, we must first identify them. This leads to our first research question: 
\begin{itemize}
  \item \textbf{RQ1: How do we identify, extract, and characterize micro-collections in social media?}
\end{itemize}
Identifying micro-collections makes it easier to accurately describe and extract them. Subsequently, it would be important to quantify the amount of micro-collections relative to conventional social media posts. As part of proposing the extraction of seeds from micro-collections, it is pertinent to verify if they are prevalent on the web.

There are currently two popular sources for automatically or semi-automatically generating seeds. The first involves extracting seeds from SERPs (e.g., Google). The second involves extracting seed URIs from tweets surfaced by hashtags or text queries on Twitter. We propose a third source for extracting seeds - extracting seeds from micro-collections. Therefore, it is important that we compare the new source to the previous popular sources. Such comparison could enable us understand if these sources are similar, and such information would be highly informative to future collection building processes. This leads to our second research question:
\begin{itemize}
  \item \textbf{RQ2: Do seeds from micro-collections differ from seeds from SERPs?}
\end{itemize}
\section{Methodology}
Here we explain considerations made in the selection of our dataset topics, the dataset generation process, the measures extracted from the dataset and how they informed our research questions.
\subsection{Topic selection}
A central objective of our research was to outline the characteristics of, and differences between, collections generated by scraping SERPs (\textbf{P$_1$A$_1$} post class - Table \ref{tab:postClasses}) and micro-collections (\textbf{MC} post class). Therefore, the choice of queries was not arbitrary. Instead, we developed a temporal classification system (partly informed by Gossen et al. \cite{gossen2017extracting}) of real world stories and events based on three temporal (Table \ref{tab:datasetQueries}) attributes: \textit{Expectation}, \textit{Recurrence}, and \textit{Occurrence definition} - Start and End date definitions. A story can be described by a combination of different states of the temporal attributes. 

For the expectation attribute, an event may be \textit{expected} or \textit{unexpected}. For example, the \textit{Ebola outbreak} event was unexpected. Thus we classify this event as an unexpected event. For the recurrence attribute, an event may occur repeatedly at regular or non-regular intervals. For example, the FIFA World Cup tournaments recur at four-year intervals, thus we consider this event a recurring event. Ebola outbreaks in general may also be considered a recurring event, even though they occur at irregular intervals. For the occurrence definition attribute, an event may have a defined or undefined start and end date. For example, the \textit{MSD Shooting} event started and ended the same day (February 14, 2018), but the \textit{Flint water crisis} event started in April 2014, and is still ongoing (no end definition).

Following the specification of the temporal classification system, we selected five topics (Table \ref{tab:datasetQueries}) specified by the following queries and hashtags (for Twitter):
\begin{enumerate}
  \item ``ebola virus outbreak'' (\#ebolavirus)
  \item ``flint water crisis'' (\#FlintWater)
  \item ``stoneman douglas high school shooting'' (\#MSDStrong)
  \item ``2018 world cup'' (\#WorldCup)
  \item ``2018 midterm elections'' (\#election2018)
\end{enumerate}
In addition to text queries, for Twitter, we selected hashtag queries for each topic to discern if seeds generated with text-based queries differ from those extracted with hashtag queries.
\footnotesize
\begin{table}
   \setlength{\tabcolsep}{0.6pt}

   \centering
   \caption{Post class counts (Class), Social media posts (Posts), and URI counts (URIs) for dataset generated by extracting URIs from post classes (\textbf{P$_1$A$_1$}, \textbf{P$_n$A$_1$}, and \textbf{P$_n$A$_n$}) of Reddit, Twitter, Twitter Moments, and Scoop.it. The Micro-collection (MC) post class is formed by combining posts in \textbf{P$_n$A$_1$} and \textbf{P$_n$A$_n$} post classes.}
   \begin{tabular}{|c!{\vrule width 2pt}c|c|c!{\vrule width 2pt}c|c|c!{\vrule width 2pt}c|c|c|}
          \hline      
          
           &
          \multicolumn{3}{c!{\vrule width 2pt}}{  } & 
          \multicolumn{6}{c|}{ \textbf{Micro-collections (MC)} } \\ \hline

           &
          \multicolumn{3}{c!{\vrule width 2pt}}{ \textbf{P$_1$A$_1$ Counts} } & 
          \multicolumn{3}{c!{\vrule width 2pt}}{ \textbf{P$_n$A$_1$ Counts} } & 
          \multicolumn{3}{c|}{ \textbf{P$_n$A$_n$ Counts} } \\ \hline

           &
          \textbf{ Class } &
          \textbf{ Posts } &
          \textbf{ URIs }  &
          \textbf{ Class } &
          \textbf{ Posts } &
          \textbf{ URIs }  &
          \textbf{ Class } &
          \textbf{ Posts } &
          \textbf{ URIs } \\ \hline

          \makecell{Reddit\\Relevance\\\includegraphics[width=0.5cm]{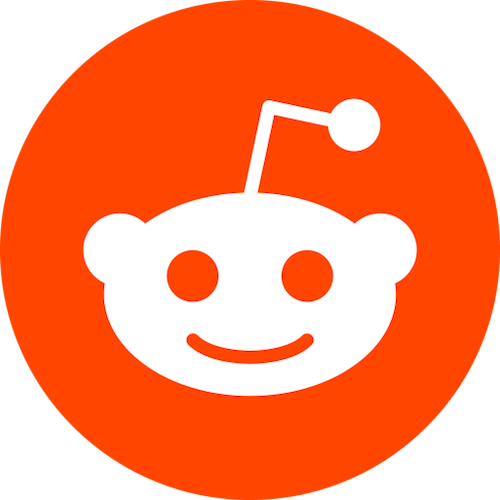}} &
          766     &
          766     &
          1,776   &

          56      &
          115     &
          206     &
          
          542     &
          36,124  &
          3,387    
          \\ \hline

          \makecell{Reddit\\Top\\\includegraphics[width=0.5cm]{redditLogo}} &
          
          931     &
          931     &
          10,857  &
          
          37      &
          177     &
          319     &
          
          1,021    &
          100,006  &
          18,992    
          \\ \hline

          \makecell{Reddit\\New\\\includegraphics[width=0.5cm]{redditLogo}} &
          854     &
          854     &
          8,056   &

          26      &
          68      &
          1,062   &

          340       &
          9,298     &
          6,412      
          \\ \hline

          \makecell{Reddit\\Comments\\\includegraphics[width=0.5cm]{redditLogo}} &
          834     &
          834     &
          8,381   &

          53      &
          423     &
          691     &

          1,077   &
          117,378 &
          18,781   
          \\ \hline

          \makecell{Twitter\\Top\\\includegraphics[width=0.5cm]{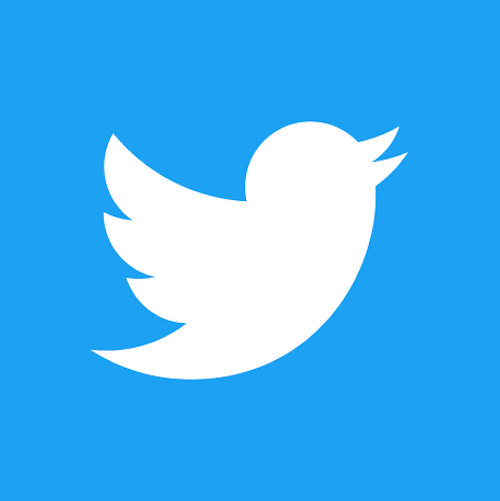}} &
          2,936    &
          2,936    &
          3,548    &

          540      &
          4,983    &
          3,026    &

          4,009    &
          79,347   &
          12,457    
          \\ \hline

          \makecell{Twitter\\Latest\\\includegraphics[width=0.5cm]{twitterLogo}} &
          2,341    &
          2,341    &
          2,792    &
          
          639      &
          6,366    &
          3,628    &
         
          4,471    &
          82,499   &
          13,576    
          \\ \hline

          \makecell{Twitter\\Moments\\\includegraphics[width=0.5cm]{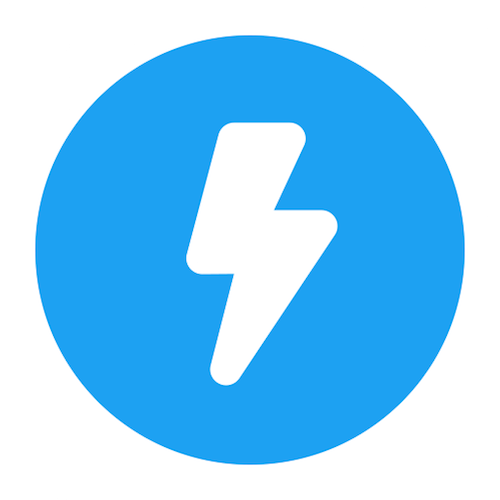}} &
          NA      &
          NA      &
          NA      &

          NA      &
          NA      &
          NA      &

          73      &
          1,285   &
          621      
          \\ \hline

          \makecell{Scoop.it\\\includegraphics[width=0.5cm]{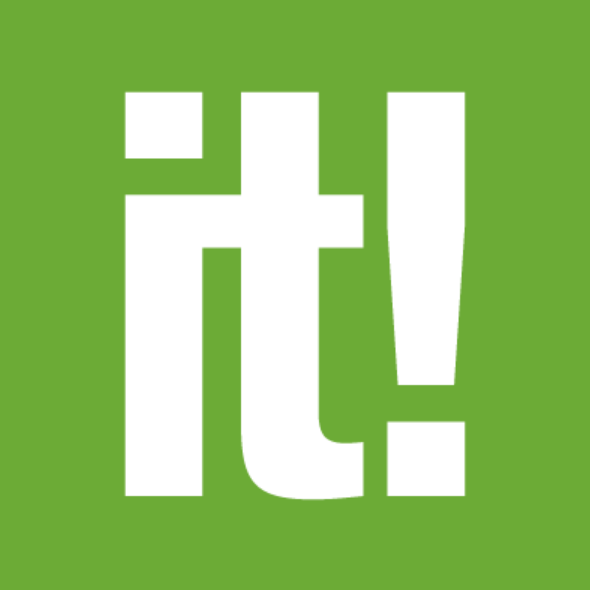}} &
          1,533    &
          1,533    &
          1,533    &

          33      &
          1,083   &
          343     &

          NA      &
          NA      &
          NA       
          \\ \hline

          \textbf{Subtotal}  &
          10,195    &
          10,195    &
          36,943    &

          1,384     &
          13,215    &
          9,275     &

          11,533    &
          425,937   &
          74,226     
          \\ \hline \hline

          \textbf{Total}  &
          \multicolumn{3}{c!{\vrule width 2pt}}{ \textbf{Class: 23,112} } &
          \multicolumn{3}{c!{\vrule width 2pt}}{ \textbf{Posts: 449,347} } &
          \multicolumn{3}{c|}{ \textbf{URIs: 120,444} } \\ \hline

          
   \end{tabular}
   \label{tab:dataset}
\end{table}
\normalsize

\subsection{Dataset generation and segmentation of social media posts into post classes}
\label{sec:datasetSegmentation}
For Reddit, we issued all five queries to four Reddit SERPs (Relevance, Top, New, and Comments), and extracted posts from the SERPs. For each query we extracted a maximum of 500 posts and recursively extracted a maximum of 500 comment replies from each post extracted from the SERP.

For Twitter, similar to Reddit, we issued all five text and hashtag queries to the two Twitter SERPs (Top and Latest), and extracted tweets from the SERPs with the use of the Local Memory Project \cite{nwala2017local} \textit{local news generator} \cite{chromeExtension}. For each query, we extracted a maximum of 500 tweets and recursively extracted a maximum of 500 tweet replies for each tweet extracted from the SERP.

For Reddit and Twitter, the posts directly visible from the SERP were assigned to the \textbf{P$_1$A$_1$} post class. We use the term ``post'' in order to be general. Different social media platforms have different names for posts, for example, on Twitter, a post is called a tweet. Posts with replies were assigned either to the \textbf{P$_n$A$_1$} or \textbf{P$_n$A$_n$} class depending on the number of authors. Posts from the SERP with a reply or a contiguous set of replies exclusively authored by a single user were assigned to the \textbf{P$_n$A$_1$} post class. Finally, posts with a reply or a series of replies authored by multiple users were assigned to the \textbf{P$_n$A$_n$} post class. The \textbf{P$_1$A$_n$} micro-collection post class is rare and not available in Twitter, Reddit, or Scoop.it. However, our gold standard data was extracted from Wikipedia references which belong to \textbf{P$_1$A$_n$}.

For Twitter Moments, we issued all five queries to Google with (``site:twitter.com/i/moments'') in order to restrict the search results to links from Twitter Moments. Next, we extracted Twitter Moments URIs from the first two pages of the Google default SERP. Next, we dereferenced URIs and extracted the tweets. Tweets from Twitter Moments are authored by multiple users, and thus assigned the \textbf{P$_n$A$_n$} label.

In addition to the extraction of posts from well-known social media (Reddit and Twitter), we considered a lesser known social media Scoop.it (\url{https://www.scoop.it/}). Scoop.it is a content curation social media service that enables users to bookmark a single URI (\textit{scoop}) or multiple URIs (\textit{topics}). For Scoop.it, we issued all five queries to the Scoop.it SERPs (Scoops and Topics), and extracted posts (scoops) from the SERPs. The scoops visible from the \textit{Scoops} SERP were assigned to the \textbf{P$_1$A$_1$} post class. For a single dataset topic, the scoops found in the \textit{Topic} SERP were assigned to the \textbf{P$_n$A$_n$} post class since they are authored by multiple users.

From all social media posts, we extracted the URIs to create collections corresponding to the post class from which the URIs were extracted. Social media posts often link to intra-site posts (e.g., tweet URI in a tweet). We dereferenced and extracted seeds from such intra-site URIs, and substituted them with the extracted seeds.

\subsection{Gold standard dataset generation}
\label{sec:gsPrep}
The following steps were taken in order to generate the gold standard dataset to facilitate measuring precision of URI collections extracted from the various post classes.

First, we selected a corresponding Wikipedia page for the five topics (Table \ref{tab:datasetQueries}). Second, we extracted the URIs from the references section of each Wikipedia page. Third, we dereferenced the URIs from each reference corresponding to a topic (e.g., Flint water crisis) and removed the HTML boilerplate leaving only the plaintext documents (stopwords removed). The set of plaintext documents were concatenated into one document. Fourth, for each topic, we created a collection vector consisting of the normalized Term Frequency (TF) weights of the concatenated document.

\subsection{Primitive measures extraction}
\label{sec:primitiveMeasures}

We counted the number of URIs (HTML, non-HTML, and both) per topic, per social media source, and per post class (Table \ref{tab:dataset}). Additionally, we extracted the distribution of posts with URIs by counting the number of posts with a specified number of links for a given social media source (e.g., Reddit) to facilitate probability distribution calculation (Table \ref{tab:proburi}). The distribution answers questions such as: ``for Reddit posts with links, how many posts had 1 link or 2 links?'' Subsequently, the following measures were extracted from the dataset to address our research questions.

\subsubsection{\textbf{Probability distribution of posts with links}}
\label{sec:probDist}\textcolor{white}{.}\\
For all topics $T$ (e.g., \textit{World cup}), given the set of post classes $C \in \{\textbf{P$_1$A$_1$}, MC, \textbf{P$_1$A$_n$}, \textbf{P$_n$A$_1$}, \textbf{P$_n$A$_n$}\}$, given a social media seed source $s$ (e.g., Reddit), the probability $P(p_c^s=k)$ of the event that a post $p_c^s$ of post class $c \in C$ with a URI, has $k$ URIs (e.g., 1 URI) is calculated using Eqn. \ref{eqn:probCalc0}. $P(p_{P_1A_1}^{Reddit} = 1)$ reads: ``What is the probability of the event that a Reddit \textbf{P$_1$A$_1$} post with a URI has one (i.e., $k = 1$) URI?''

The general probability $P(p_{All}^s=k)$ of the event that a post $p_{All}^s$ with a URI from social media $s$ of any post class, has $k$ URIs is calculated using Eqn. \ref{eqn:probCalc1}. In Eqn. \ref{eqn:probCalc0} \& \ref{eqn:probCalc1}, if $c = P_1A_1$, and $t = 1$, $|c_1|$ represents the count of \textbf{P$_1$A$_1$} posts for the first ($t=1$) topic.

\setlength{\multicolsep}{0pt}
\begin{multicols}{2}
  \begin{equation}
    P(p_c^s=k) = \sum_{t=1}^{|T|} \frac{p_{c_t}^s=k}{|c_t|}
    \label{eqn:probCalc0}
  \end{equation}

  \begin{equation}
      P(p_{All}^s=k) = \sum_{t=1}^{|T|} \sum_{c \in C} \frac{p_{c_t}^s=k}{|c_t|}
    \label{eqn:probCalc1}
  \end{equation}
\end{multicols}
\subsubsection{\textbf{Precision of the URIs in post class collections}}\textcolor{white}{.}\\
Given a candidate collection of seed URIs $C$ to be evaluated, the URIs may be extracted from a single post (\textbf{P$_1$A$_1$}) or multiple posts (e.g., \textbf{P$_n$A$_1$}) from a social media site (e.g., Reddit). We calculated the precision of $C$ as follows. First, the URIs in $C$ were processed in the same manner as the gold standard (Section \ref{sec:gsPrep}), i.e., dereferenced and boilerplate removed, and $|C|$ plaintext documents concatenated. Second, a document collection matrix $M$ was created from $C$ and its corresponding gold standard (e.g., Flint water crisis gold standard). The first row of matrix consisted of the gold standard vector, and the second row of the matrix consisted of the vector of $C$ (document to be evaluated). The columns represent the normalize TF weights. Third, cosine similarity was calculated between the pair of rows. If the similarity exceeded relevance threshold of an empirically learned threshold of 0.25, $C$ was declared relevant, otherwise, it was declared non-relevant.

For a given topic (e.g., Flint water crisis) and SERP vertical (e.g., Twitter-Top), a URI or multiple URIs may be extracted from a post authored by a single (\textbf{P$_1$A$_1$}) or multiple (\textbf{P$_n$A$_1$}, \textbf{P$_n$A$_n$}) users. Each group of URIs extracted from a post has an associated precision value (Relevant URIs / Total URIs). The average precision metric for a post class (e.g., \textbf{P$_1$A$_1$}) is an average over all the precision value of all posts in the post class. It provides answers to questions such as: ``what is the average precision of the URIs in the \textbf{P$_1$A$_1$} post class?'' For non-HTML URIs we evaluated precision by extracting text from the post that embedded the URI.

\subsubsection{\textbf{Age distribution of relevant webpages per post class}}
The distribution of ages is an aggregation of the ages of the relevant webpages in a given post class of a given social media. The age of a webpage was calculated by finding the difference between the publication date of a webpage and the date the post containing the webpage URI was retrieved. The publication dates of webpages were extracted with CarbonDate \cite{salaheldeen2013carbon} which estimates the creation date of webpages based on information polled from multiple sources such as the document timestamps, web archives, backlinks, etc. Publication dates of webpages may potentially provide useful information about the kinds of events discussed. For example, the Democratic Republic of Congo in Central Africa has been grappling with another Ebola outbreak (2017-Present). Therefore, webpages published before 2017 are not expected to discuss the 2017 outbreak.
\begin{figure*}[h]
    \centering
    \subfloat[Ebola Virus Outbreak Precision Distribution]{{ \fbox{\includegraphics[width=0.4\textwidth, height=0.43\textwidth]{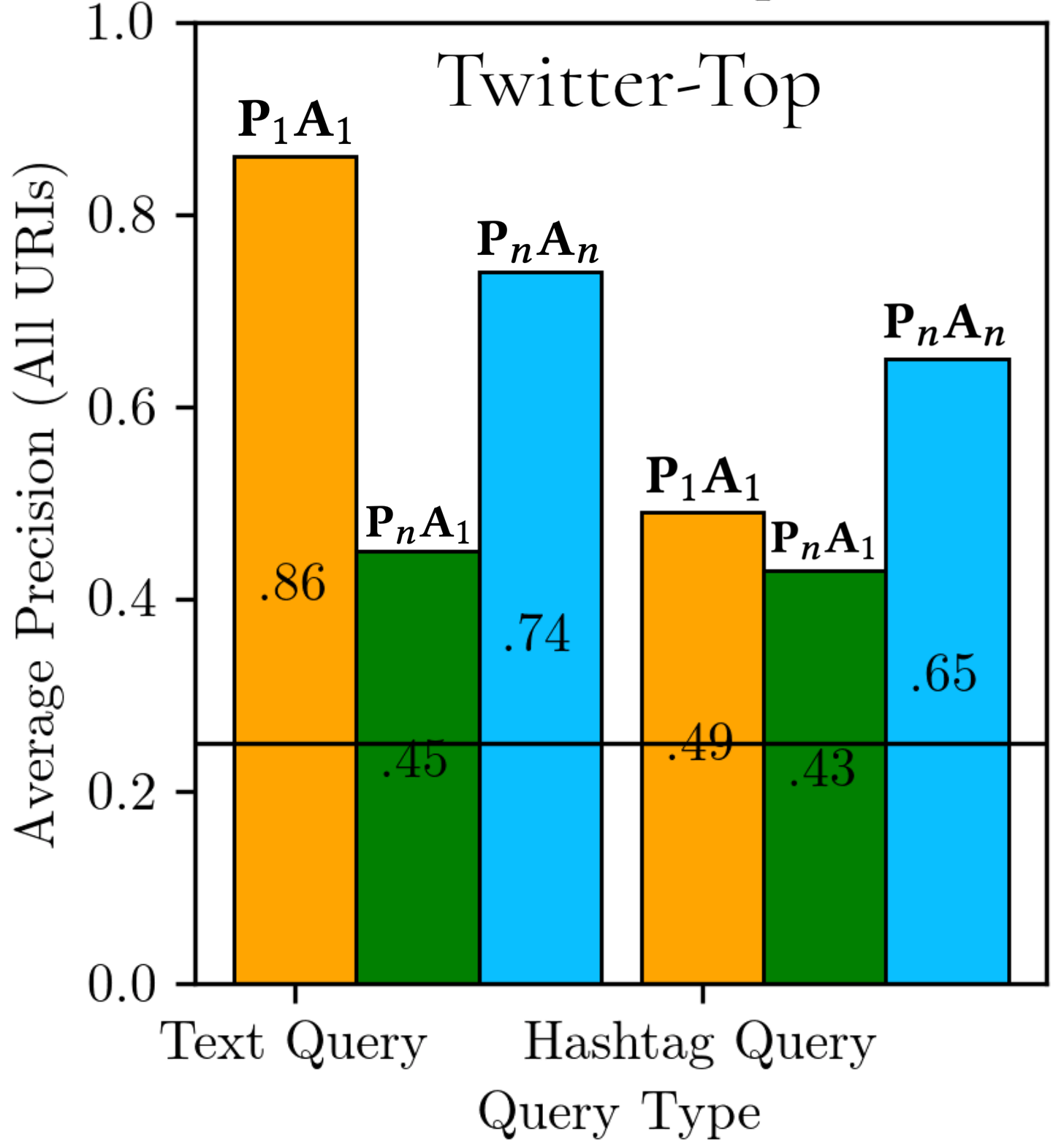}} }}
    \,
    \subfloat[Ebola Virus Outbreak Age Distribution]{{ \fbox{\includegraphics[width=0.4\textwidth, height=0.43\textwidth]{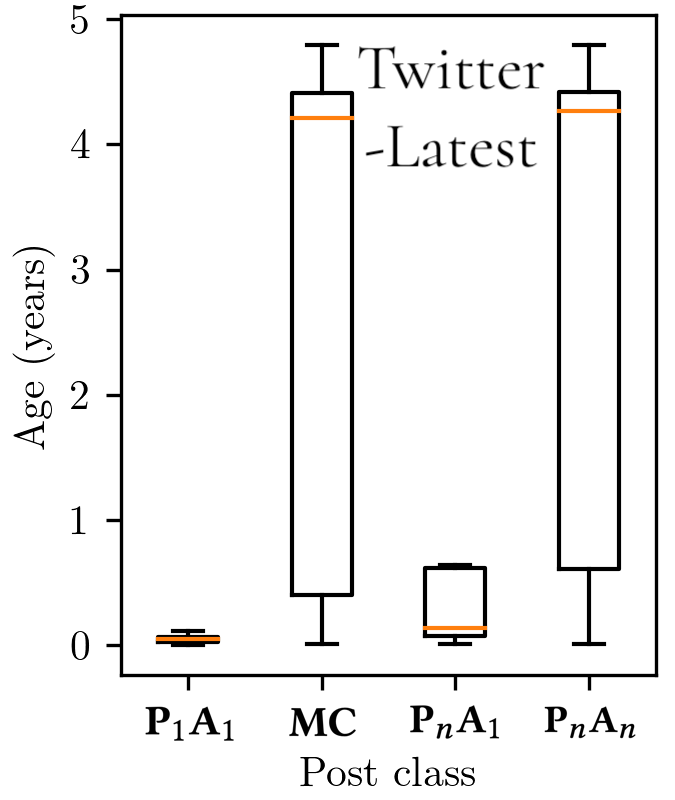}} }}
    
    \caption{a: \textbf{P$_1$A$_1$} seeds produced webpages with a higher precision than \textbf{P$_n$A$_n$} for text but not hashtag queries. The black line marks the relevance threshold. b: MCs produced older webpages in the Twitter-Latest vertical for the older topics. See also Appendices 8 and 9 for additional figures of the age distribution of URIs per post class, per social media.}%
    \label{fig:genericPlots}%
\end{figure*}
\subsubsection{\textbf{Distribution of hostname diversity per post class}}\textcolor{white}{.}\\
\label{subsec:uriDiv}
Given a collection of URIs $C$ for a given post class of a given social media, the hostname diversity \cite{NwalaHT2018} of $C$ is a single value ($d \in [0, 1]$) that reports whether $C$ consists of URIs from a single host ($d = 0.0$, e.g., \textit{www.cnn.com}) or distinct hosts ($d = 1.0$, e.g., \textit{www.cnn.com} and \textit{www.foxnews.com}). It answers questions such as: ``how diverse are the hosts in the Reddit \textbf{P$_1$A$_1$} post class?''

\subsubsection{\textbf{Overlap between Google collections and post class}}
We measured the overlap between URIs extracted from Google and URIs extracted from a combination of social media and post class. This was done in order to determine how easy it was to find the URIs scraped from social media micro-collections. Extracting seeds from micro-collections requires more effort than scraping Web search engine SERPs. For example, generating a collection of URIs of the \textbf{P$_n$A$_1$} or \textbf{P$_n$A$_n$} post class requires independently dereferencing each social media post and extracting the replies from the post. Therefore, if the URIs discovered from micro-collections are easily discoverable via a search engine such as Google, it does not justify the extra effort of extracting seeds from micro-collections.
\footnotesize
\begin{table}
   \centering
   \setlength{\tabcolsep}{0.6pt}
   
   \caption{Probability (e.g., $P(p_{P_1A_1}^{Reddit} = 1)$ = \maxColor{\textbf{0.63}}) of the event that a social media post from a given post class (e.g., Reddit \textbf{P$_1$A$_1$}) has $k$ HTML URIs (e.g., $k=1$). See also Appendix 1 for additional figures of the distribution of the number of links in a post, per post class, per social media.}
   \begin{tabular}{|c!{\vrule width 1pt}c|c|c|c|c!{\vrule width 1pt}c|c|c|c|c!{\vrule width 1pt}c|c|c|c|}
      \hline
      \multicolumn{1}{|c!{\vrule width 1pt}}{ }
      & \multicolumn{5}{c!{\vrule width 1pt}}{ \myfbox{white}{\includegraphics[width=0.5cm]{redditLogo}} }        
      & \multicolumn{5}{c!{\vrule width 1pt}}{ \myfbox{white}{\includegraphics[width=0.5cm]{twitterLogo}} } 
      & \multicolumn{4}{c|}{ \myfbox{white}{\includegraphics[width=0.5cm]{scoopitLogo}} } \\ \hline

      \multicolumn{1}{|c!{\vrule width 1pt}}{ \textbf{k} } & P$_1$A$_1$ & MC & P$_n$A$_1$ & P$_n$A$_n$ & All & P$_1$A$_1$ & MC & P$_n$A$_1$ & P$_n$A$_n$ & All & P$_1$A$_1$ & MC & P$_n$A$_1$ & All \\ \hline
      
      \multicolumn{1}{|c!{\vrule width 1pt}}{ \makecell{1} }      &  \maxColor{\textbf{.63}}        &  .23                      & \maxColor{\textbf{.43}}   & .22                       &   .37                       &  \maxColor{\textbf{.98}}  &  \maxColor{\textbf{.69}}  & \maxColor{\textbf{.60}} & \maxColor{\textbf{.70}}   &  \maxColor{\textbf{.75}}  &  \maxColor{\textbf{1.00}}     &  .21                       &  .21                      &  \maxColor{\textbf{.97}}  \\ \hline
      \multicolumn{1}{|c!{\vrule width 1pt}}{ \makecell{2} }      &  .11                            &  \minColor{\textbf{.12}}  & .13                       & \minColor{\textbf{.12}}   &   \minColor{\textbf{.12}}   &  .02                      &  .17                      & .17                     & .17                       &  .14                      &  .00                          &  \minColor{\textbf{.00}}   &  \minColor{\textbf{.00}}  &  \minColor{\textbf{.00}}  \\ \hline
      \multicolumn{1}{|c!{\vrule width 1pt}}{ \makecell{3-4} }    &  \minColor{\textbf{.06}}        &  .15                      & \minColor{\textbf{.09}}   & .15                       &   \minColor{\textbf{.12}}   &  \minColor{\textbf{.00}}  &  .08                      & \minColor{\textbf{.11}} & .08                       &  .06                      &  .00                          &  .12                       &  .12                      &  .01                      \\ \hline
      \multicolumn{1}{|c!{\vrule width 1pt}}{ \makecell{5+} }     &  .20                            &  \maxColor{\textbf{.50}}  & .35                       & \maxColor{\textbf{.51}}   &   \maxColor{\textbf{.39}}   &  \minColor{\textbf{.00}}  &  \minColor{\textbf{.07}}  & .12                     & \minColor{\textbf{.06}}   &  \minColor{\textbf{.05}}  &  .00                          &  \maxColor{\textbf{.67}}   &  \maxColor{\textbf{.67}}  &  .03                      \\ \hline
    \end{tabular}
   \label{tab:proburi}
\end{table}
\begin{table}
   \centering
   \setlength{\tabcolsep}{0.6pt}
   
   \caption{Conditional probability (e.g., $P(relevant | p_{P_1A_1}^{Reddit} = 1)$ = \maxColor{\textbf{0.64}}) of the event that the URIs in a social media post from a given post class (e.g., Reddit P$_1$A$_1$) are relevant, given that the post has $k$ (e.g., $k = 1$) HTML URIs. Column markers: \minColor{minimum} and \maxColor{maximum}. See also Appendix 10 for additional figures of the average precision per count of links in a post, per post class, per social media. For the \textbf{P$_1$A$_1$}, $k = 5+$ Twitter cell, the probability was calculated for just one post with eight HTML URIs.}
   \begin{tabular}{|c!{\vrule width 1pt}c|c|c|c|c!{\vrule width 1pt}c|c|c|c|c!{\vrule width 1pt}c|c|c|c|}
      \hline
      \multicolumn{1}{|c!{\vrule width 1pt}}{ }
      & \multicolumn{5}{c!{\vrule width 1pt}}{ \myfbox{white}{\includegraphics[width=0.5cm]{redditLogo}} }        
      & \multicolumn{5}{c!{\vrule width 1pt}}{ \myfbox{white}{\includegraphics[width=0.5cm]{twitterLogo}} } 
      & \multicolumn{4}{c|}{ \myfbox{white}{\includegraphics[width=0.5cm]{scoopitLogo}} } \\ \hline

      \multicolumn{1}{|c!{\vrule width 1pt}}{ \textbf{k} } & P$_1$A$_1$ & MC & P$_n$A$_1$ & P$_n$A$_n$ & All & P$_1$A$_1$ & MC & P$_n$A$_1$ & P$_n$A$_n$ & All & P$_1$A$_1$ & MC & P$_n$A$_1$ & All \\ \hline

      \multicolumn{1}{|c!{\vrule width 1pt}}{ \makecell{1} }      &  .64                            &   .54                     &   .54                     &   .54                     &    .60                      &   .63                     &   .60                     & .49                     &  \maxColor{\textbf{ .61}} &  \maxColor{\textbf{ .61}} &  \maxColor{\textbf{ .76}}     &\minColor{\textbf{ .00}}      &    \minColor{\textbf{.00}}&  \maxColor{\textbf{ .76}}     \\ \hline
      \multicolumn{1}{|c!{\vrule width 1pt}}{ \makecell{2} }      & \maxColor{\textbf{ .80}}        &  \maxColor{\textbf{ .59}} &  \maxColor{\textbf{ .57}} &  \maxColor{\textbf{ .59}} &   \maxColor{\textbf{ .65}}  &   .50                     &  \maxColor{\textbf{ .61}} &\maxColor{\textbf{.64}}  &   .60                     &   .60                     &   NA                          &\minColor{\textbf{ .00}}      &    \minColor{\textbf{.00}}&  \minColor{\textbf{ .00}}     \\ \hline
      \multicolumn{1}{|c!{\vrule width 1pt}}{ \makecell{3-4} }    &  .62                            &  \minColor{\textbf{ .45}} &  \minColor{\textbf{ .50}} &  \minColor{\textbf{ .44}} &   \minColor{\textbf{ .48}}  & \minColor{\textbf{.33}}   &   .46                     & .51                     &   .45                     &   .46                     &   NA                          & .50                          &    .50                    &   .50                         \\ \hline
      \multicolumn{1}{|c!{\vrule width 1pt}}{ \makecell{5+} }     & \minColor{\textbf{ .51}}        &   .50                     &   .53                     &   .50                     &    .50                      &\maxColor{\textbf{ 1.00}}&  \minColor{\textbf{ .42}}&\minColor{\textbf{.46}}  &  \minColor{\textbf{ .41}} &  \minColor{\textbf{ .42}} &   NA                         & \maxColor{\textbf{.59}}      &   \maxColor{\textbf{ .59}}&   .59                         \\ \hline
    \end{tabular}
   \label{tab:probReluri}
\end{table}
\normalsize
\section{Results and Discussion}
Recall the post class (Table \ref{tab:postClasses}) acronyms and their respective meanings and examples: \textbf{P$_1$A$_1$} (e.g., a tweet) - single \textbf{P}ost from a single \textbf{A}uthor, \textbf{P$_1$A$_n$} (e.g., Wikipedia reference) - single \textbf{P}ost from multiple \textbf{A}uthors, \textbf{P$_n$A$_1$} (e.g., twitter thread) - multiple \textbf{P}osts by a single \textbf{A}uthor, and \textbf{P$_n$A$_n$} (e.g., twitter conversation) - multiple \textbf{P}osts from multiple \textbf{A}uthors.

To address the first research question, we identified micro-collections (\textbf{MC} = \textbf{P$_n$A$_1$} $\cup$ \textbf{P$_n$A$_n$}) as the collection of social media posts that show some properties of collection building\footnote{Some \textbf{P$_1$A$_1$} posts which are visible to SERP scrapers could be added to \textbf{MC} if they contain links above the median number of links, calculated from the same pool of social media posts. However, we did not make such a distinction in our study.}. Next, we extracted the \textbf{P$_n$A$_1$} and \textbf{P$_n$A$_n$} post classes by identifying social media posts with replies (comments) and extracted the parent post as well as the child posts.

Following the identification and extraction of micro-collections, to address the second research question, we characterized \textbf{MC}s and compared seeds extracted from them to seeds extracted from SERPs (\textbf{P$_1$A$_1$}). Here we present the results for each of the respective measures introduced in Section \ref{sec:primitiveMeasures}, and Appendices 1 - 10 includes additional figures for these metrics.
\subsection{\textbf{URI and post counts per post class}}
Micro-collections (\textbf{MC}s) are prevalent on the Web and outnumber (12,917 vs. 10,195) conventional SERP posts (\textbf{P$_1$A$_1$}). Also, in general, \textbf{MC}s produced more URIs (Appendices 2 - 4) than conventional SERP posts (\textbf{P$_1$A$_1$}). Additionally, \textbf{MC}s produced more non-HTML URIs than \textbf{P$_1$A$_1$} across all topics. In fact, the total number of \textbf{P$_1$A$_1$} non-HTML URIs were between 19\% to 44\% the size of \textbf{MC}s. These findings are potentially consequential for curators interested in enriching their collections with non-HTML resources.

From Table \ref{tab:dataset}, for all topics in the Reddit SERPs except (Reddit-New), \textbf{P$_n$A$_n$} mostly produced the largest count of URIs (41,160), next to \textbf{P$_1$A$_1$} (51\% \textbf{P$_n$A$_n$}), next to \textbf{P$_n$A$_1$} (3\% \textbf{P$_n$A$_n$}): \textbf{P$_n$A$_n$} > \textbf{P$_1$A$_1$} > \textbf{P$_n$A$_1$}. The relatively low number of Reddit \textbf{P$_n$A$_1$} posts and URIs shows that it is a rare phenomenon for a Reddit user to reply to his/her initial post especially since Reddit does not impose any size restriction on the length of posts. For the Reddit-New SERP, \textbf{P$_1$A$_1$} had more URIs (8,056) than \textbf{P$_n$A$_n$} (80\% \textbf{P$_1$A$_1$}): \textbf{P$_1$A$_1$} > \textbf{P$_n$A$_n$} > \textbf{P$_n$A$_1$}. This is likely due to the fact that in the \textit{New} SERP, \textbf{P$_n$A$_n$} do not get sufficient opportunity to increase because they must compete with newer posts, since the SERP is in ``newest first'' order. Consequently, before \textbf{P$_n$A$_n$} sufficiently grow, they are pushed down (rank demotion) by newer \textbf{P$_1$A$_1$} posts, and do not get sufficient exposure, leading to fewer replies which leads to a reduced \textbf{P$_n$A$_n$} size.

The results show a high degree of inter/extra-user engagement on Twitter, and thus for Post and URI Counts (Table \ref{tab:dataset}), \textbf{P$_n$A$_n$} > \textbf{P$_n$A$_1$} > \textbf{P$_1$A$_1$}. In contrast, Scoop.it showed lesser user engagement, and thus: \textbf{P$_1$A$_1$} > \textbf{P$_n$A$_1$}.

\subsection{\textbf{Probability distribution of posts with links}}
From Table \ref{tab:proburi}, unsurprisingly, the probability of the event that a social media post with a URI of a given post class (\textbf{P$_1$A$_1$} - \textbf{P$_n$A$_n$}) had more than one HTML URI ($k > 1$) seemed to correlate with whether the social media platform restricts the size of posts. For example, due to the character limit imposed on tweets, the probability of the event that a tweet with a URI has only 1 HTML URI is 0.98 ($P(p_{P_1A_1}^{Twitter} = 1) = 0.98$). On the other hand, single tweets with 3+ HTML URIs are rare. We observed three tweets with 3 or 4 HTML URIs (out of 3,501 tweets).
\subsection{\textbf{Precision of post class URIs}}

Table \ref{tab:probReluri} shows the conditional probability of the event that the URIs contained in a post of a given post class are relevant, given that the post has a specified count of URIs ($k$). Across almost all $k$ per post class, we see that the seeds generated from \textbf{P$_1$A$_1$} posts had a higher probability (maximum: 1, median: 0.63, minimum: 0.33) of being relevant than \textbf{MC} (0.61, 0.5, 0.0). For example, for Reddit when $k=2$, \textbf{P$_1$A$_1$} - 0.80, while \textbf{MC} - 0.59. This shows that \textbf{P$_1$A$_1$} posts benefit from SERP filters; \textbf{P$_1$A$_1$} posts are posts directly returned by SERPs and their text often matches a subset of the query. This indicates that a match between a query and a post text lends some relevance to the URI extracted from the post. However, given the fact that \textbf{MC}s do not all benefit from SERP filters since the vast majority of \textbf{MC}s are not extracted directly from the SERP, but from the reply or comment threads, the 0.5 median precision value indicates that comments and replies possess quality URIs. See also Appendices 5 - 7 for additional figures of the average precision of URIs per post class, per social media.

In general, \textbf{P$_1$A$_1$} post URIs (all URIs, HTML, and non-HTML) had the highest average precision compared to \textbf{P$_n$A$_1$} and \textbf{P$_n$A$_n$} for Reddit, Scoop.it, and Twitter posts extracted with text queries. For tweets extracted with hashtags, \textbf{P$_n$A$_n$} posts had the highest average precision compared to \textbf{P$_n$A$_1$} and \textbf{P$_1$A$_1$}. 

For Reddit, \textbf{P$_1$A$_1$} > \textbf{P$_n$A$_1$} > \textbf{P$_n$A$_n$}: across all topics, \textbf{P$_1$A$_1$} posts had the highest average precision (all URIs) 80\% of the time than \textbf{P$_n$A$_1$} and \textbf{P$_n$A$_n$}. The Maximum, Median, and Minimum (MMM) average precision values were 0.88, 0.59, and 0.15, respectively. Next, \textbf{P$_n$A$_1$} posts had a higher average precision than \textbf{P$_n$A$_n$} 70\% of the time, MMM - (0.88, 0.50, 0.00), for \textbf{P$_n$A$_n$} - (0.70, 0.42, 0.07). 

For tweets exposed with text queries, \textbf{P$_1$A$_1$} > \textbf{P$_n$A$_n$} > \textbf{P$_n$A$_1$}: \textbf{P$_1$A$_1$} (0.91, 0.66, 0.45) had the highest average precision 90\% of the time than \textbf{P$_n$A$_n$} and \textbf{P$_n$A$_1$}. \textbf{P$_n$A$_n$} (0.74, 0.46, 0.28) had a higher average precision 70\% of the time than \textbf{P$_n$A$_1$} (0.58, 0.39, 0.35).

For tweets exposed with hashtags, \textbf{P$_n$A$_n$} > \textbf{P$_n$A$_1$} > \textbf{P$_1$A$_1$}: \textbf{P$_n$A$_n$} (0.65, 0.29, 0.27) posts had the highest average precision 60\% of the time than \textbf{P$_n$A$_1$} and \textbf{P$_1$A$_1$}. \textbf{P$_n$A$_1$} (0.45, 0.39, 0.21) posts had a higher average precision than \textbf{P$_1$A$_1$} (0.50, 0.26, 0.11) 70\% of the time. For example, from Fig. \ref{fig:genericPlots}a, the average precision for \textbf{P$_1$A$_1$} URIs in the Twitter-Top vertical for the \textit{Ebola virus outbreak} topic was 0.86 (\textbf{P$_n$A$_n$} - 0.74) for posts extracted with the text query ``ebola virus outbreak.'' However, \textbf{P$_n$A$_n$} outperformed (0.65) \textbf{P$_1$A$_1$} (0.49) when the query used to extract posts was the hashtag ``\#ebolavirus.''

For Scoop.it, \textbf{P$_1$A$_1$} > \textbf{P$_n$A$_1$}: \textbf{P$_1$A$_1$} (0.87, 0.78, 0.55) posts had a higher average precision than \textbf{P$_n$A$_1$} (0.80, 0.55, 0.27) 100\% of the time. Similar to Twitter \textbf{P$_1$A$_1$}, Scoop.it \textbf{P$_1$A$_1$} are derived directly from the SERP, and thus benefit from SERP filtering. \textbf{P$_n$A$_1$} do not benefit from SERP filtering since they are not extracted directly from the SERP.

\subsection{\textbf{Age distribution of relevant webpages}}
We compared the ages of \textbf{P$_1$A$_1$} and \textbf{MC} post class URIs, by focusing on the older topics  (\textit{Ebola virus outbreak} and \textit{Flint water crisis}) for social media that supports \textbf{P$_1$A$_1$}, \textbf{P$_n$A$_1$}, and \textbf{P$_n$A$_n$} - Reddit and Twitter. \textbf{MC} posts consistently produce older webpages in the Twitter-Latest vertical. A possible explanation for this is: \textbf{P$_1$A$_1$} tweets (extracted directly from the Twitter-Latest SERP) are highly likely to be new tweets if the topic is ongoing. Even though new tweets can include URIs of old stories, for ongoing news stories such as those we considered, new tweets are likely to include the URIs of the latest developments. We observed that the Twitter-Latest \textbf{P$_1$A$_1$} tweets were created within days from the query issue dates, and thus more likely to produce new URIs for both topics. In contrast, \textbf{MC}s are extracted from conversations that can mix new and old tweets; a new tweet can reply to an old tweet that contains old URIs. Therefore, Twitter-Latest \textbf{MC}s produced a mix of tweets created within days and years from the query issue dates. See also Appendices 8 and 9 for additional figures of the age distribution of URIs per post class, per social media.

For the Reddit-Top/Relevance/Comments SERPs for \textit{Ebola virus outbreak}, \textbf{MC}s and \textbf{P$_1$A$_1$} produced older webpages with similar distributions. For example, for \textit{Ebola virus outbreak} both post classes had a median webpage age of ~4.3 years.

As expected, the Reddit-New, for both topics, \textbf{MC}s and \textbf{P$_1$A$_1$} produced the newest webpages compared to other Reddit SERPs with median age $<$ 1 year.

\textbf{P$_1$A$_1$} and \textbf{MC} posts from Twitter-Top produced webpages with similar age distributions. For example, for \textit{Flint water crisis} both post classes
had a median webpage age $<$ 5 months. In contrast, in the Twitter-Latest vertical, for both topics \textbf{MC}s  produced older webpages than \textbf{P$_1$A$_1$}. For example, \textbf{MC}s for Ebola virus outbreak produced older webpages (median: 4.2 years) than those from \textbf{P$_1$A$_1$} (19 days) (Fig. \ref{fig:genericPlots}b).

\subsection{\textbf{Distribution of hostname diversity}}
For Reddit, the \textbf{P$_n$A$_1$} posts produced the highest hostname diversity. For Twitter, \textbf{P$_1$A$_1$} posts produced the highest hostname diversity.

For Reddit, \textbf{P$_n$A$_1$} > \textbf{P$_1$A$_1$} > \textbf{P$_n$A$_n$}: across all topics, \textbf{P$_n$A$_1$} posts had the highest hostname diversity (HTML URIs) 95\% of the time than \textbf{P$_1$A$_1$} and \textbf{P$_n$A$_1$}. The Maximum, Median, and Minimum (MMM) hostname diversity values were 1.0, 0.55, and 0.0, respectively. Next, \textbf{P$_1$A$_1$} posts had more diverse hostnames than \textbf{P$_n$A$_n$} 61\% of the time, MMM - (0.6, 0.33, 0.11), for \textbf{P$_n$A$_n$} - (0.55, 0.28, 0.1). 

For Twitter, \textbf{P$_1$A$_1$} > \textbf{P$_n$A$_n$} > \textbf{P$_n$A$_1$}: \textbf{P$_1$A$_1$} (0.70, 0.60, 0.43) produced more diverse hostnames 74\% of the time than \textbf{P$_n$A$_n$} and \textbf{P$_n$A$_1$}. Similarly, \textbf{P$_n$A$_n$} (0.61, 0.45, 0.39) produced more diverse hostnames 79\% of the time than \textbf{P$_n$A$_1$} (0.74, 0.37, 0.31). Scoop.it did not produce enough URIs for two topics, as a result had fewer \textbf{P$_n$A$_1$} to derive a fair comparison with \textbf{P$_1$A$_1$}.

Reddit and Twitter had \textbf{P$_1$A$_1$} > \textbf{P$_n$A$_n$} in common. This is not unexpected; hostname diversity rewards unique hosts, and given that the \textbf{P$_1$A$_1$} collection is smaller than \textbf{P$_n$A$_n$}, it is more likely for \textbf{P$_1$A$_1$} to fill in the hostname slots with additional different hosts than \textbf{P$_n$A$_n$}. However, for Twitter \textbf{P$_n$A$_1$} had the lowest diversity unlike Reddit for the following reasons. First, \textbf{P$_n$A$_1$} is the set of all threads authored by the same user. These threads on Twitter, especially those from News (e.g., \texttt{@nytimes}, \texttt{@vice}) and non-News organizations (e.g., \texttt{@splcenter}, \texttt{@TurnoutPAC}) tend to link to webpages within their websites, leading to a lower hostname diversity. This phenomenon was most prominent in the 2018 \textit{world cup} and \textit{midterm elections} topics.
\subsection{\textbf{Overlap: Google collections vs. post classes}}
All post classes showed small amount of overlap with the collections of URIs returned from the first 10 pages of Google for the respective dates the post class URIs were extracted. This highlights the fluidity of the Google SERP. Thus, URIs extracted from \textbf{MC} and \textbf{P$_1$A$_1$} collections are not easily discoverable.

Reddit \textbf{P$_1$A$_1$} and \textbf{MC} posts had overlap $<$ 0.1 85\% of the time. Their MMM overlap were: 0.13, 0.04, and 0.1, respectively. Twitter \textbf{P$_1$A$_1$} posts had overlap (0.09, 0.02, 0.0) $<$ 0.1 100\% the time. Similarly, Twitter \textbf{MC} posts had overlap (0.13, 0.04, 0.0) $<$ 0.1 80\% of the time.
\subsection{\textbf{Recommendations for generating seeds}}
Considering the results presented, it is clear that collections generated from social media SERPs (\textbf{P$_1$A$_1$}) are different from collections generated from micro-collections (\textbf{MC}s), and both post classes yield seeds not easily discoverable by scraping Google. Consider the following highlights and how they could affect decisions made in generating seeds from social media.

\textbf{MC}s are more prevalent and produce more seeds than \textbf{P$_1$A$_1$}. This means seed generation that prioritizes quantity would benefit from extracting seeds from \textbf{MC}s. \textbf{P$_1$A$_1$} produced higher quality URIs for all social media SERP combinations except with seeds generates with hashtags. The poorer precision performance of hashtag queries compared to text queries shows that hashtags can be used as a vehicle for spreading non-relevant content, especially when the hashtag is popular. However, when users reply to a tweet that contains a link and a hashtag (the composition of \textbf{P$_n$A$_n$} set), it is likely they are responding to a relevant tweet. Replies may serve as a quality check. Therefore, \textbf{MC}s produced more relevant URIs when hashtags were used to surface tweets. Consequently, seed generation that prioritizes quality would benefit from extracting seeds from \textbf{P$_1$A$_1$}, but for Twitter, if hashtags are used, \textbf{MC}s should be considered first.

\textbf{MC}s consistently produced older webpages than \textbf{P$_1$A$_1$} posts for the Twitter-Latest vertical because \textbf{MC}s included older tweets. Consequently, if seed generation from the Twitter-Latest vertical intends to extract older stories, \textbf{MC}s should be prioritized. Finally, we showed that \textbf{P$_1$A$_1$} produced more diverse hostnames than \textbf{MC}s for Twitter unlike Reddit. Therefore, seed generation that intends to include different hosts should consider \textbf{P$_1$A$_1$}, instead of Twitter \textbf{P$_n$A$_1$}, since it showed a low level of hostname diversity due to reuse of the same domains, which is a common practice especially among news organizations.

\section{Future work and Conclusions}
We populated the \textbf{MC} set with posts from \textbf{P$_n$A$_1$} and \textbf{P$_n$A$_n$} exclusively. However, we believe posts from \textbf{P$_1$A$_1$} with a higher than normal median number of URIs may be added to the \textbf{MC} set. The median number of URIs in social media posts for all social media/SERP combination was 1, 86\% of the time. Therefore, we may add to \textbf{MC}, \textbf{P$_1$A$_1$} posts with URIs above the median. Also, our precision evaluation was biased to favor pages rich in text. Consequently, we observed false negative URIs: relevant URIs marked as non-relevant, a problem we plan to address. Finally, we plan to expand the set of metrics for comparing seeds to include metrics that account for the authority of hosts. For example, such metric would assign a higher authority weight to a CDC (Centers for Disease Control and Prevention) webpage than an obscure webpage for the same \textit{Ebola virus} topic.

The seed URIs that form the building blocks of Web archive collections are often hand-selected by curators. Manual selection produces high quality URIs but it does not scale and requires domain knowledge. Due to a shortage of curators amidst an abundance of unfolding events, Google and Twitter have been widely adopted for automatically scraping seeds. In this work, we introduced a cross-platform vocabulary (post class: \textbf{P$_1$A$_1$}, \textbf{P$_1$A$_n$}, \textbf{P$_n$A$_1$}, and \textbf{P$_n$A$_n$}) for describing social media posts. It is common practice to scrape social media such as Twitter for (\textbf{P$_1$A$_1$}) seeds, we introduced an overlooked source of social media posts - micro-collections (\textbf{MC}s), and showed that seeds generated from \textbf{MC}s are more prevalent and different from their \textbf{P$_1$A$_1$} counterpart across multiple dimensions. Finally, we provided recommendations for curators generating seeds from these sources. For example, a seed generation that prioritizes quantity may target \textbf{MC}s first, while a precision priority favors \textbf{P$_1$A$_1$}. Our study outlines how to compare seeds generated from different venues on social media and our findings may inform the decisions made during seed generation. Our research dataset comprising of 120,444 links extracted from 449,347 social media posts, as well as the source code for the application utilized to generate the collections are publicly available \cite{jcdl2019Repo}.

\section*{Acknowledgments}
This work was supported in part by IMLS LG-71-15-0077-15.
\bibliographystyle{ACM-Reference-Format}
\clearpage
\balance
\bibliography{NwalaJCDL2019}

\clearpage
\begin{figure*}
   \captionsetup{font=Large}
   \centering
   \caption*{APPENDIX 1\\Distribution of the number of links in a post, per post class, per social media.}
\end{figure*}

    \begin{figure*}
      \centering
      \fbox{\includegraphics[width=0.7\textwidth]{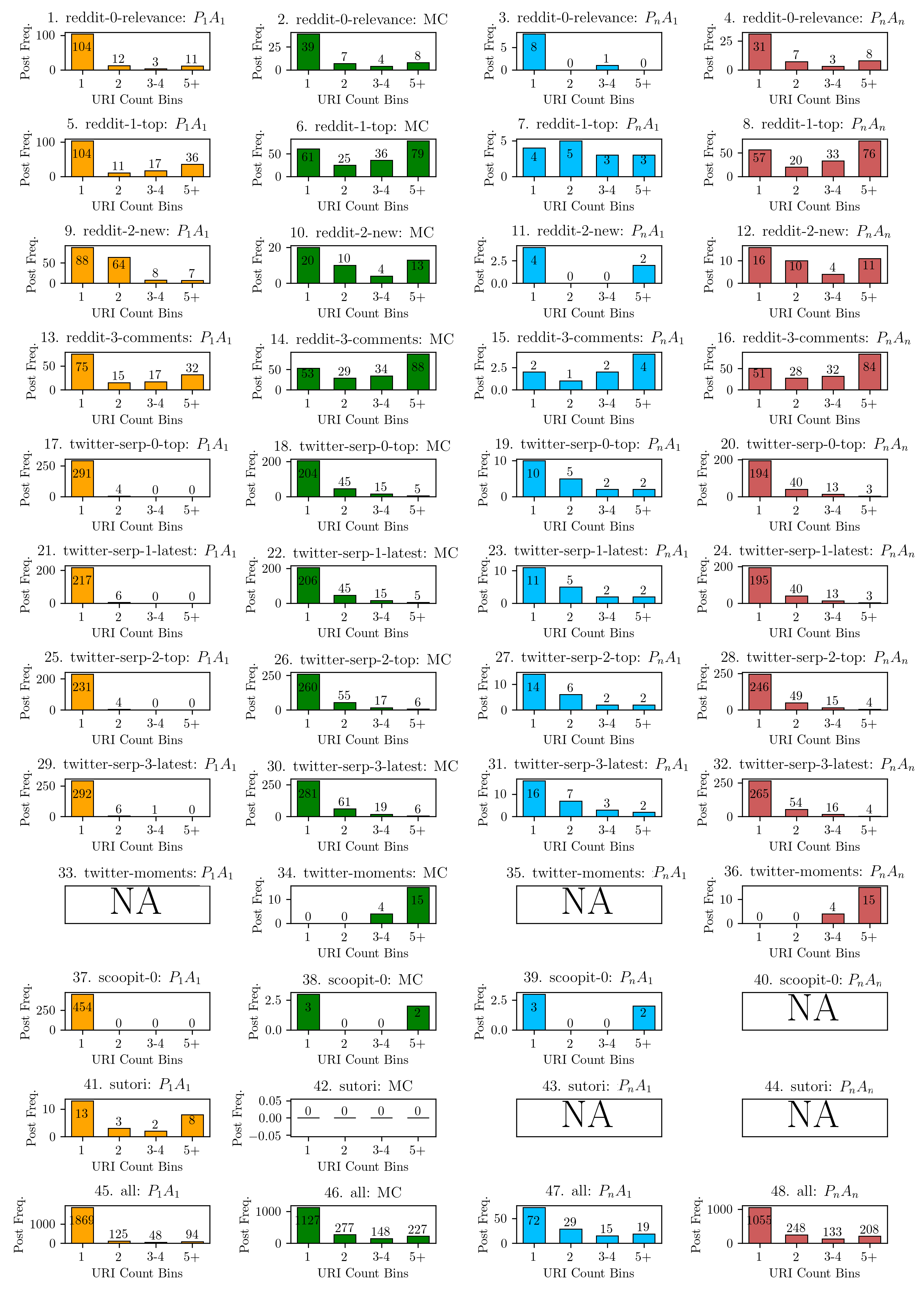}}  
      \caption{Distribution of the number of links in a post, per post class, per social media for \textit{Ebola Virus Outbreak}. A single sub-figure (e.g., sub-figure 1) reads as follows: There were 104 Reddit (\textit{relevance} vertical) posts with 1 URI, 12 Reddit posts with 2 URIs, etc. The remaining figures in this appendix are to be read similarly. \textit{twitter-serp-2-top} and \textit{twitter-serp-3-latest} represent collections generated by issuing hashtag (\texttt{\#ebolavirus}) queries.}
    \end{figure*}

    \begin{figure*}
      \centering
      \fbox{\includegraphics[width=0.8\textwidth]{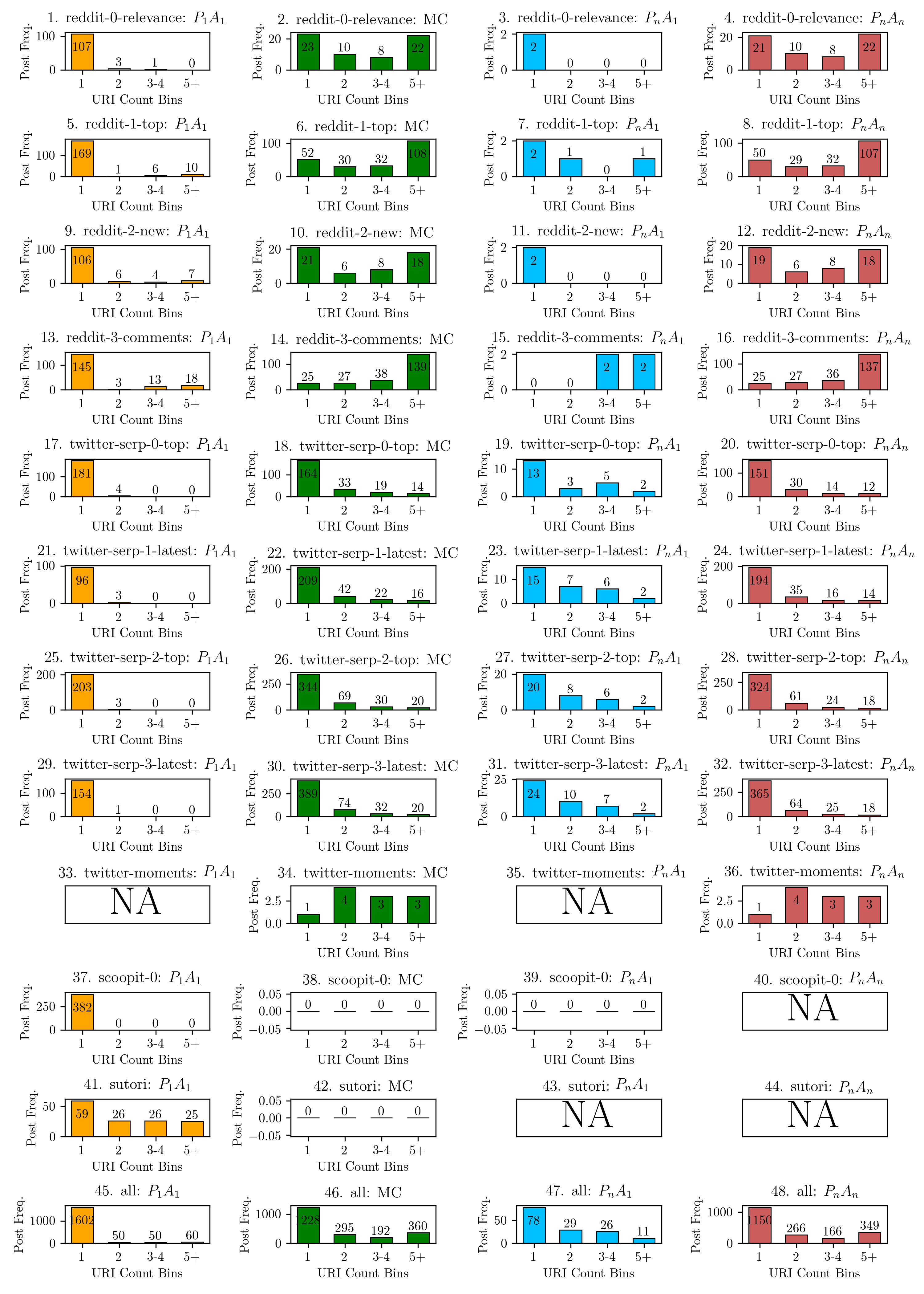}}  
      \caption{Distribution of the number of links in a post, per post class, per social media for \textit{Flint Water Crisis}}
    \end{figure*}

    \begin{figure*}
      \centering
      \fbox{\includegraphics[width=0.8\textwidth]{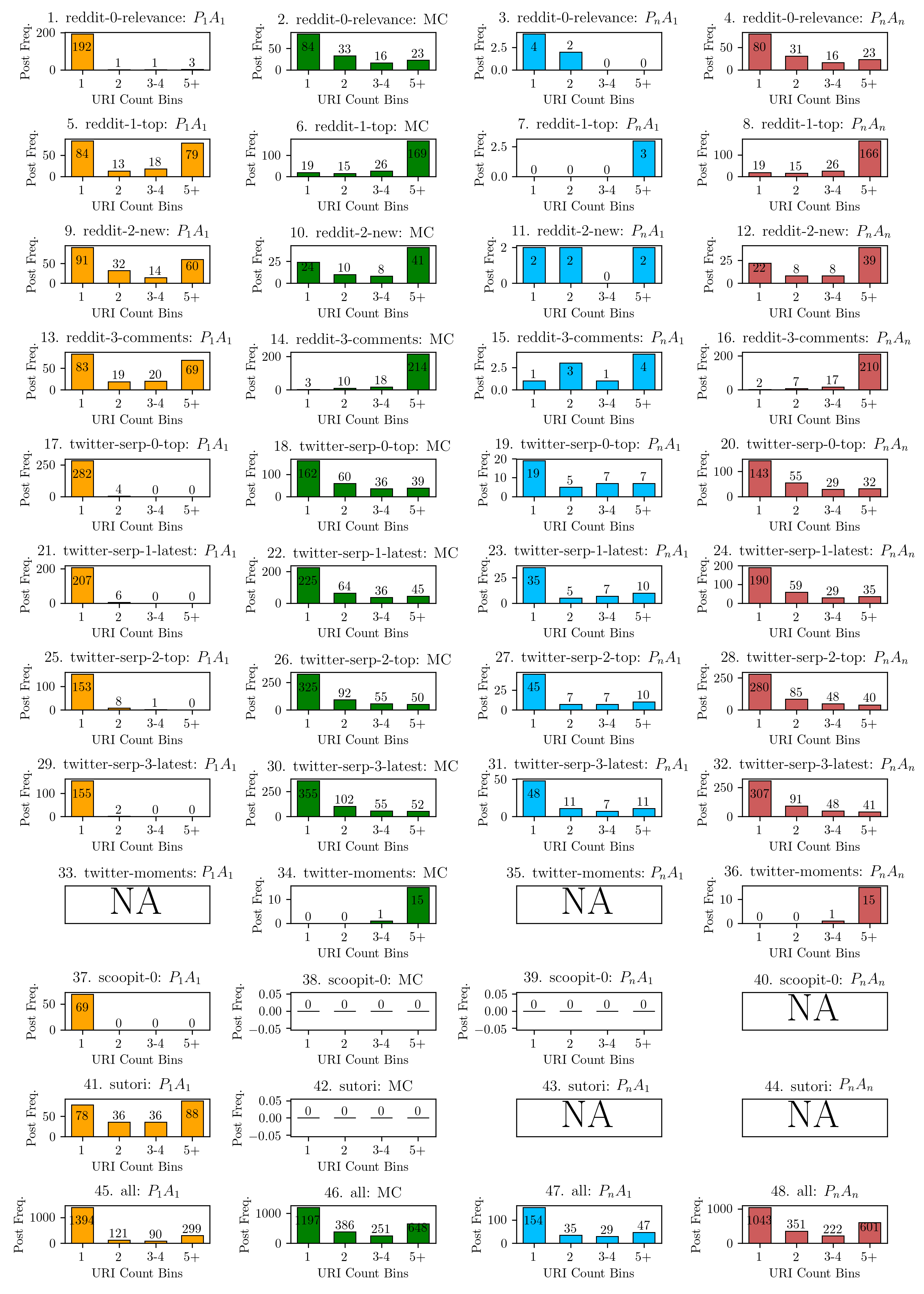}}  
      \caption{Distribution of the number of links in a post, per post class, per social media for \textit{MSD Shooting}}
    \end{figure*}

    \begin{figure*}
      \centering
      \fbox{\includegraphics[width=0.8\textwidth]{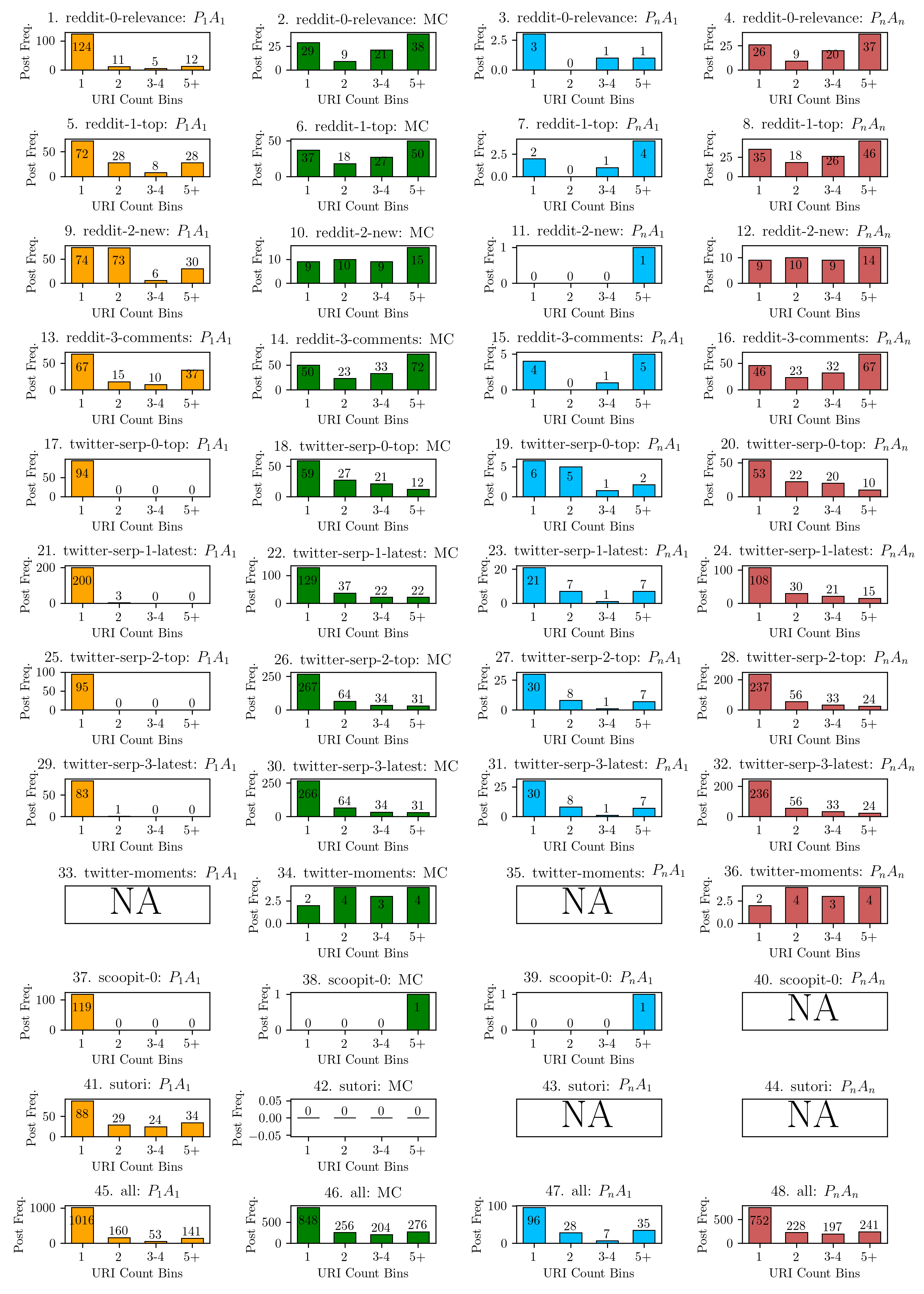}}  
      \caption{Distribution of the number of links in a post, per post class, per social media for \textit{2018 World Cup}}
    \end{figure*}

    \begin{figure*}
      \centering
      \fbox{\includegraphics[width=0.8\textwidth]{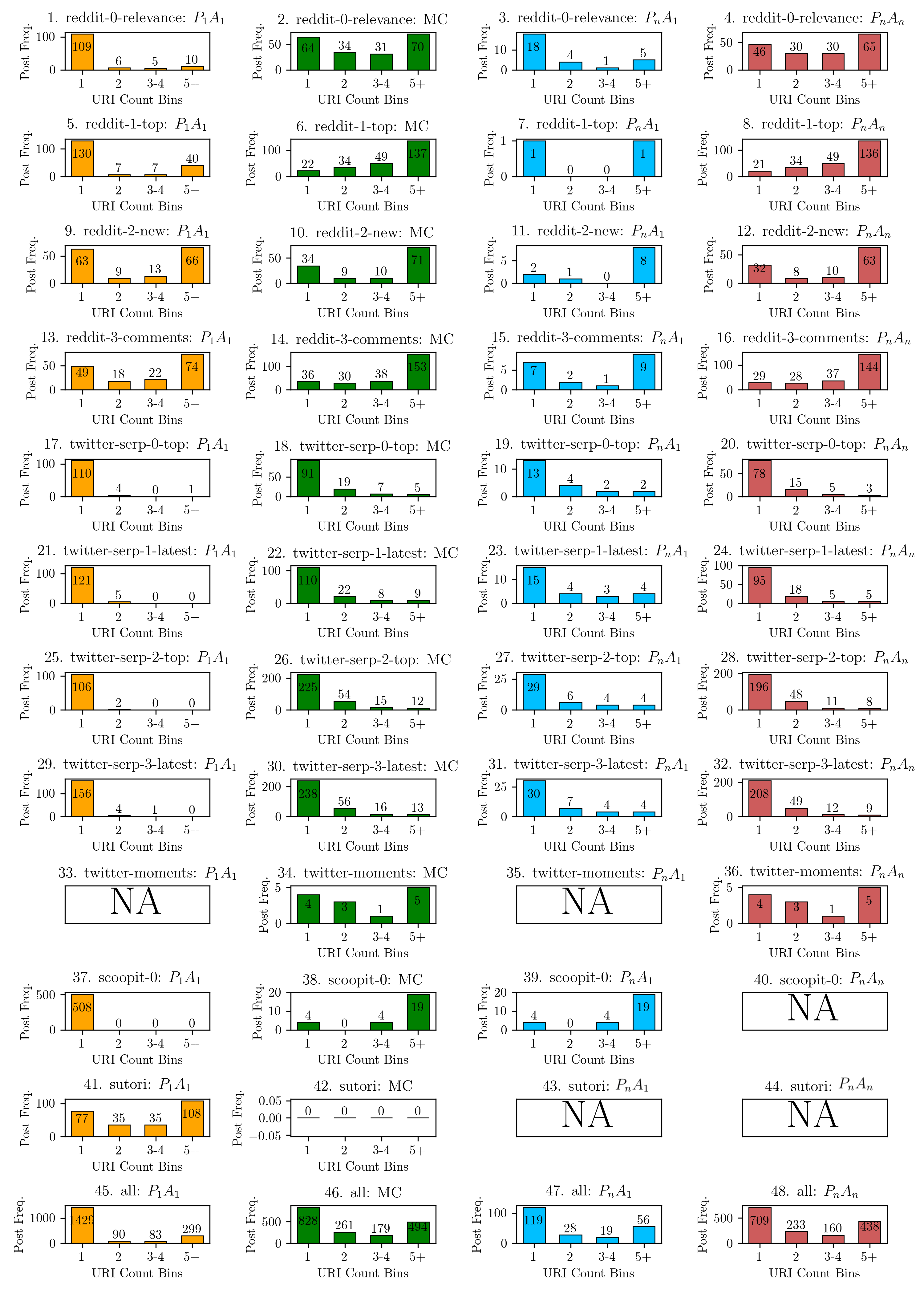}}  
      \caption{Distribution of the number of links in a post, per post class, per social media for \textit{2018 Midterm Elections}}
    \end{figure*}

\clearpage
\begin{figure*}
   \captionsetup{font=Large}
   \centering
   \caption*{APPENDIX 2\\Total number of HTML URIs per post class, per social media.}
\end{figure*}

     \begin{figure*}
      \centering
      \fbox{\includegraphics[width=0.98\textwidth]{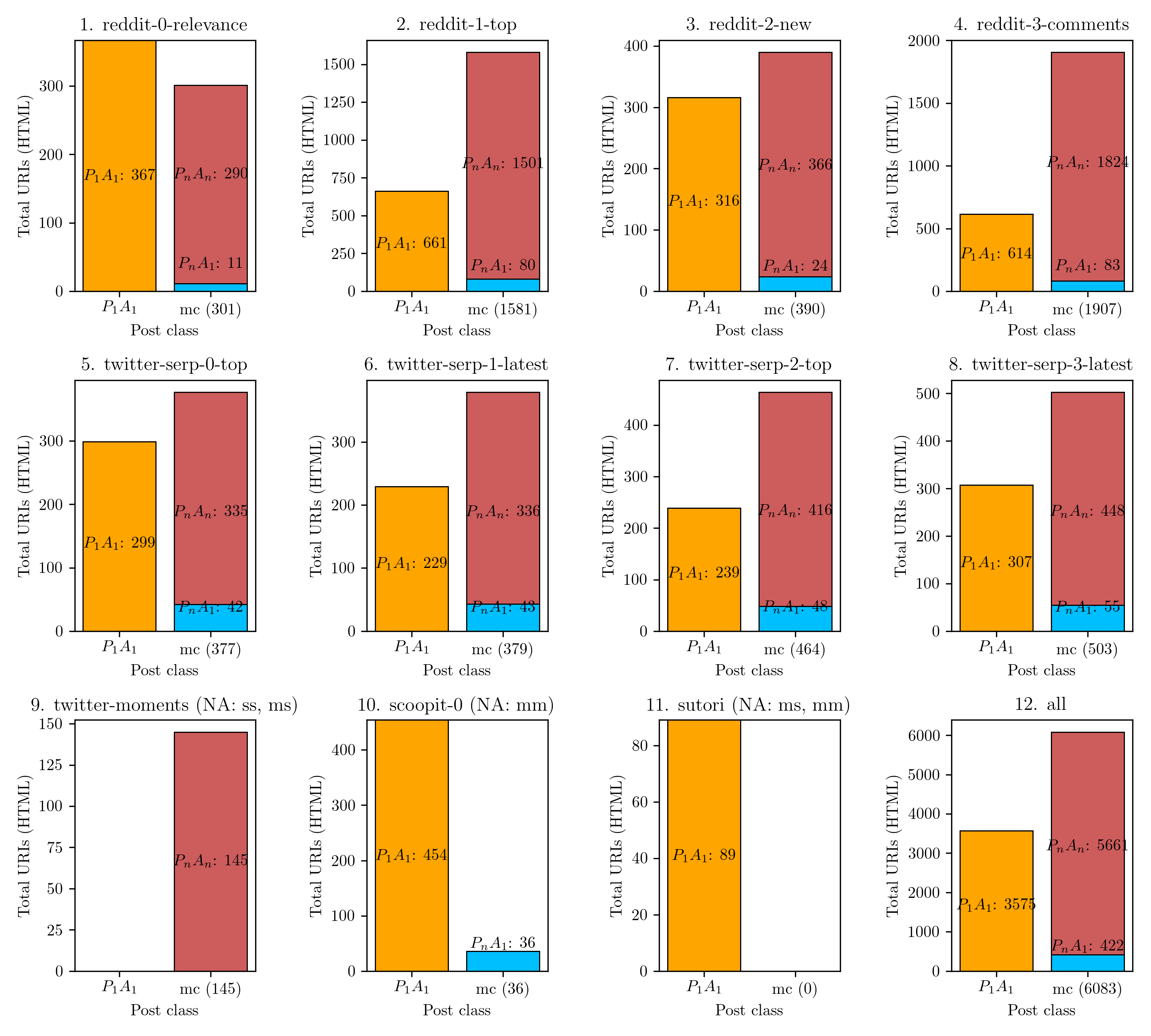}}  
      \caption{Total number of HTML URIs per post class, per social media for \textit{Ebola Virus Outbreak}. A single sub-figure (e.g., sub-figure 1) reads as follows: There were 367 URIs in the Reddit (\textit{relevance} vertical) \textbf{P$_1$A$_1$} post class and 301 (\textbf{P$_n$A$_1$}: 11 + \textbf{P$_n$A$_n$}: 290) URIs in the \textbf{MC} post class. The remaining figures in this appendix are to be read similarly. \textit{twitter-serp-2-top} and \textit{twitter-serp-3-latest} represent collections generated by issuing hashtag (\texttt{\#ebolavirus}) queries.}
      \end{figure*}

      \begin{figure*}
        \centering
        \fbox{\includegraphics[width=0.98\textwidth]{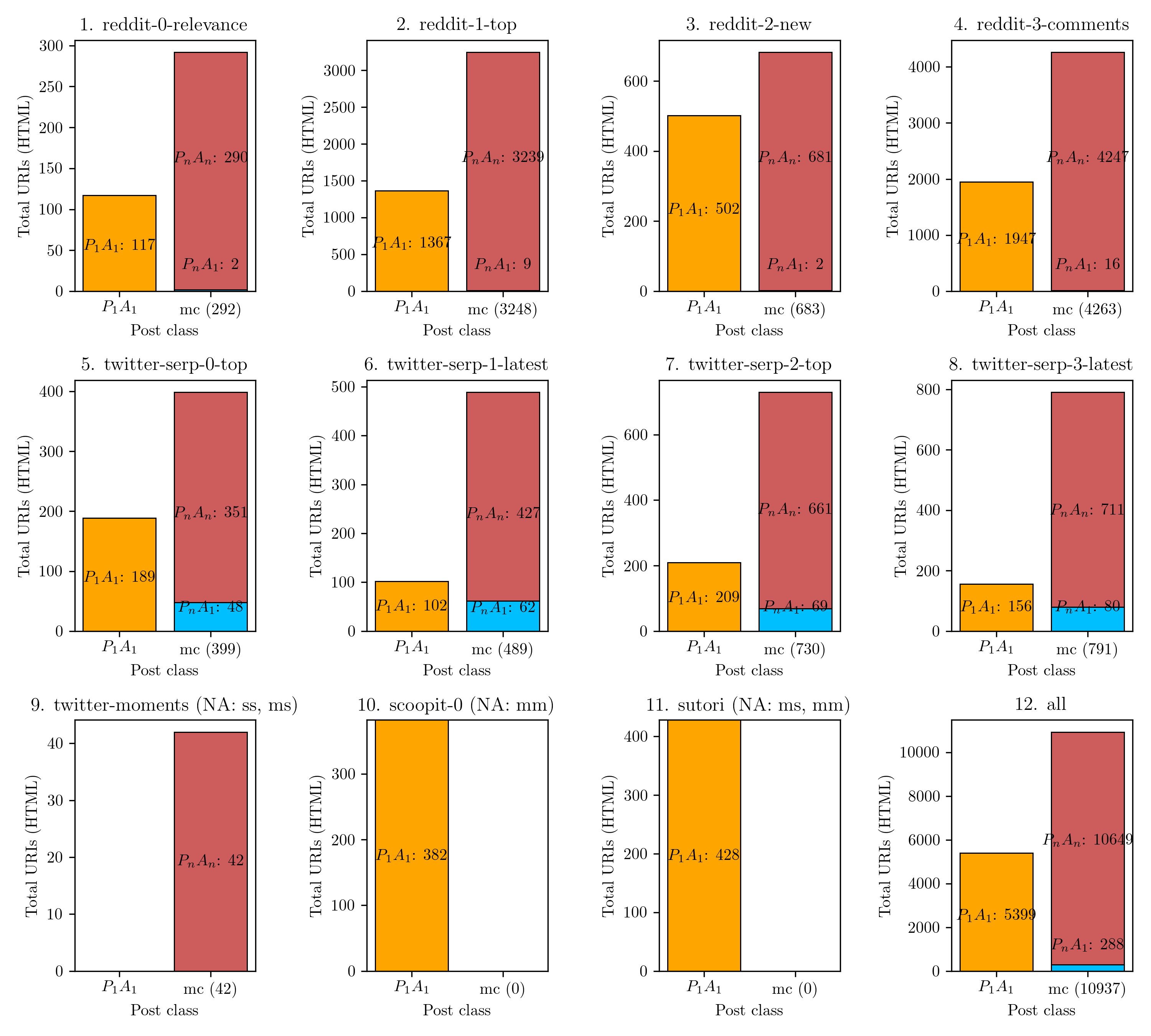}}  
        \caption{Total number of HTML URIs per post class, per social media for \textit{Flint Water Crisis}}
      \end{figure*}

      \begin{figure*}
        \centering
        \fbox{\includegraphics[width=0.98\textwidth]{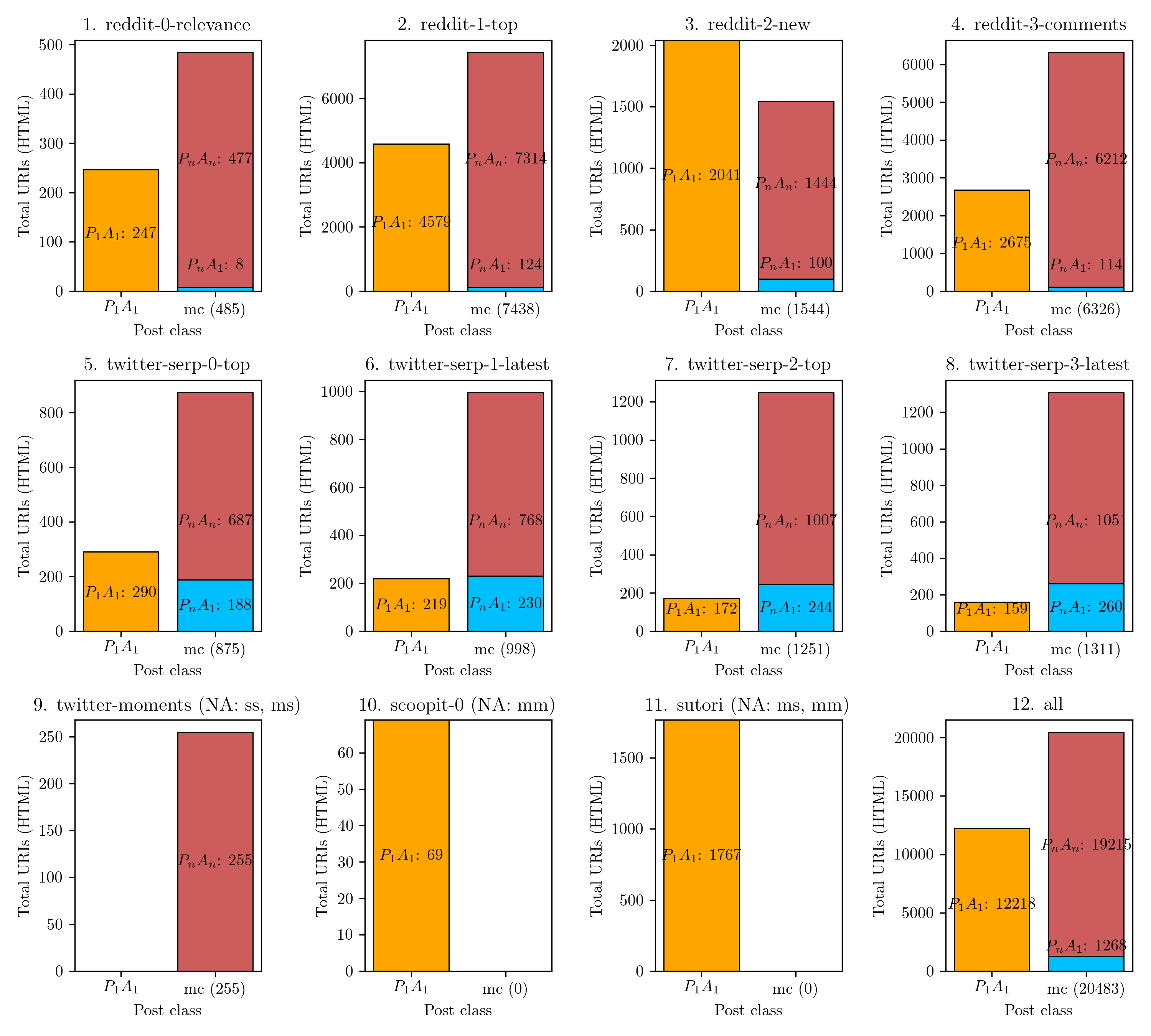}}  
        \caption{Total number of HTML URIs per post class, per social media for \textit{MSD Shooting}}
      \end{figure*}

      \begin{figure*}
        \centering
        \fbox{\includegraphics[width=0.98\textwidth]{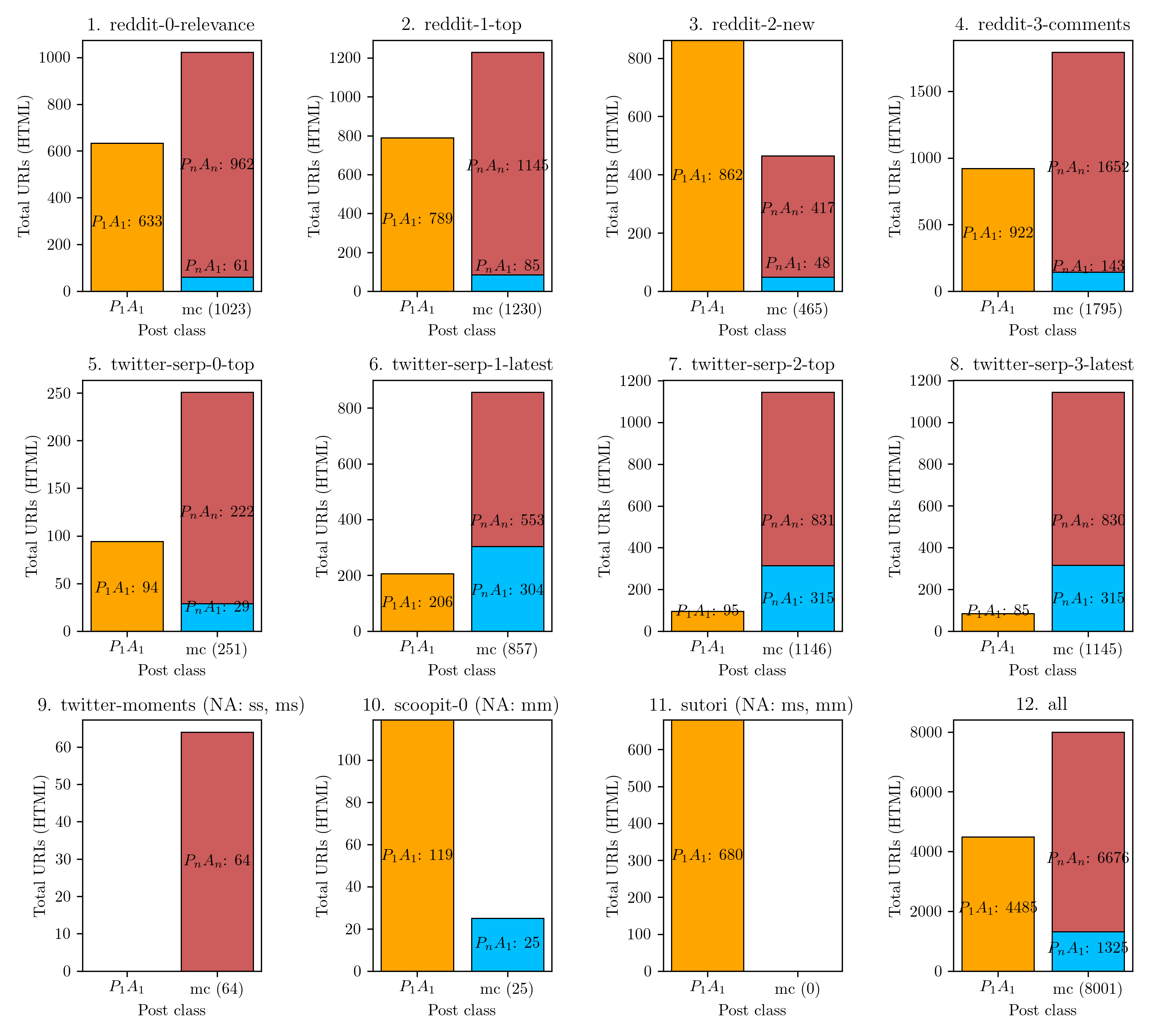}}  
        \caption{Total number of HTML URIs per post class, per social media for \textit{2018 World Cup}}
      \end{figure*}

      \begin{figure*}
        \centering
        \fbox{\includegraphics[width=0.98\textwidth]{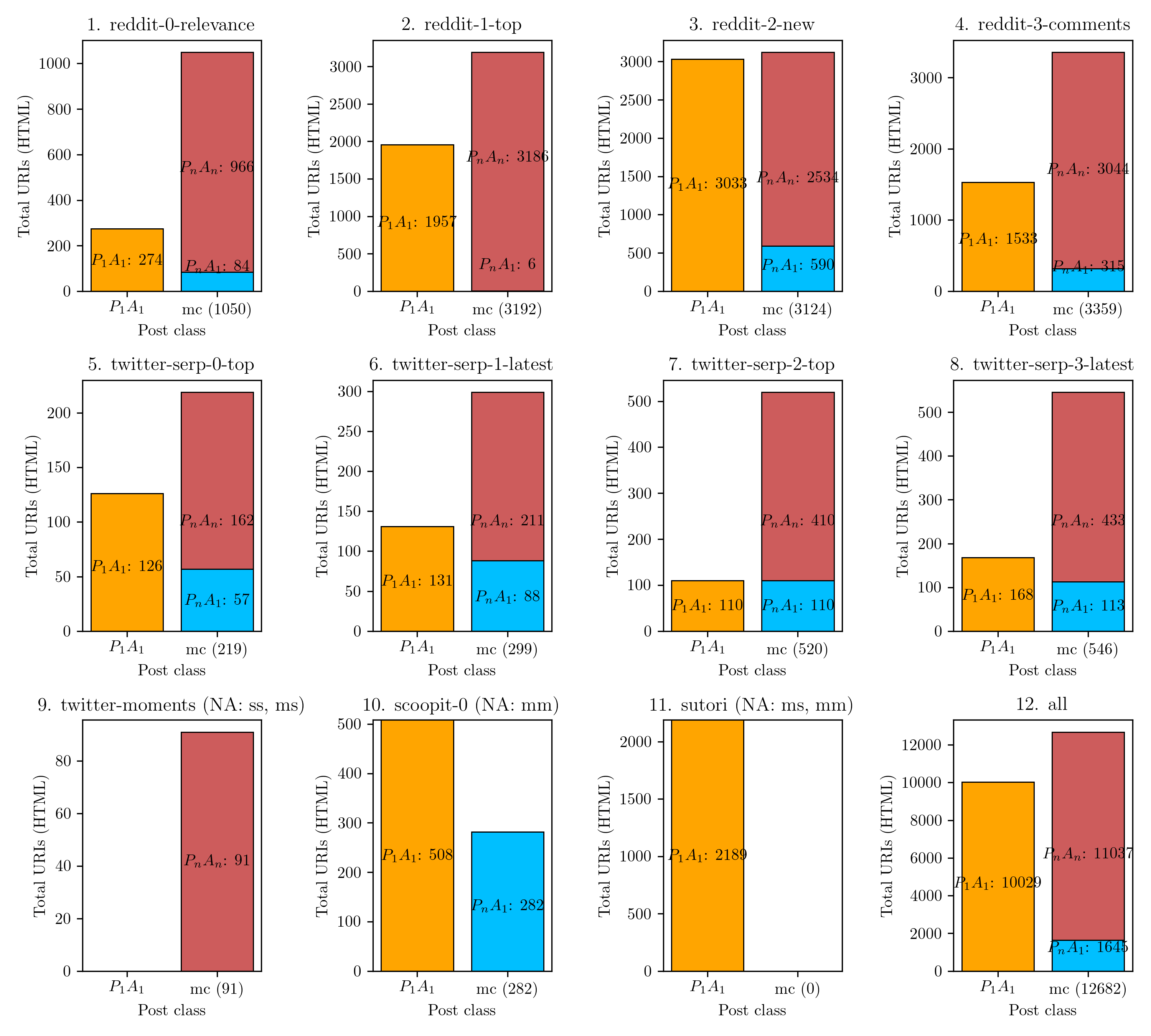}}  
        \caption{Total number of HTML URIs per post class, per social media for \textit{2018 Midterm Elections}}
      \end{figure*}

\clearpage
\begin{figure*}
   \captionsetup{font=Large}
   \centering
   \caption*{APPENDIX 3\\Total number of non-HTML URIs per post class, per social media.}
\end{figure*}

      \begin{figure*}
      \centering
      \fbox{\includegraphics[width=0.98\textwidth]{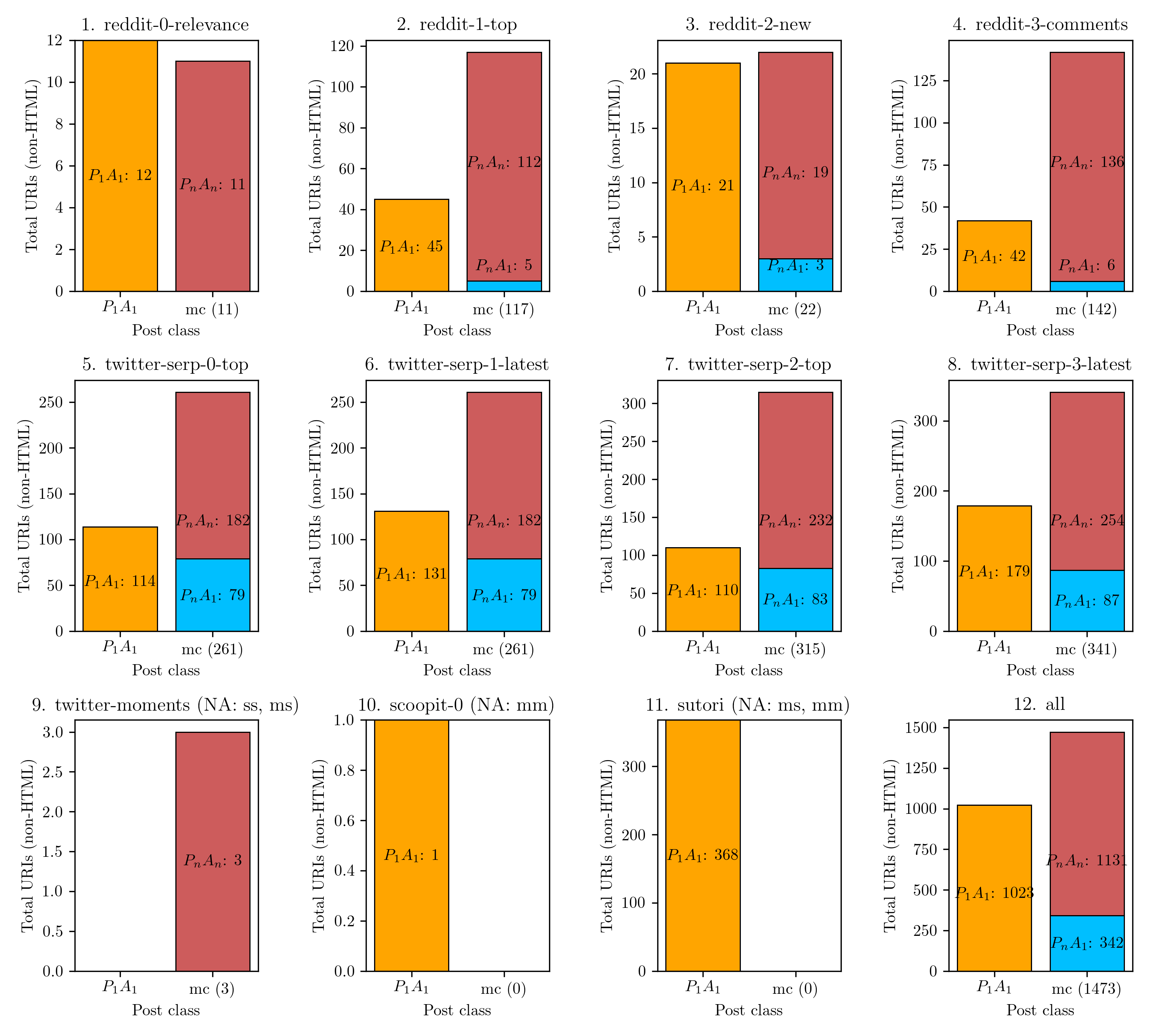}}  
      \caption{Total number of non-HTML URIs per post class, per social media for \textit{Ebola Virus Outbreak}. A single sub-figure (e.g., sub-figure 1) reads as follows: There were 12 URIs in the Reddit (\textit{relevance} vertical) \textbf{P$_1$A$_1$} post class and 11 (\textbf{P$_n$A$_1$}: 0 + \textbf{P$_n$A$_n$}: 11) URIs in the \textbf{MC} post class. The remaining figures in this appendix are to be read similarly. \textit{twitter-serp-2-top} and \textit{twitter-serp-3-latest} represent collections generated by issuing hashtag (\texttt{\#ebolavirus}) queries.}
      \end{figure*}

      \begin{figure*}
        \centering
        \fbox{\includegraphics[width=0.98\textwidth]{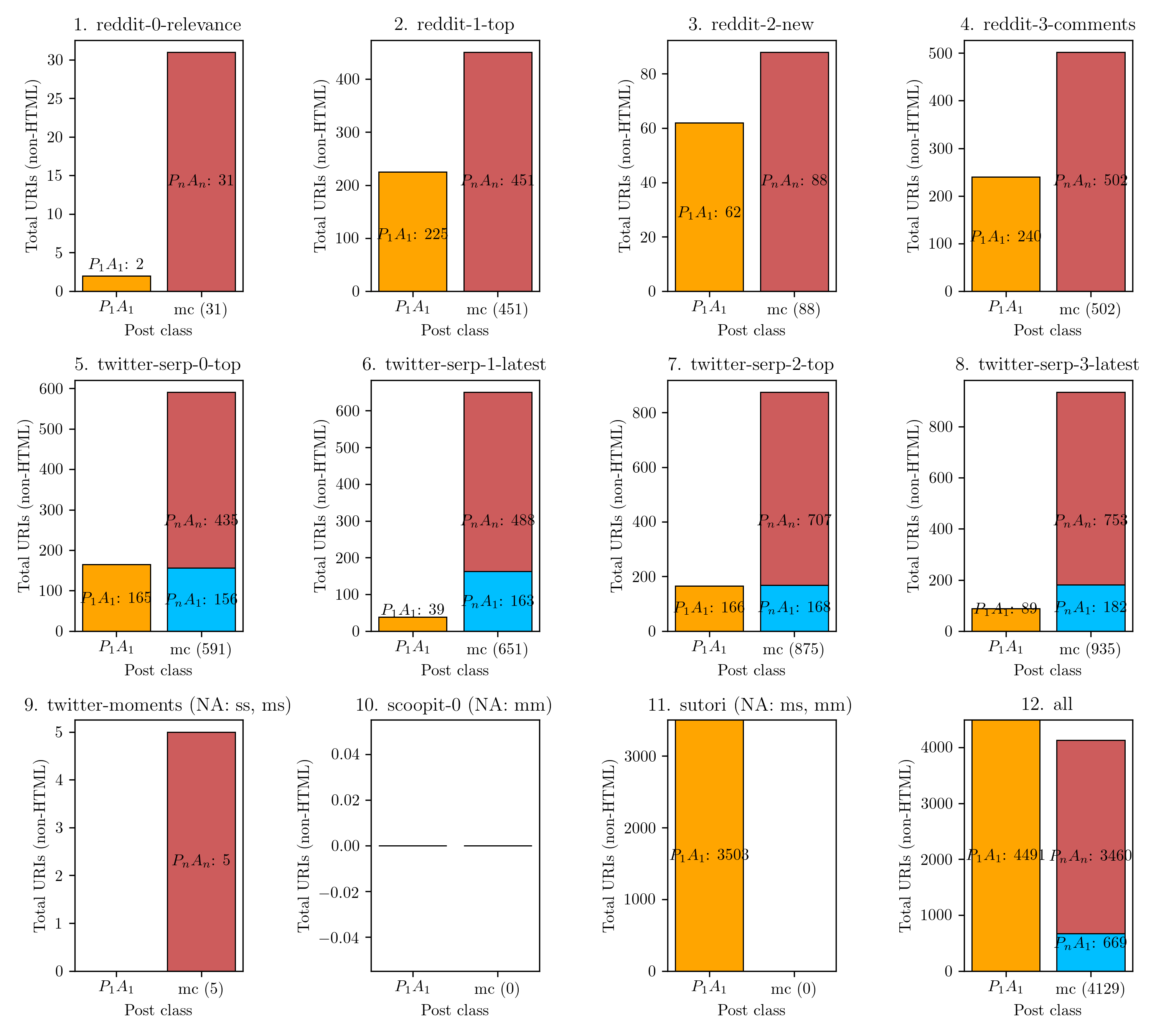}}  
        \caption{Total number of non-HTML URIs per post class, per social media for \textit{Flint Water Crisis}}
      \end{figure*}

      \begin{figure*}
        \centering
        \fbox{\includegraphics[width=0.98\textwidth]{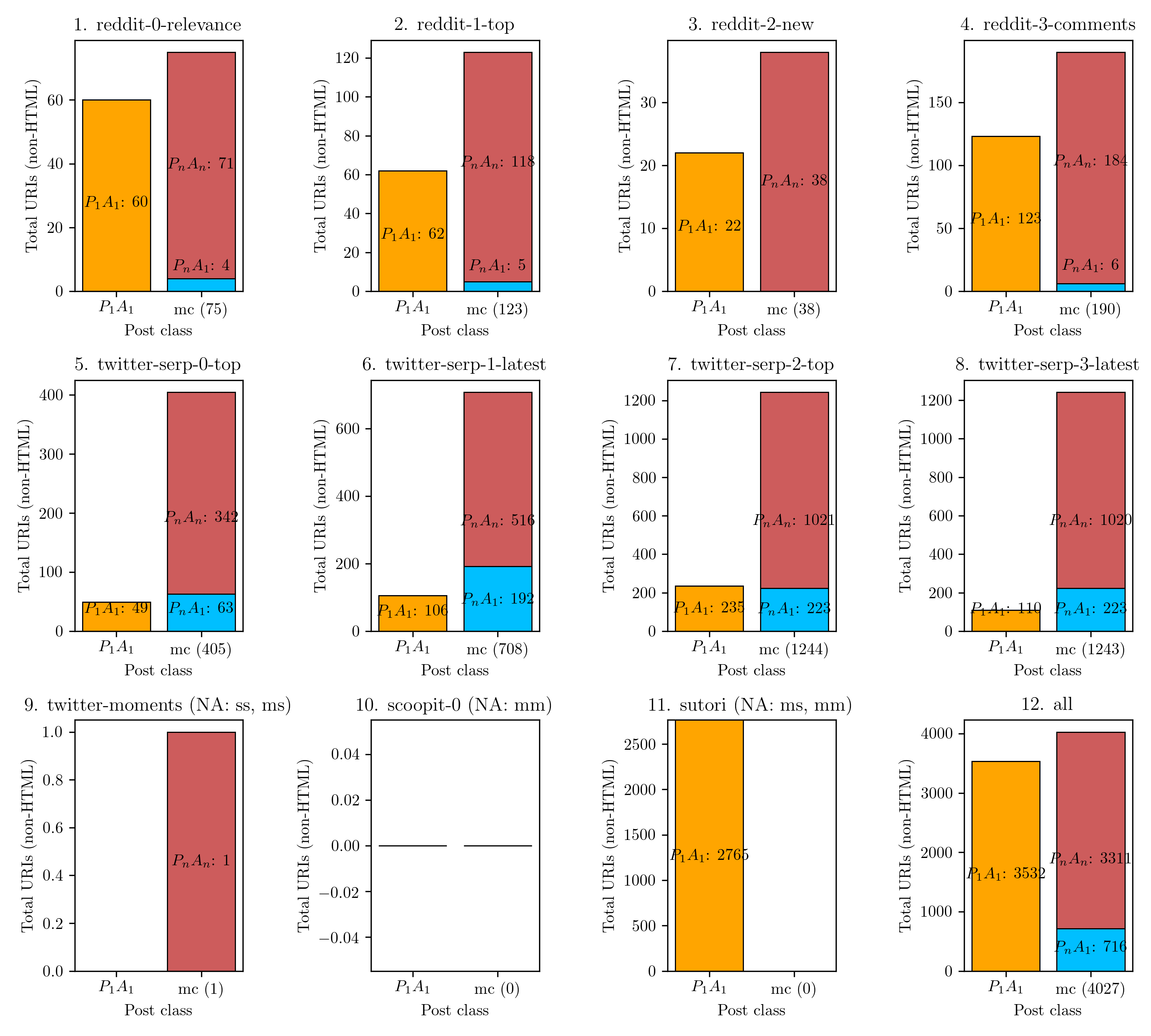}}  
        \caption{Total number of non-HTML URIs per post class, per social media for \textit{MSD Shooting}}
      \end{figure*}

      \begin{figure*}
        \centering
        \fbox{\includegraphics[width=0.98\textwidth]{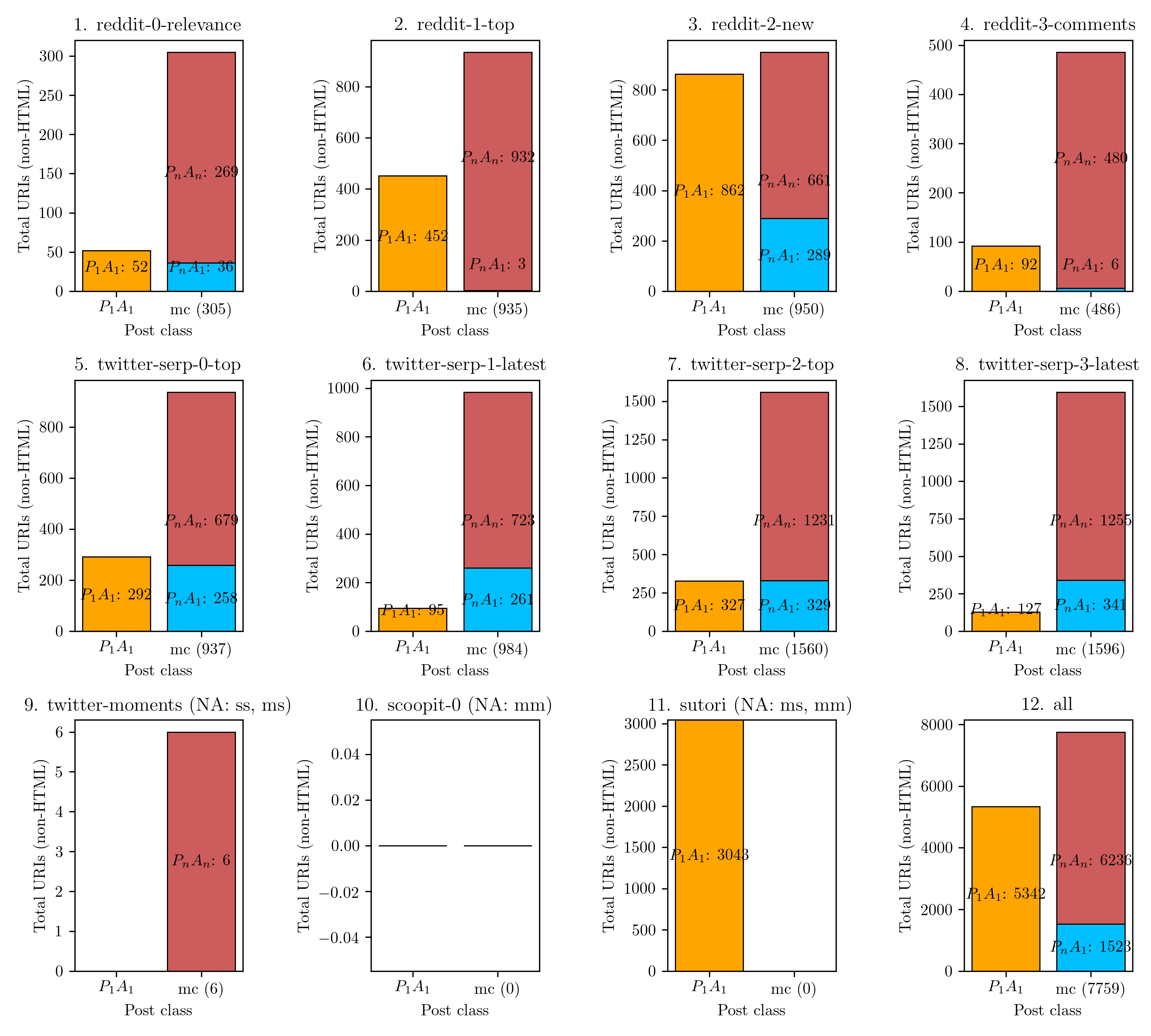}}  
        \caption{Total number of non-HTML URIs per post class, per social media for \textit{2018 World Cup}}
      \end{figure*}

      \begin{figure*}
        \centering
        \fbox{\includegraphics[width=0.98\textwidth]{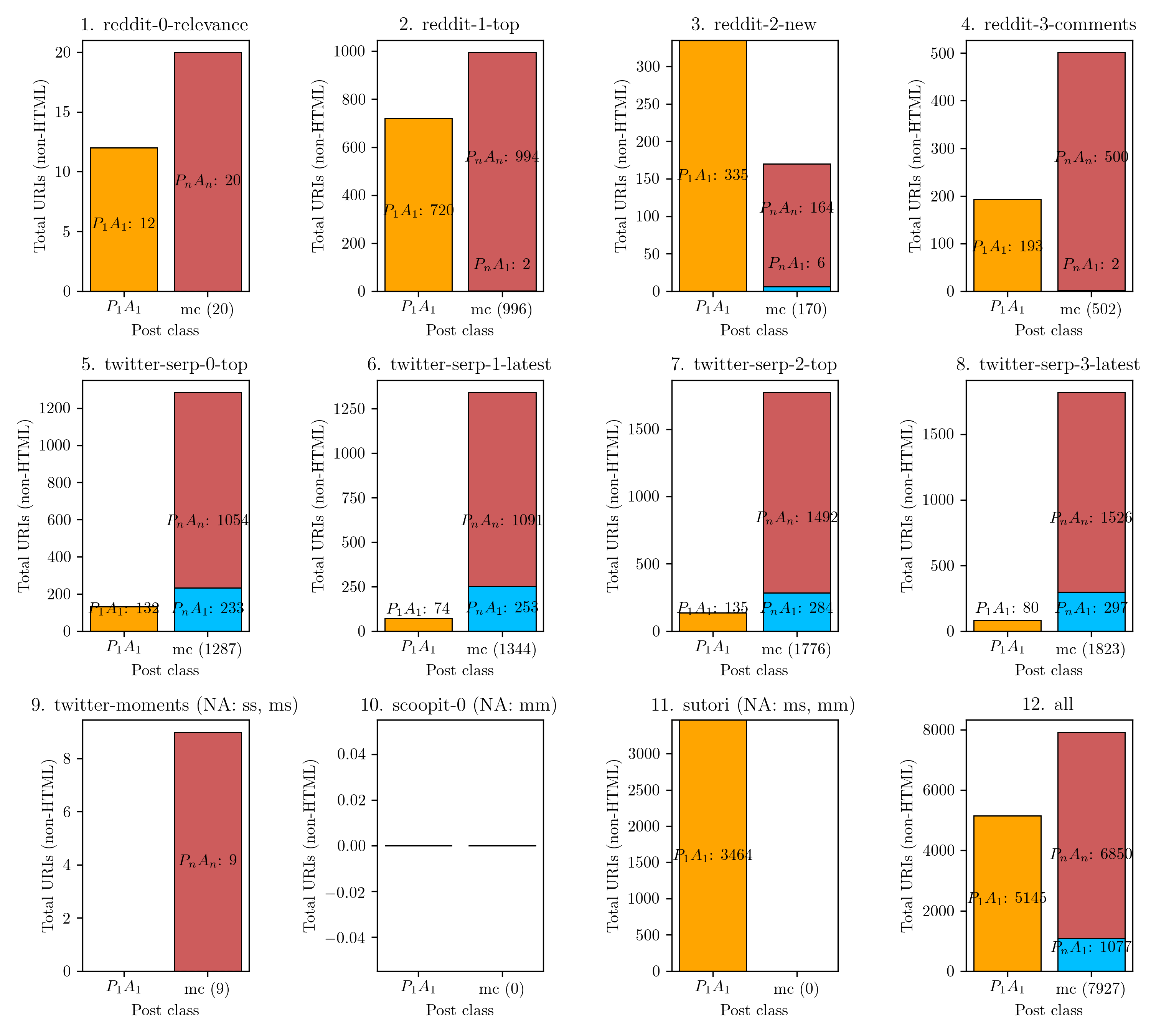}}  
        \caption{Total number of non-HTML URIs per post class, per social media for \textit{2018 Midterm Elections}}
      \end{figure*}

\clearpage
\begin{figure*}
   \captionsetup{font=Large}
   \centering
   \caption*{APPENDIX 4\\Total number URIs (HTML and non-HTML) per post class, per social media.}
\end{figure*}

      \begin{figure*}
      \centering
      \fbox{\includegraphics[width=0.98\textwidth]{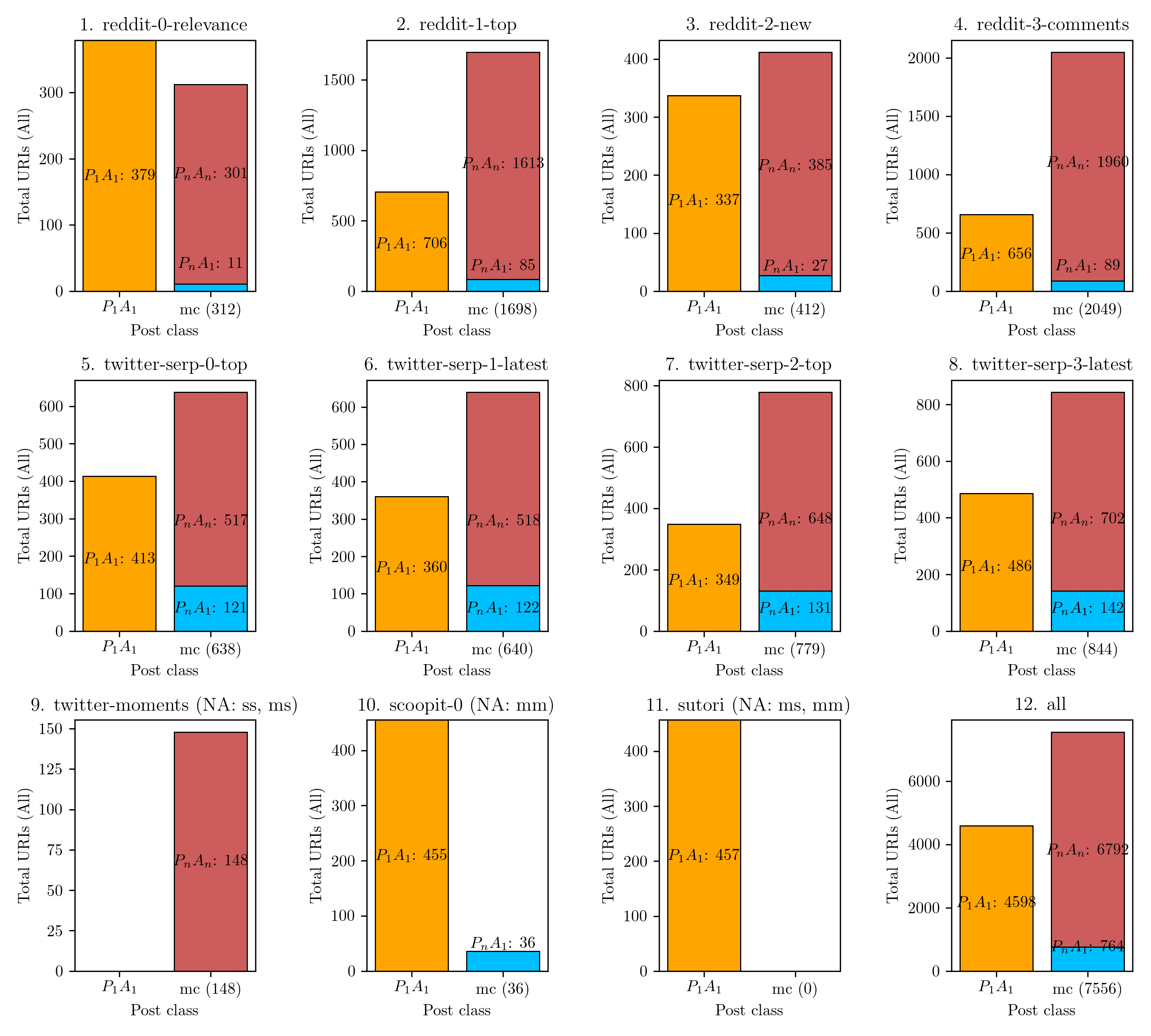}}  
      \caption{Total number of URIs (HTML and non-HTML) per post class, per social media for \textit{Ebola Virus Outbreak}. A single sub-figure (e.g., sub-figure 1) reads as follows: There were 379 URIs in the Reddit (\textit{relevance} vertical) \textbf{P$_1$A$_1$} post class and 312 (\textbf{P$_n$A$_1$}: 11 + \textbf{P$_n$A$_n$}: 301) URIs in the \textbf{MC} post class. The remaining figures in this appendix are to be read similarly. \textit{twitter-serp-2-top} and \textit{twitter-serp-3-latest} represent collections generated by issuing hashtag (\texttt{\#ebolavirus}) queries.}
      \end{figure*}

      \begin{figure*}
        \centering
        \fbox{\includegraphics[width=0.98\textwidth]{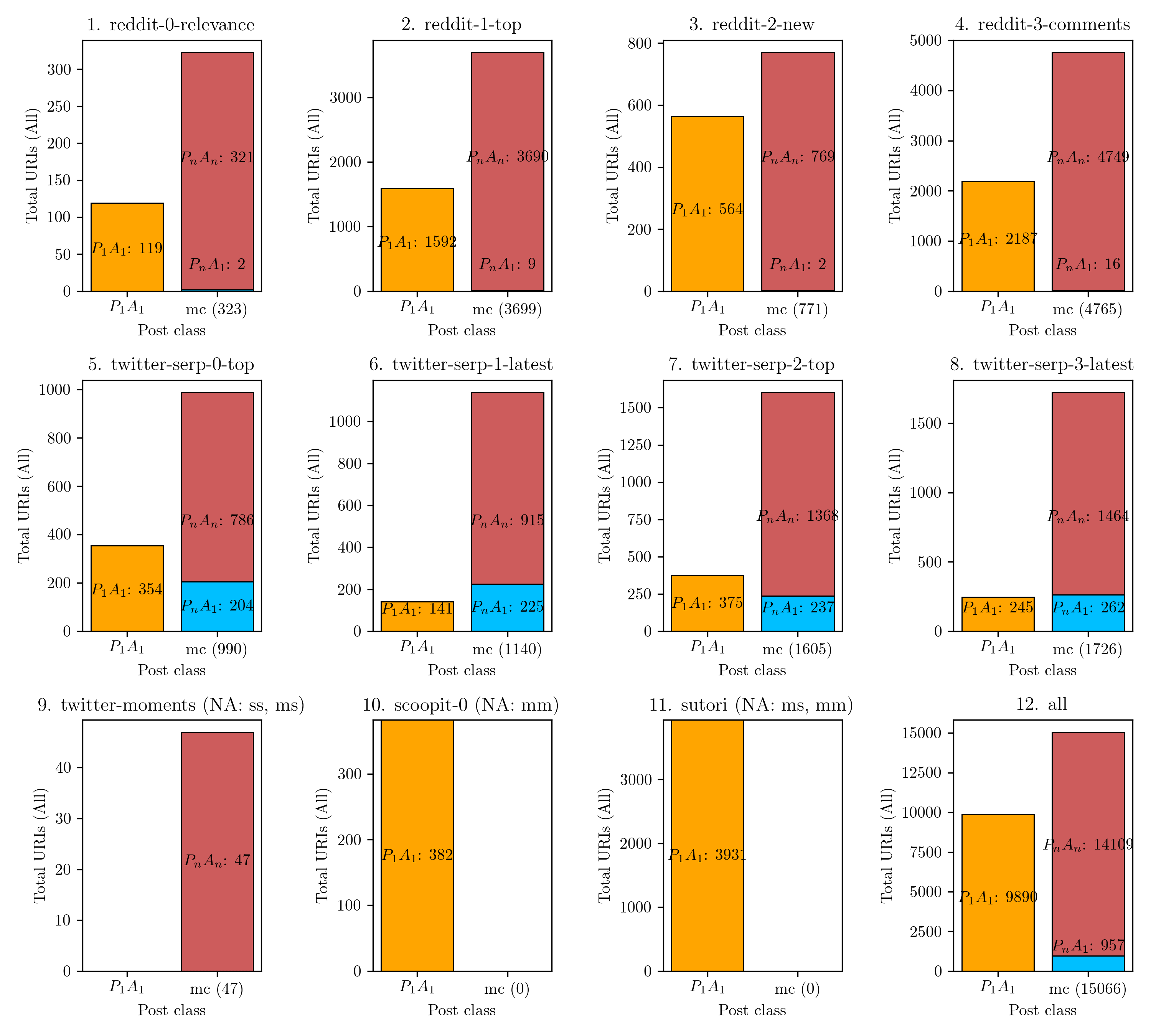}}  
        \caption{Total number of URIs (HTML and non-HTML) per post class, per social media for \textit{Flint Water Crisis}}
      \end{figure*}

      \begin{figure*}
        \centering
        \fbox{\includegraphics[width=0.98\textwidth]{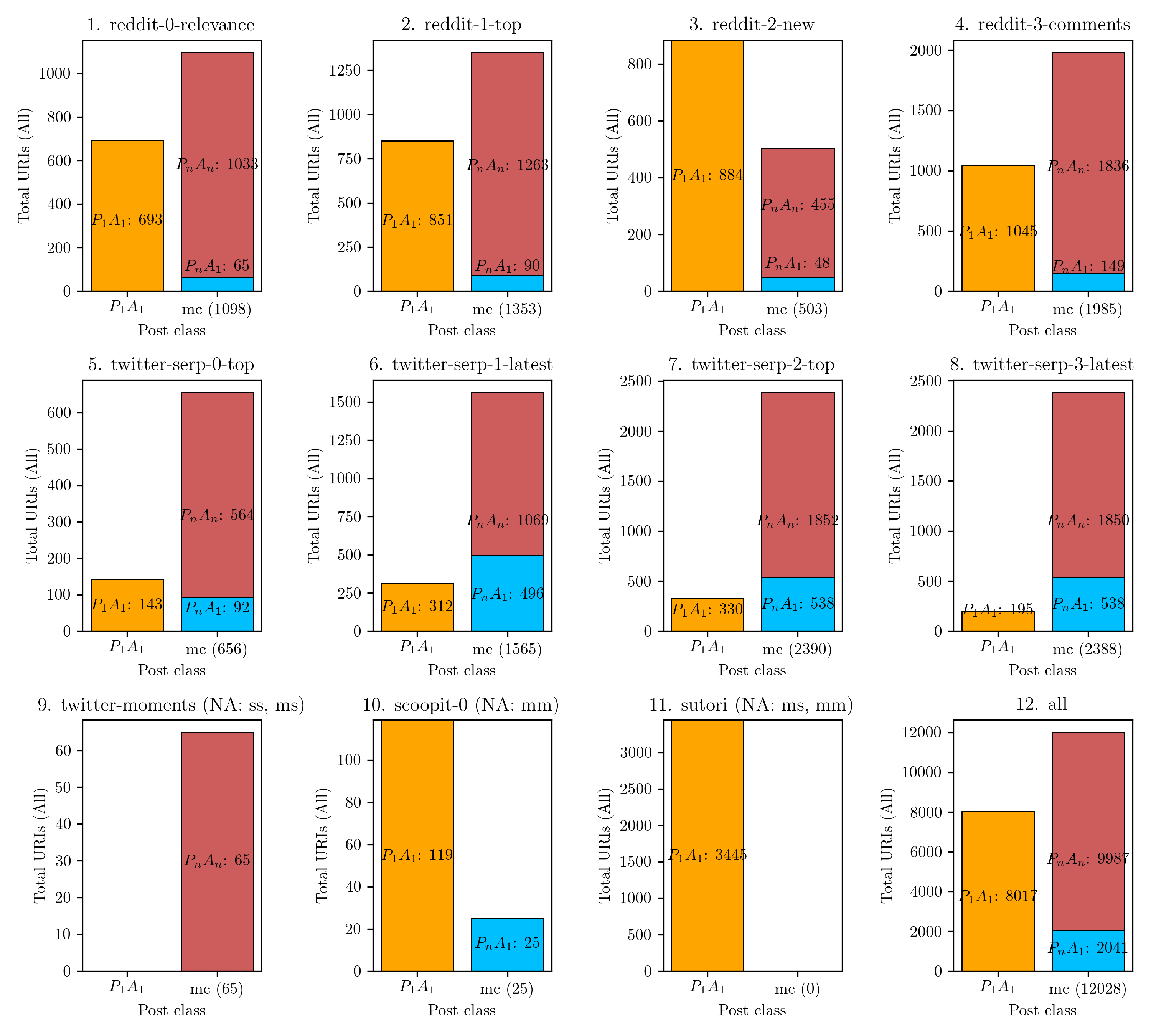}}  
        \caption{Total number of URIs (HTML and non-HTML) per post class, per social media for \textit{MSD Shooting}}
      \end{figure*}

      \begin{figure*}
        \centering
        \fbox{\includegraphics[width=0.98\textwidth]{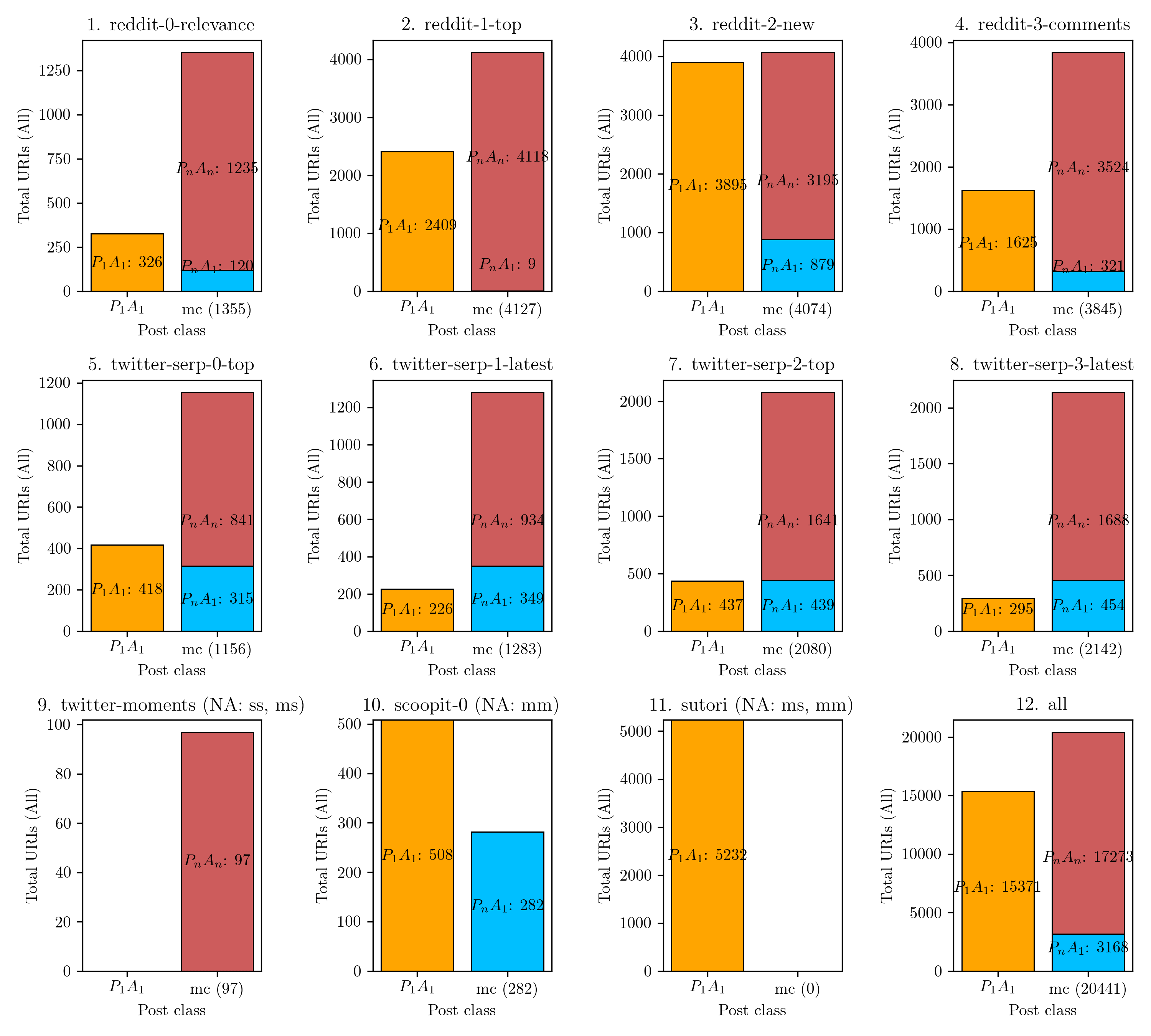}}  
        \caption{Total number of URIs (HTML and non-HTML) per post class, per social media for \textit{2018 World Cup}}
      \end{figure*} 

      \begin{figure*}
        \centering
        \fbox{\includegraphics[width=0.98\textwidth]{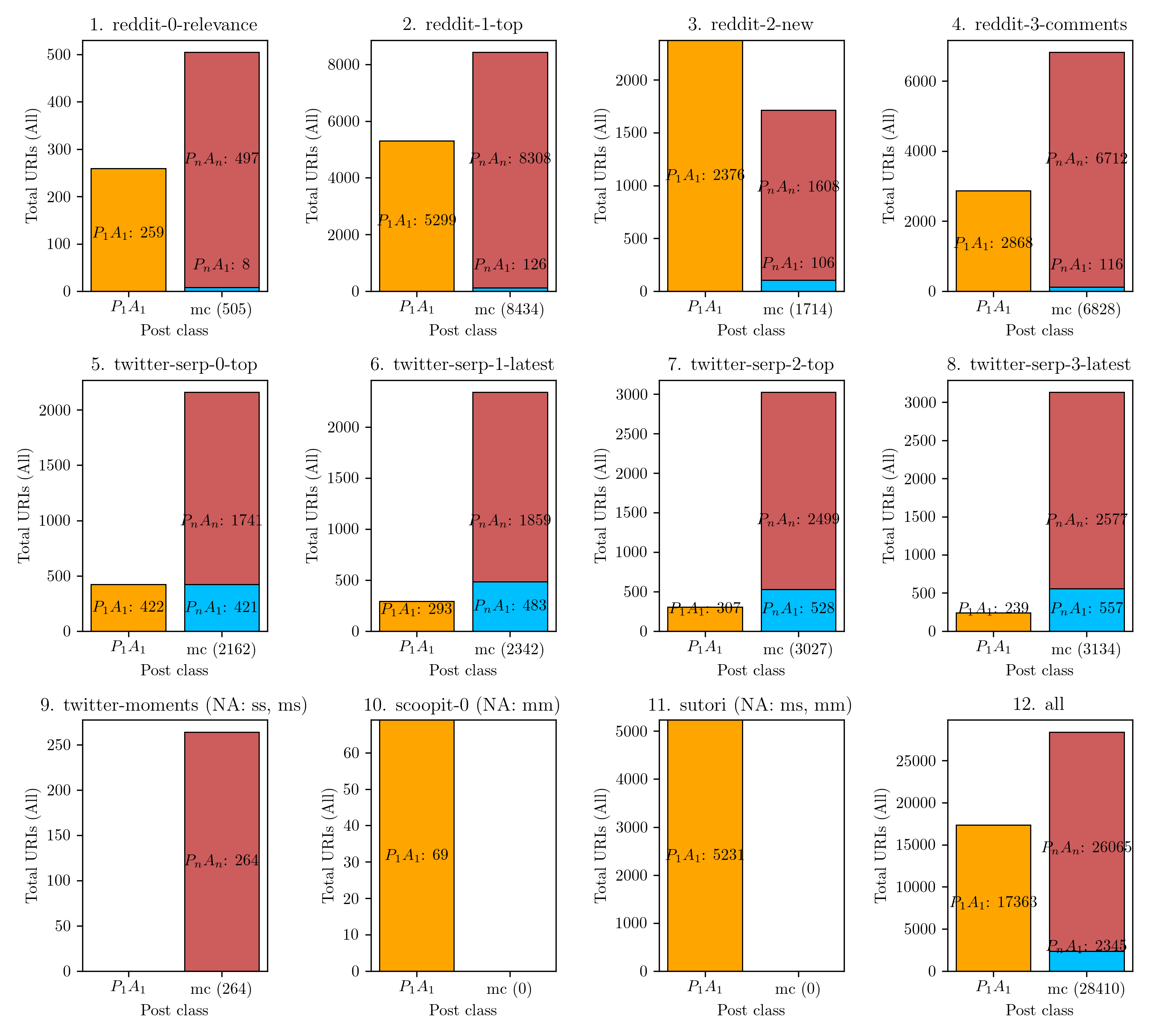}}  
        \caption{Total number of URIs (HTML and non-HTML) per post class, per social media for \textit{2018 Midterm Elections}}
      \end{figure*}

\clearpage
\begin{figure*}
   \captionsetup{font=Large}
   \centering
   \caption*{APPENDIX 5\\Average precision of HTML URIs per post class, per social media.}
\end{figure*}

      \begin{figure*}
      \centering
      \fbox{\includegraphics[width=0.98\textwidth]{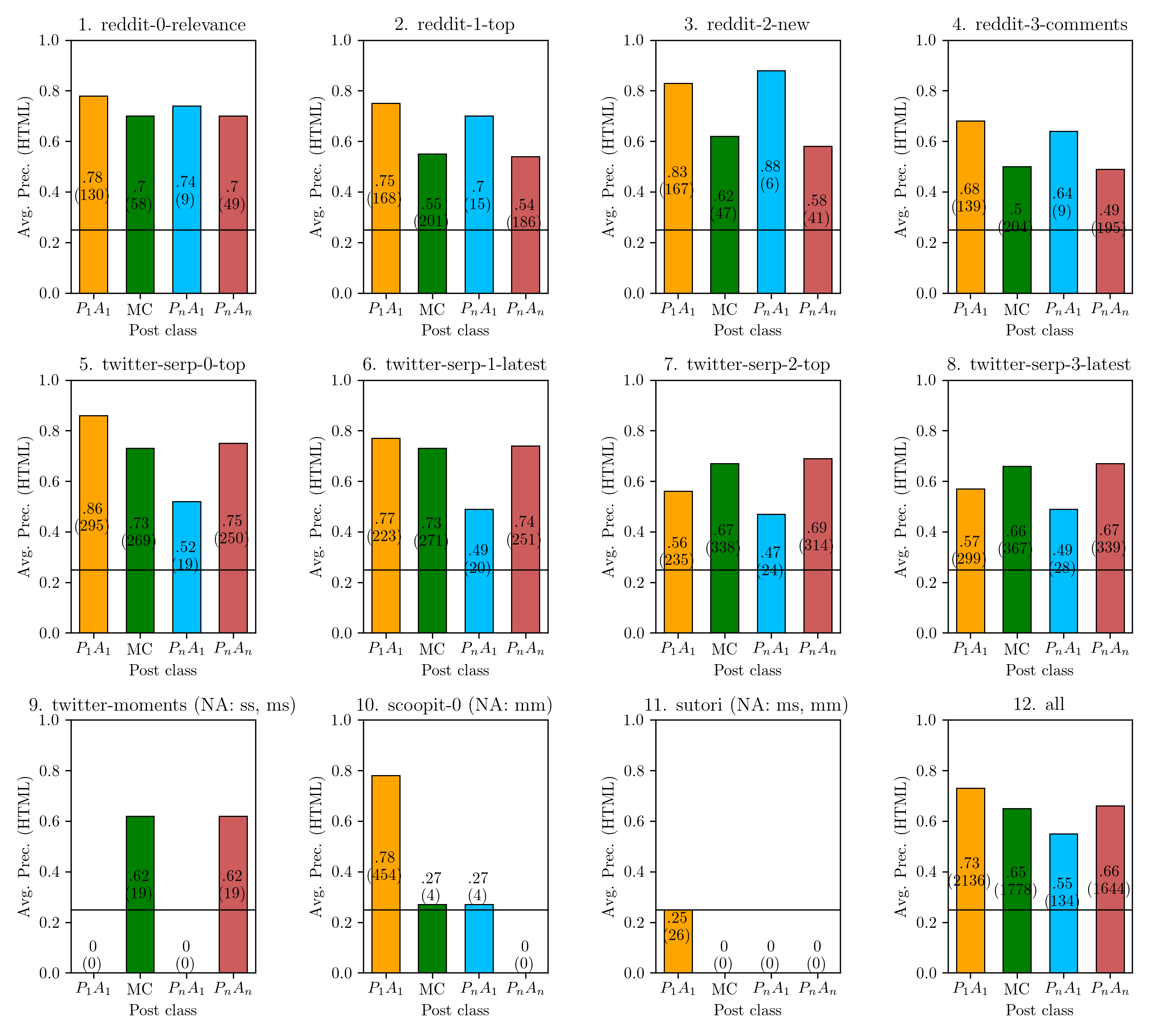}}  
      \caption{Average precision of HTML URIs per post class, per social media for \textit{Ebola Virus Outbreak}. A single sub-figure (e.g., sub-figure 1) reads as follows: The average precision for the Reddit (\textit{relevance} vertical) \textbf{P$_1$A$_1$} post class was 0.78. This average was calculate from 130 Reddit \textbf{P$_1$A$_1$} posts. Similarly, the average precision of the Reddit \textbf{MC} post class was 0.7, averaged across 58 Reddit \textbf{MC} posts. The remaining figures in this appendix are to be read similarly. \textit{twitter-serp-2-top} and \textit{twitter-serp-3-latest} represent collections generated by issuing hashtag (\texttt{\#ebolavirus}) queries.}
      \end{figure*}

      \begin{figure*}
        \centering
        \fbox{\includegraphics[width=0.98\textwidth]{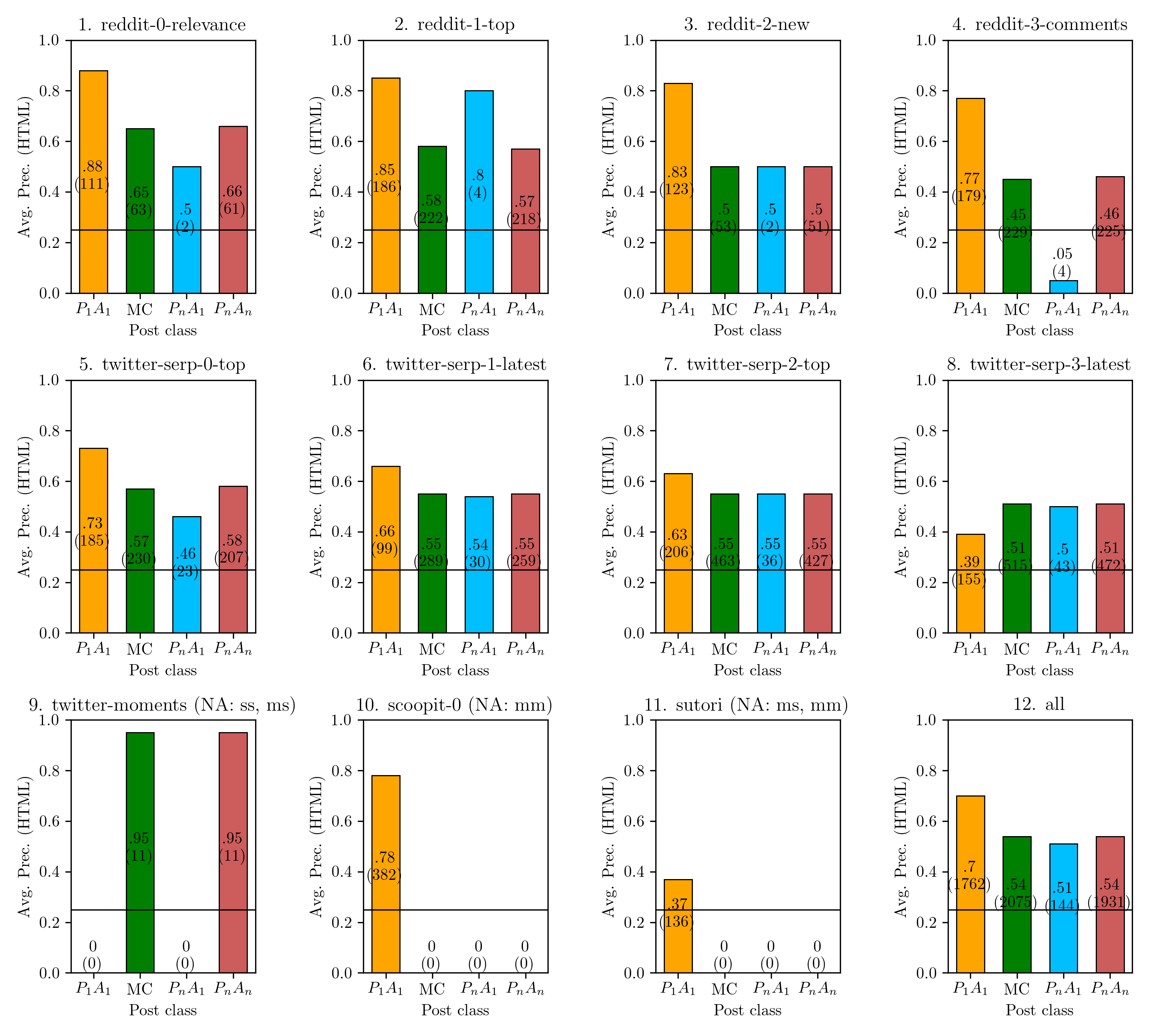}}  
        \caption{Average precision of HTML URIs per post class, per social media for \textit{Flint Water Crisis}}
      \end{figure*}

      \begin{figure*}
        \centering
        \fbox{\includegraphics[width=0.98\textwidth]{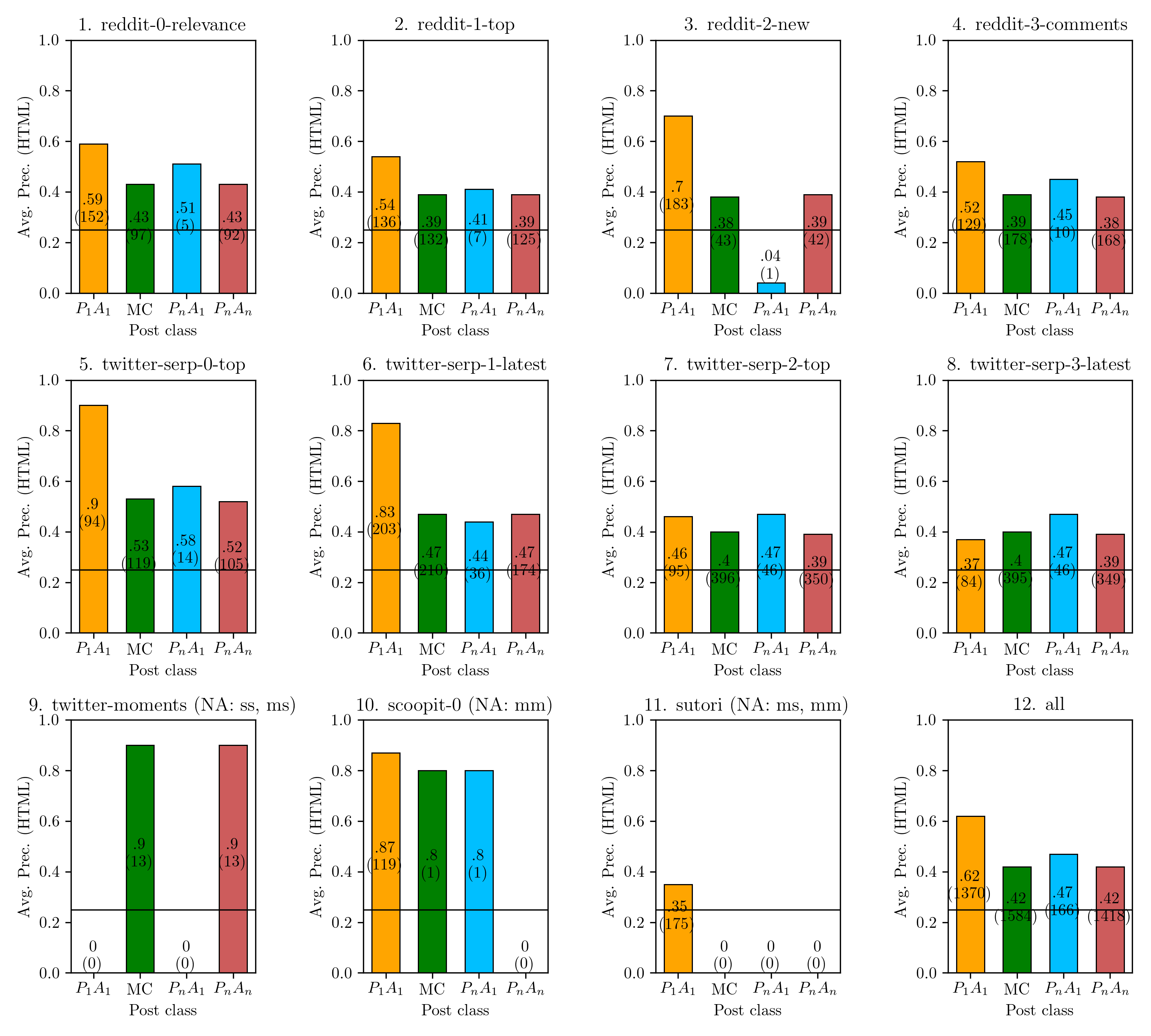}}  
        \caption{Average precision of HTML URIs per post class, per social media for \textit{MSD Shooting}}
      \end{figure*}

      \begin{figure*}
        \centering
        \fbox{\includegraphics[width=0.98\textwidth]{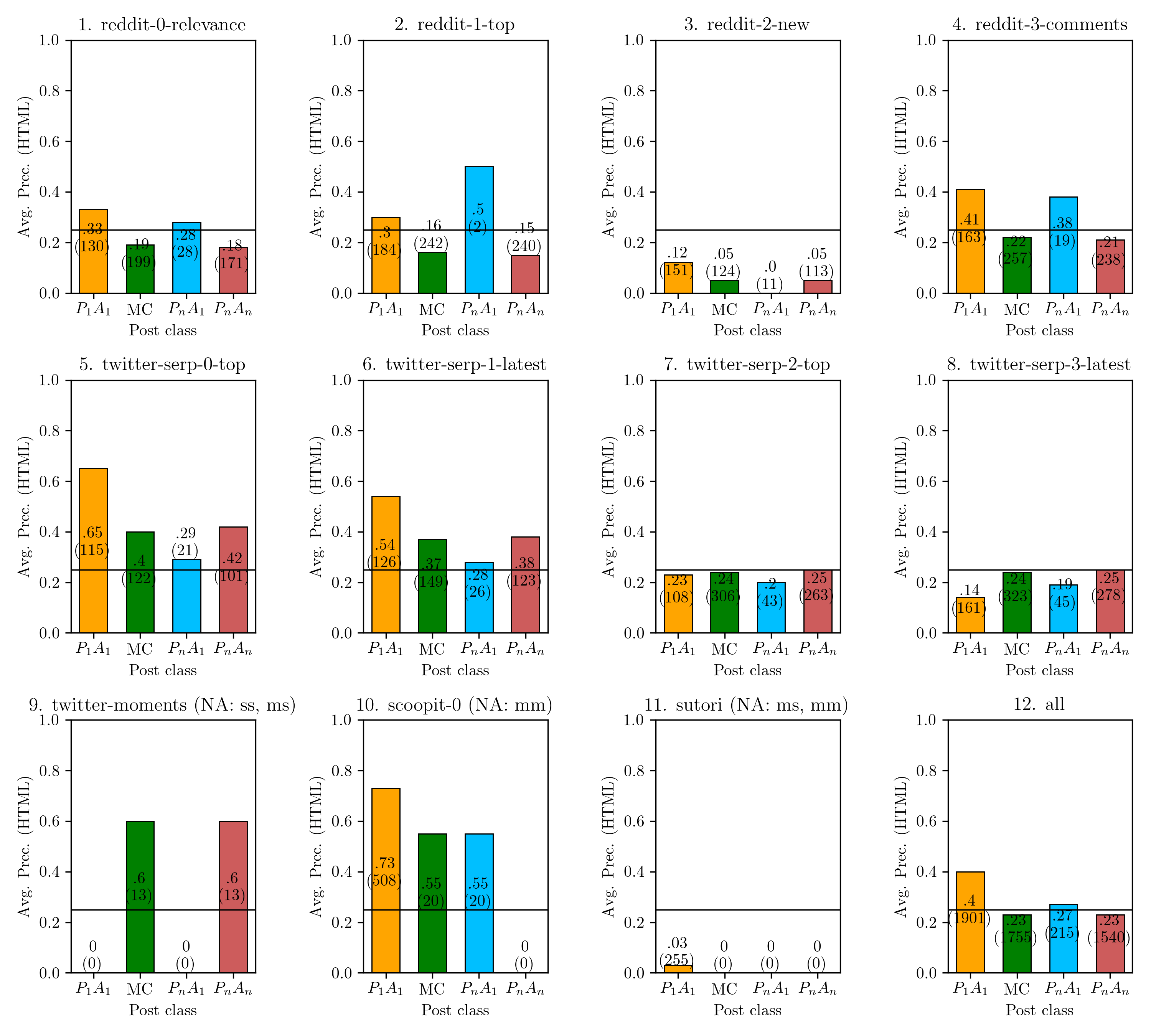}}  
        \caption{Average precision of HTML URIs per post class, per social media for \textit{2018 World Cup}}
      \end{figure*}

        \begin{figure*}
        \centering
        \fbox{\includegraphics[width=0.98\textwidth]{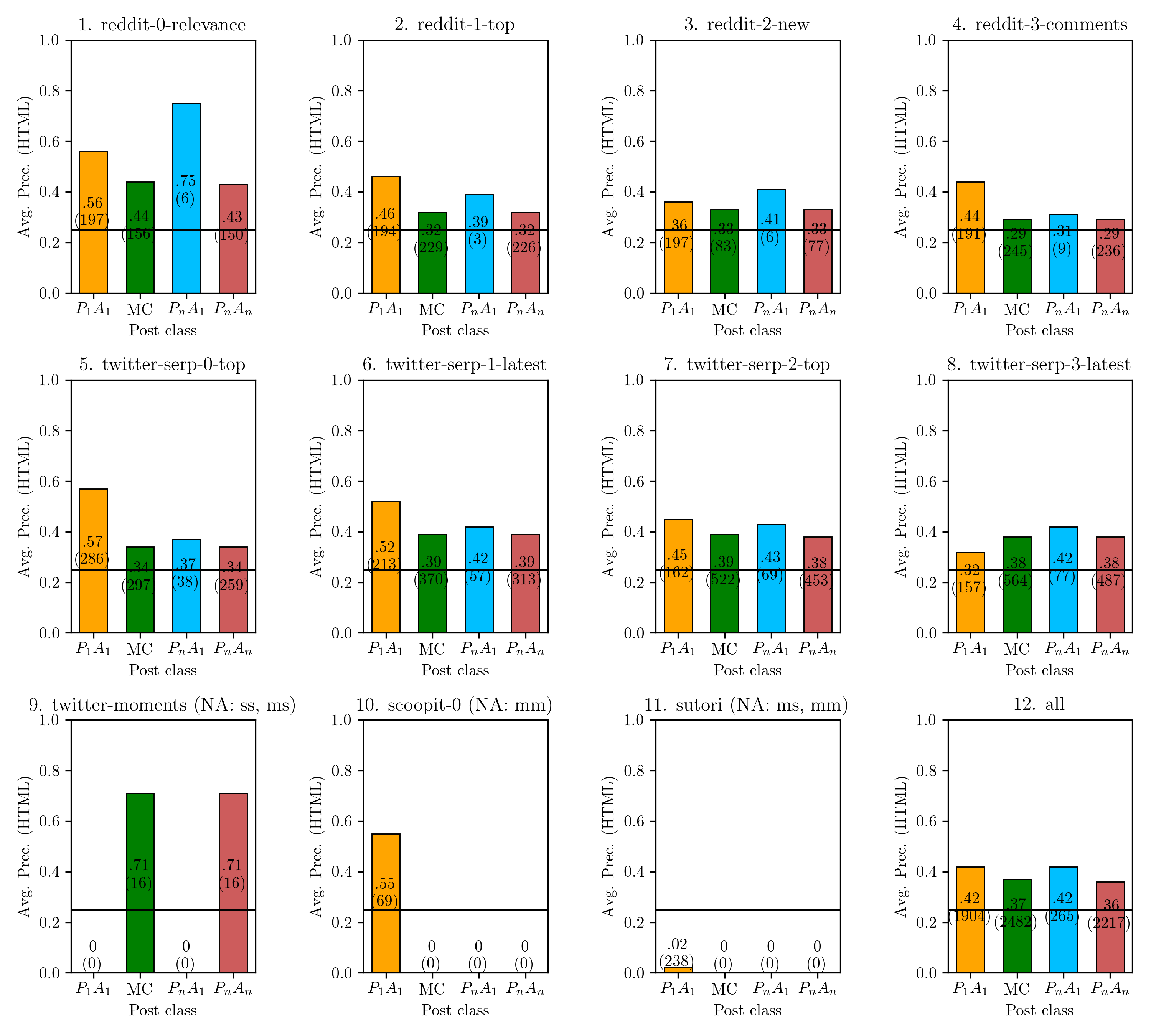}}  
        \caption{Average precision of HTML URIs per post class, per social media for \textit{2018 Midterm Elections}}
      \end{figure*}

\clearpage
\begin{figure*}
   \captionsetup{font=Large}
   \centering
   \caption*{APPENDIX 6\\Average precision of non-HTML URIs per post class, per social media.}
\end{figure*}
      
      \begin{figure*}
      \centering
      \fbox{\includegraphics[width=0.98\textwidth]{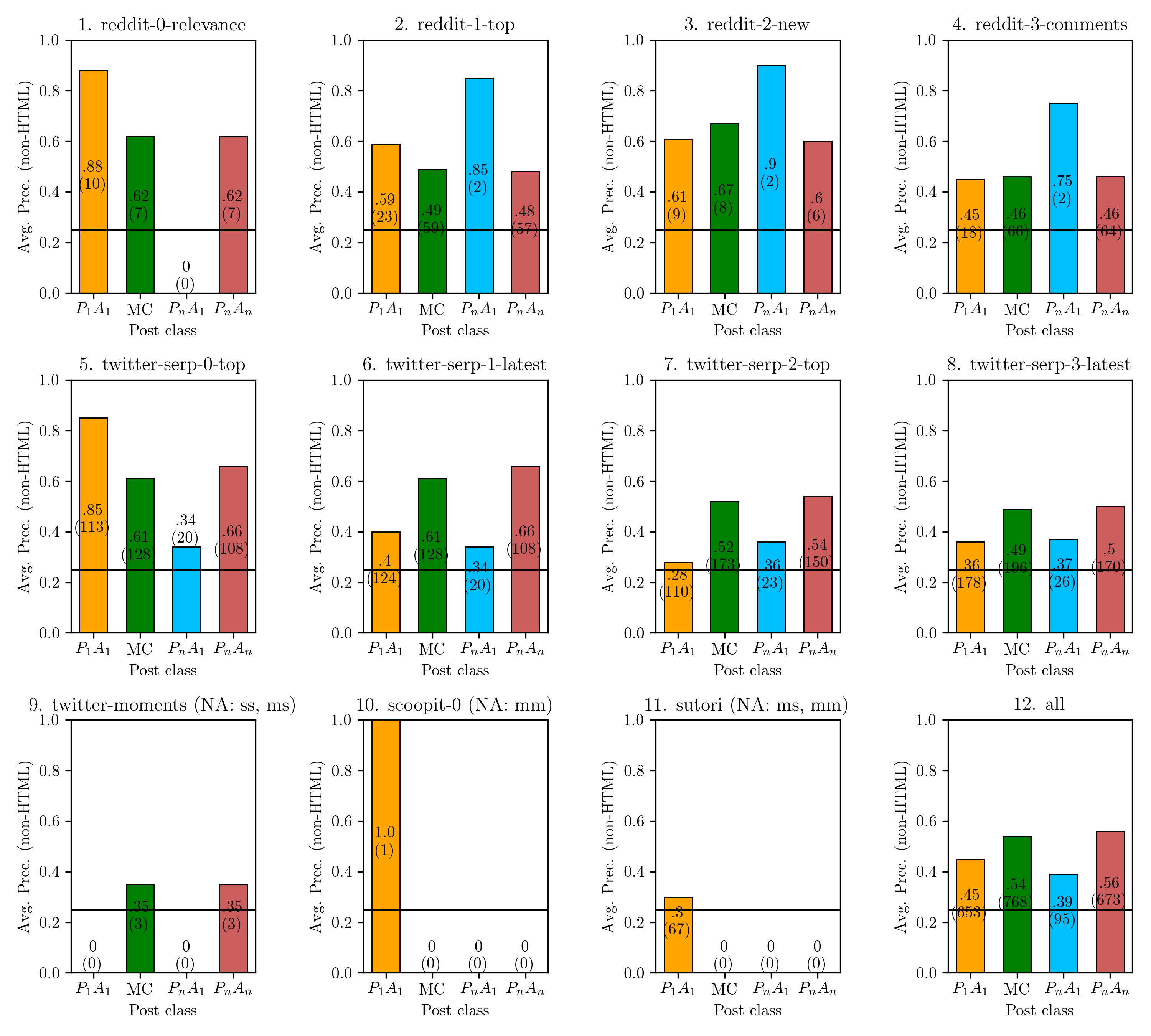}}  
      \caption{Average precision of non-HTML URIs per post class, per social media for \textit{Ebola Virus Outbreak}. A single sub-figure (e.g., sub-figure 1) reads as follows: The average precision for the Reddit (\textit{relevance} vertical) \textbf{P$_1$A$_1$} post class was 0.88. This average was calculate from 10 Reddit \textbf{P$_1$A$_1$} posts. Similarly, the average precision of the Reddit \textbf{MC} post class was 0.62, averaged across 7 Reddit \textbf{MC} posts. The remaining figures in this appendix are to be read similarly. \textit{twitter-serp-2-top} and \textit{twitter-serp-3-latest} represent collections generated by issuing hashtag (\texttt{\#ebolavirus}) queries.}
      \end{figure*}

      \begin{figure*}
        \centering
        \fbox{\includegraphics[width=0.98\textwidth]{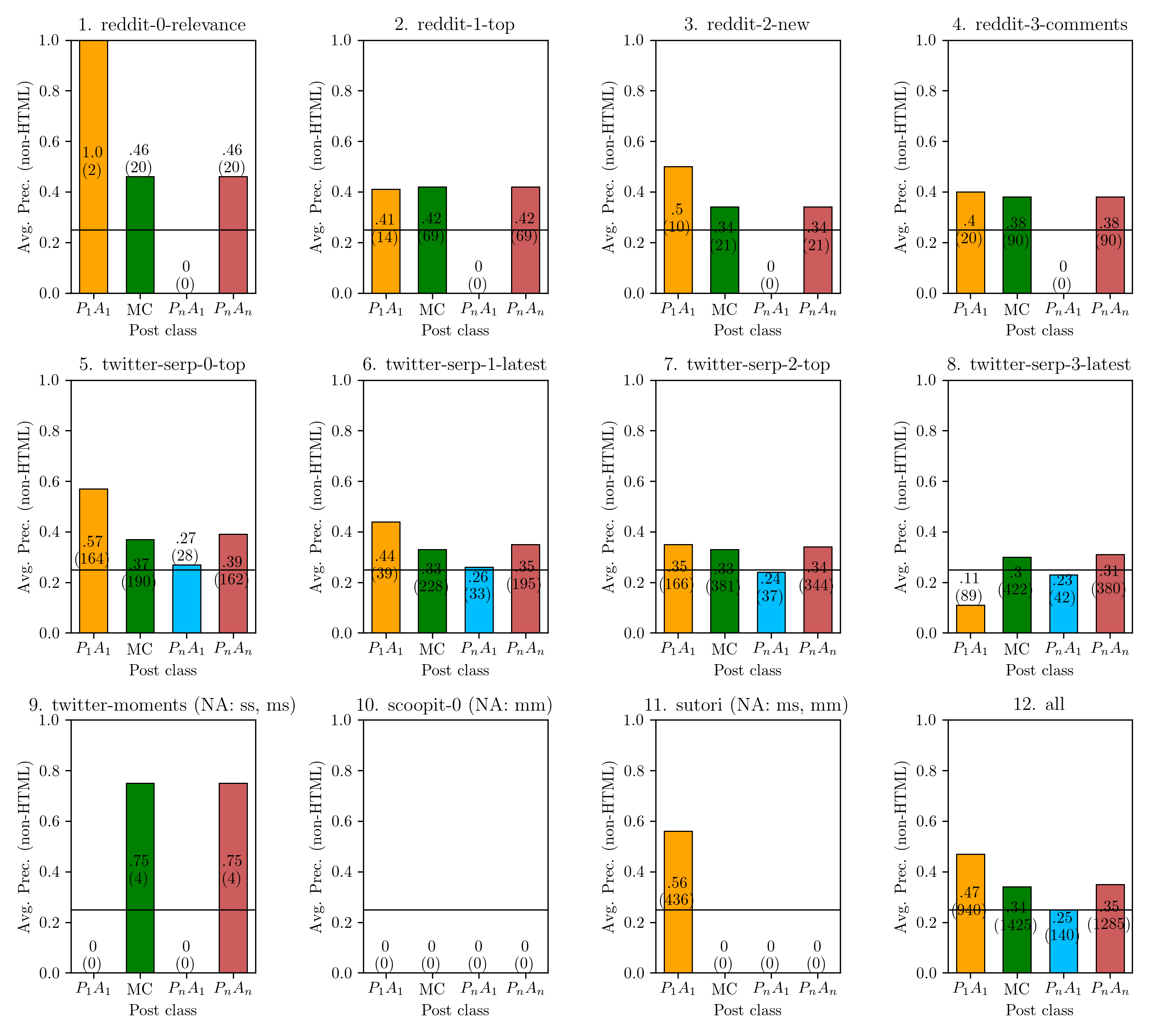}}  
        \caption{Average precision of non-HTML URIs per post class, per social media for \textit{Flint Water Crisis}}
      \end{figure*}

      \begin{figure*}
        \centering
        \fbox{\includegraphics[width=0.98\textwidth]{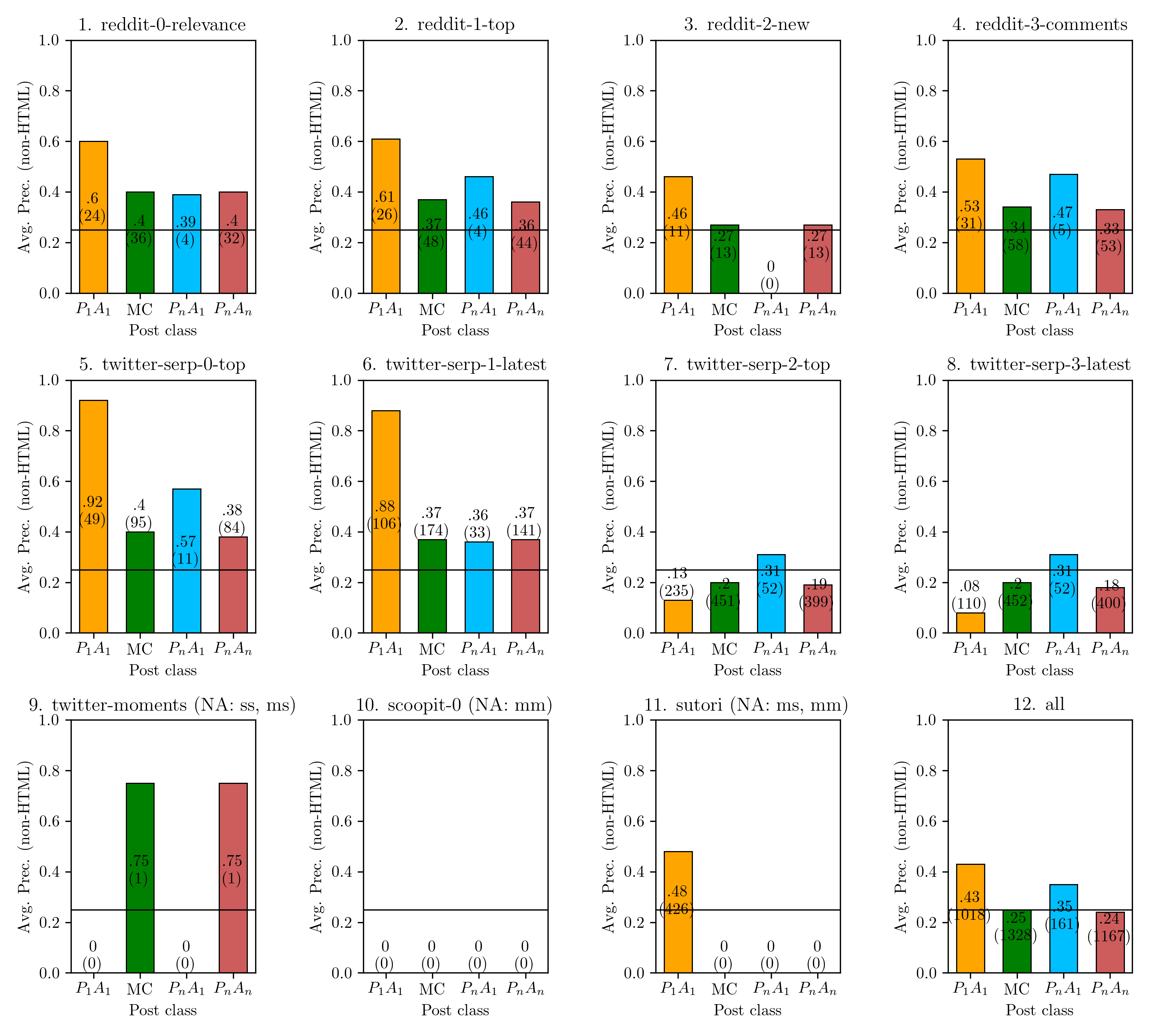}}  
        \caption{Average precision of non-HTML URIs per post class, per social media for \textit{MSD Shooting}}
      \end{figure*}

      \begin{figure*}
        \centering
        \fbox{\includegraphics[width=0.98\textwidth]{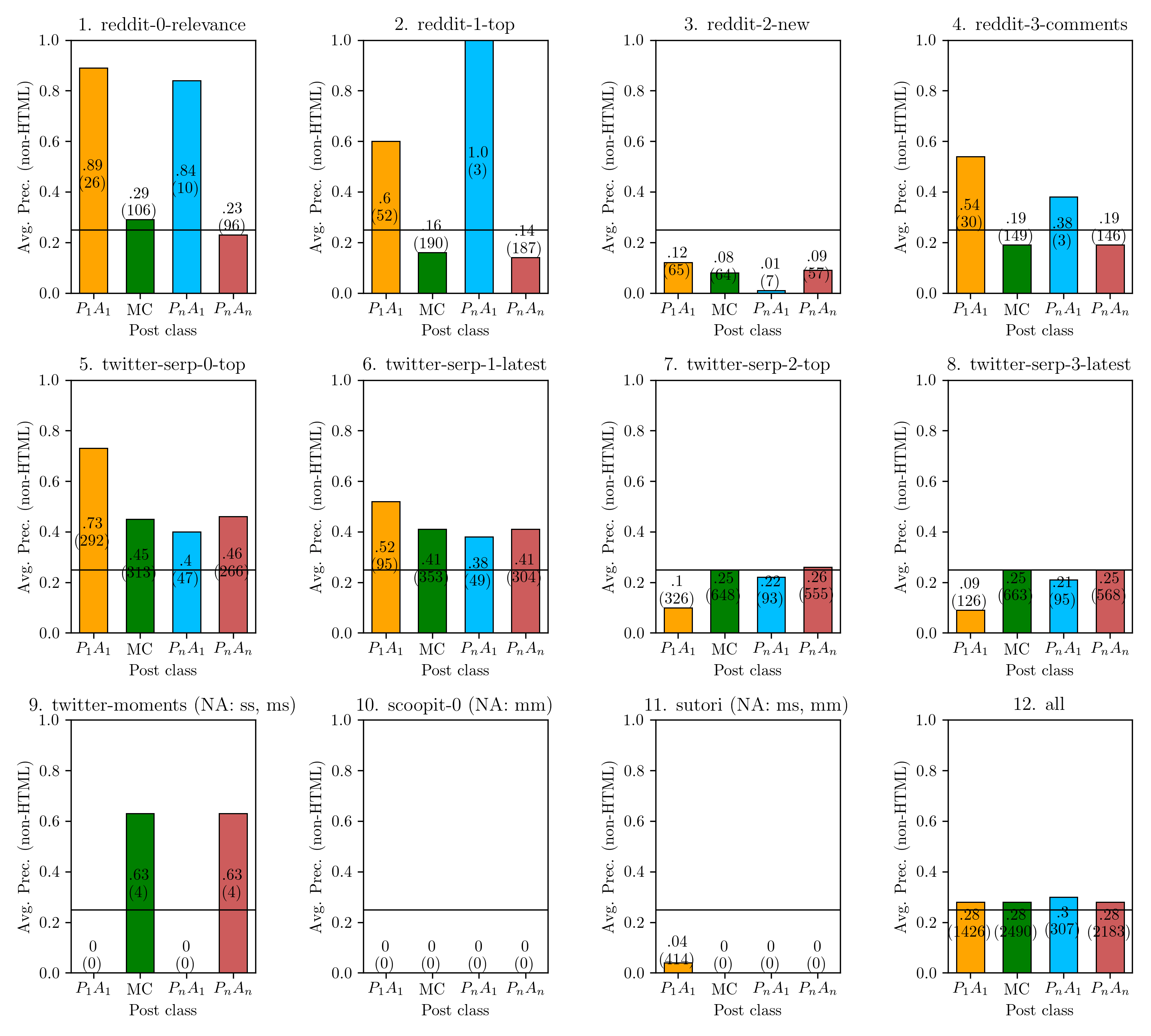}}  
        \caption{Average precision of non-HTML URIs per post class, per social media for \textit{2018 World Cup}}
      \end{figure*} 

      \begin{figure*}
        \centering
        \fbox{\includegraphics[width=0.98\textwidth]{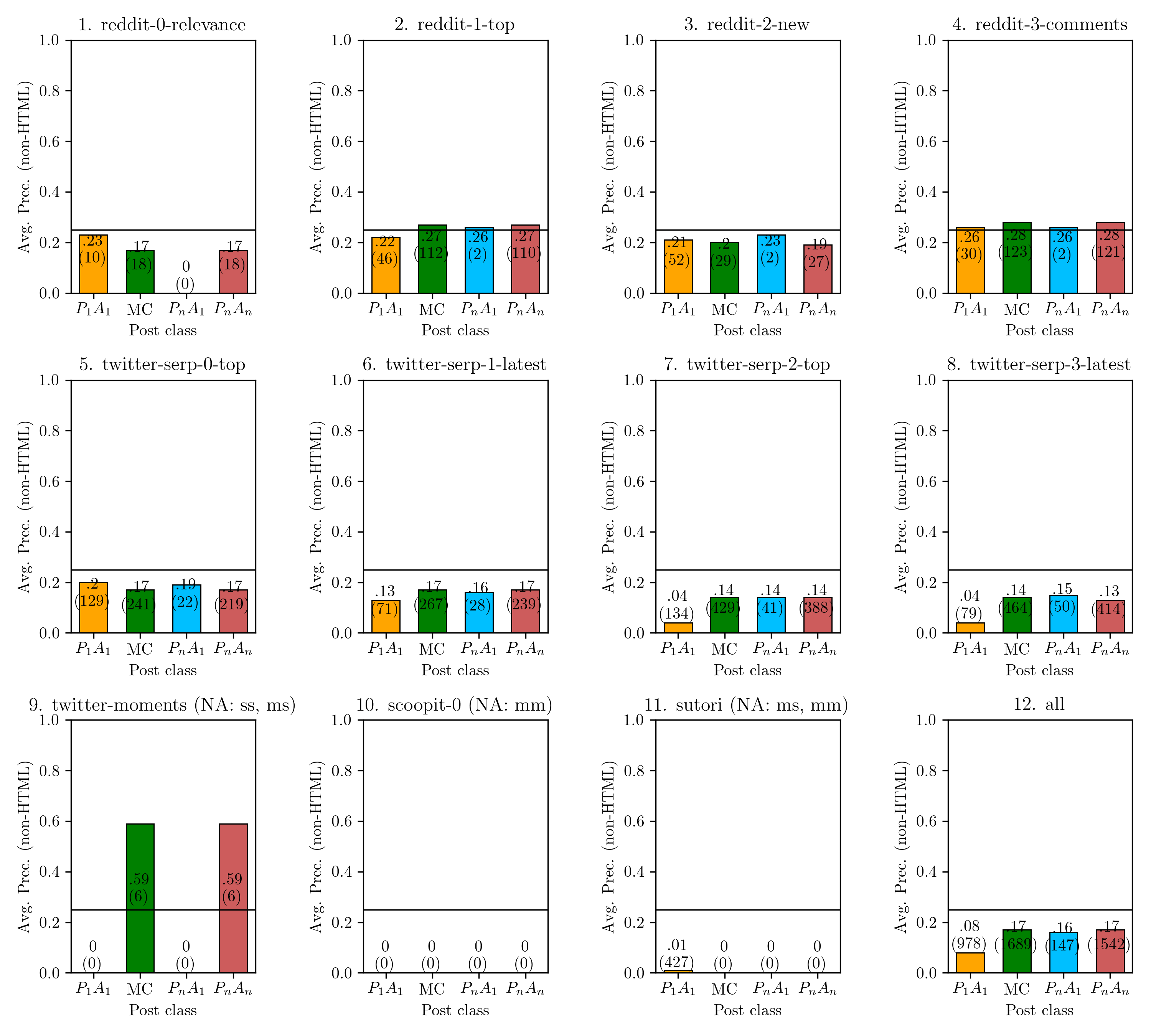}}  
        \caption{Average precision of non-HTML URIs per post class, per social media for \textit{2018 Midterm Elections}}
      \end{figure*}

\clearpage
\begin{figure*}
   \captionsetup{font=Large}
   \centering
   \caption*{APPENDIX 7\\Average precision of URIs (HTML and non-HTML) per post class, per social media.}
\end{figure*}
      
      \begin{figure*}
      \centering
      \fbox{\includegraphics[width=0.98\textwidth]{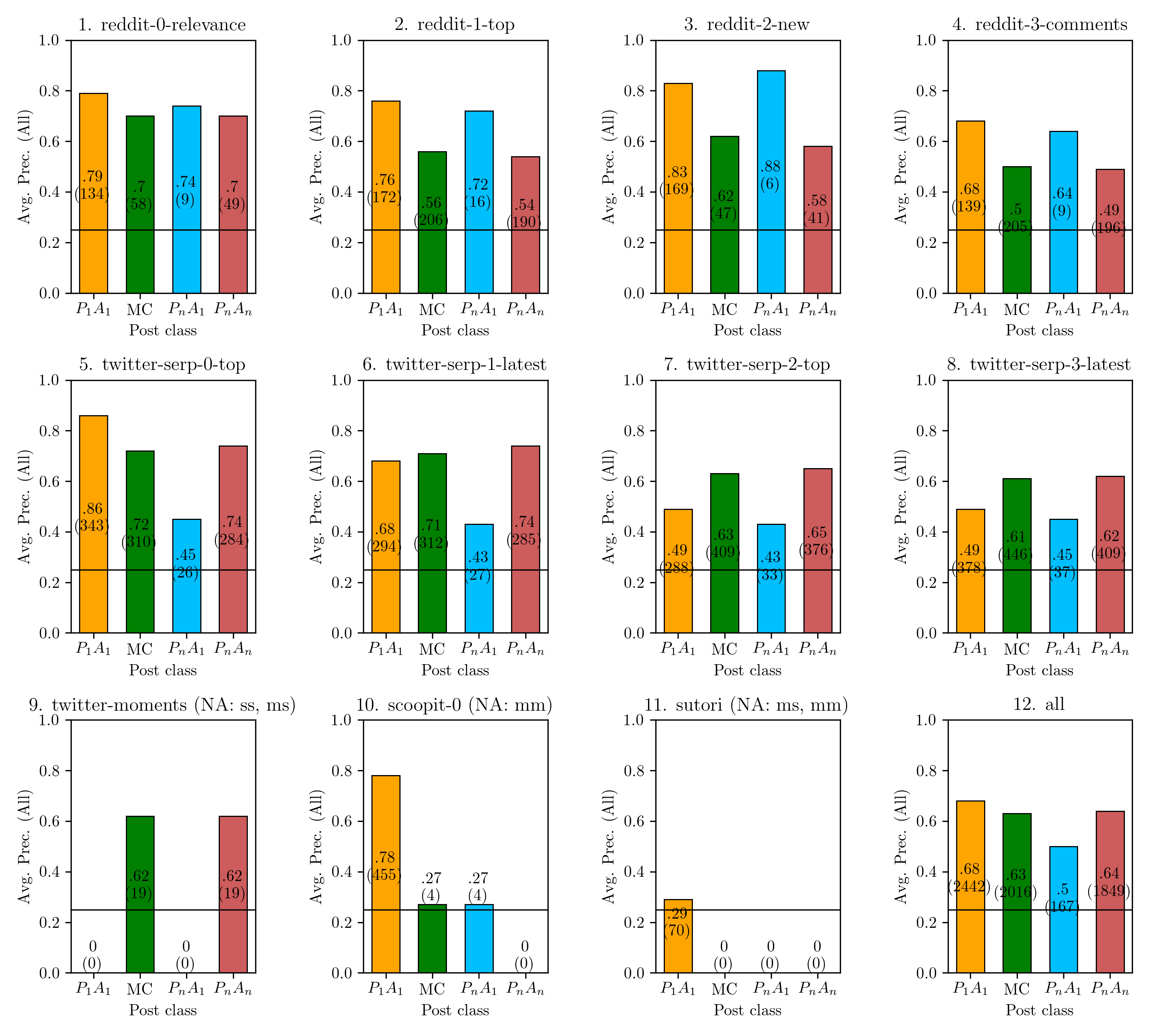}}  
      \caption{Average precision of URIs (HTML and non-HTML) per post class, per social media for \textit{Ebola Virus Outbreak}. A single sub-figure (e.g., sub-figure 1) reads as follows: The average precision for the Reddit (\textit{relevance} vertical) \textbf{P$_1$A$_1$} post class was 0.79. This average was calculated from 134 Reddit \textbf{P$_1$A$_1$} posts. Similarly, the average precision of the Reddit \textbf{MC} post class was 0.7, averaged across 58 Reddit \textbf{MC} posts. The remaining figures in this appendix are to be read similarly. \textit{twitter-serp-2-top} and \textit{twitter-serp-3-latest} represent collections generated by issuing hashtag (\texttt{\#ebolavirus}) queries.}
      \end{figure*}

      \begin{figure*}
        \centering
        \fbox{\includegraphics[width=0.98\textwidth]{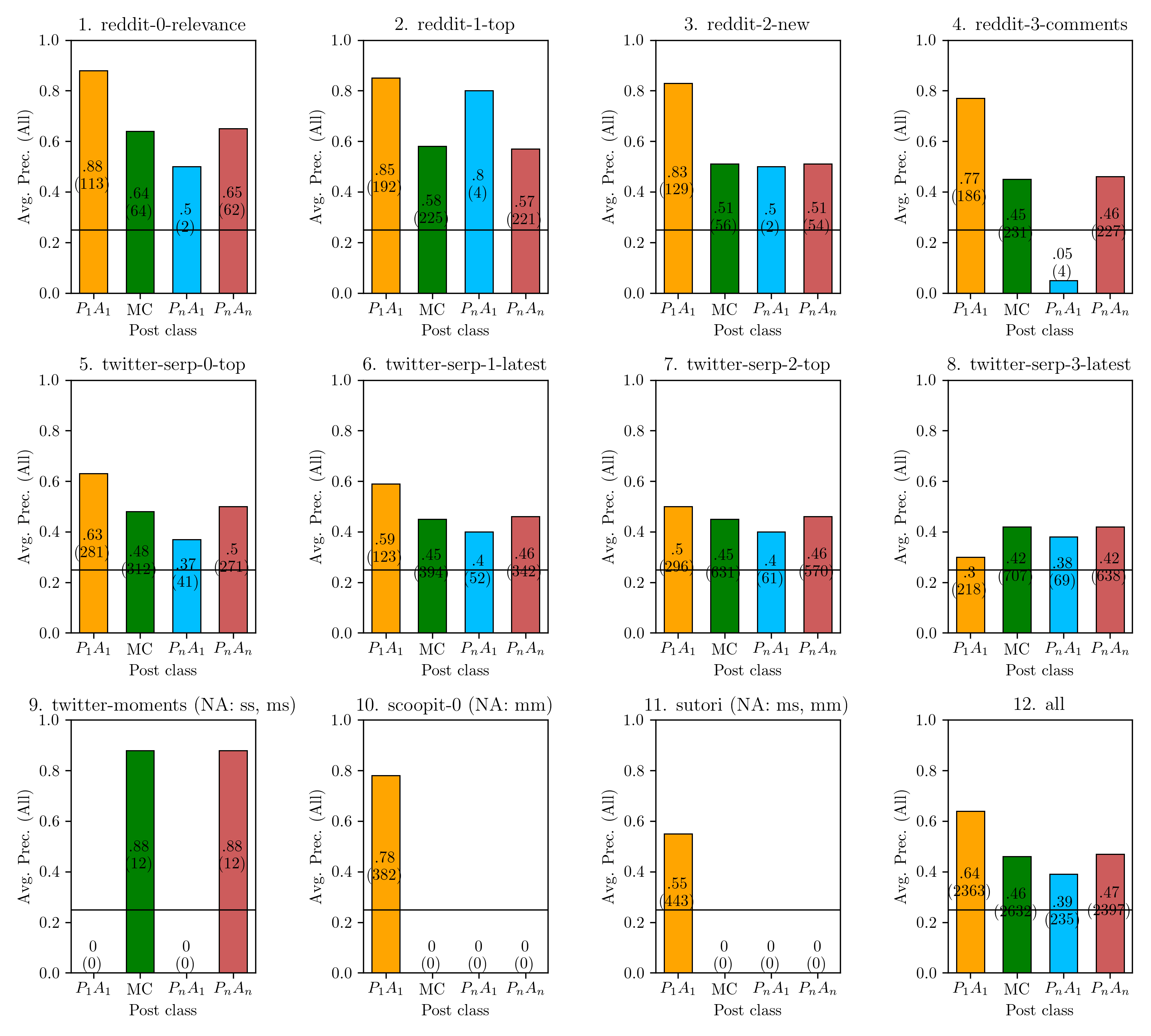}}  
        \caption{Average precision of URIs (HTML and non-HTML) per post class, per social media for \textit{Flint Water Crisis}}
      \end{figure*}

      \begin{figure*}
        \centering
        \fbox{\includegraphics[width=0.98\textwidth]{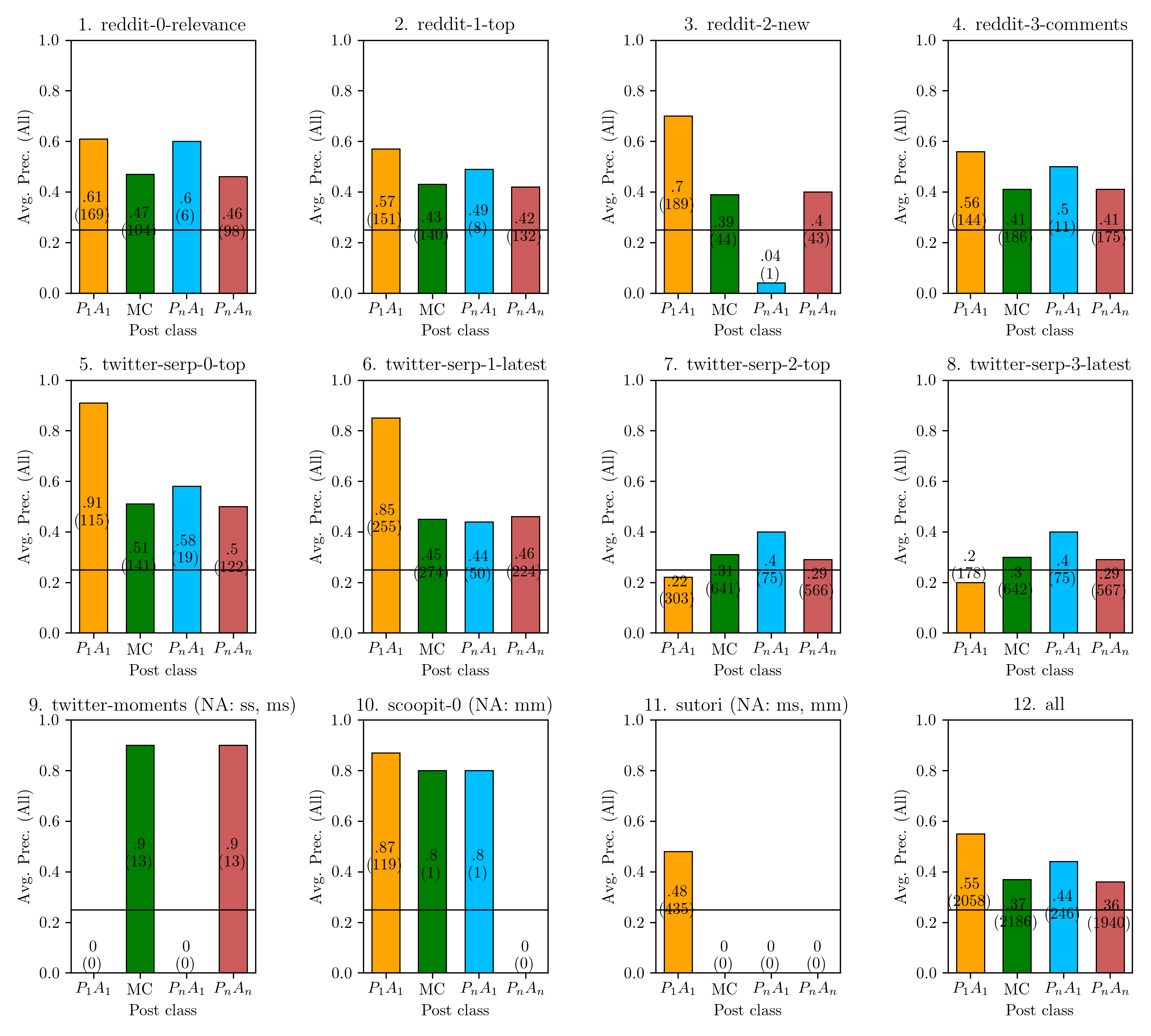}}  
        \caption{Average precision of URIs (HTML and non-HTML) per post class, per social media for \textit{MSD Shooting}}
      \end{figure*}

      \begin{figure*}
        \centering
        \fbox{\includegraphics[width=0.98\textwidth]{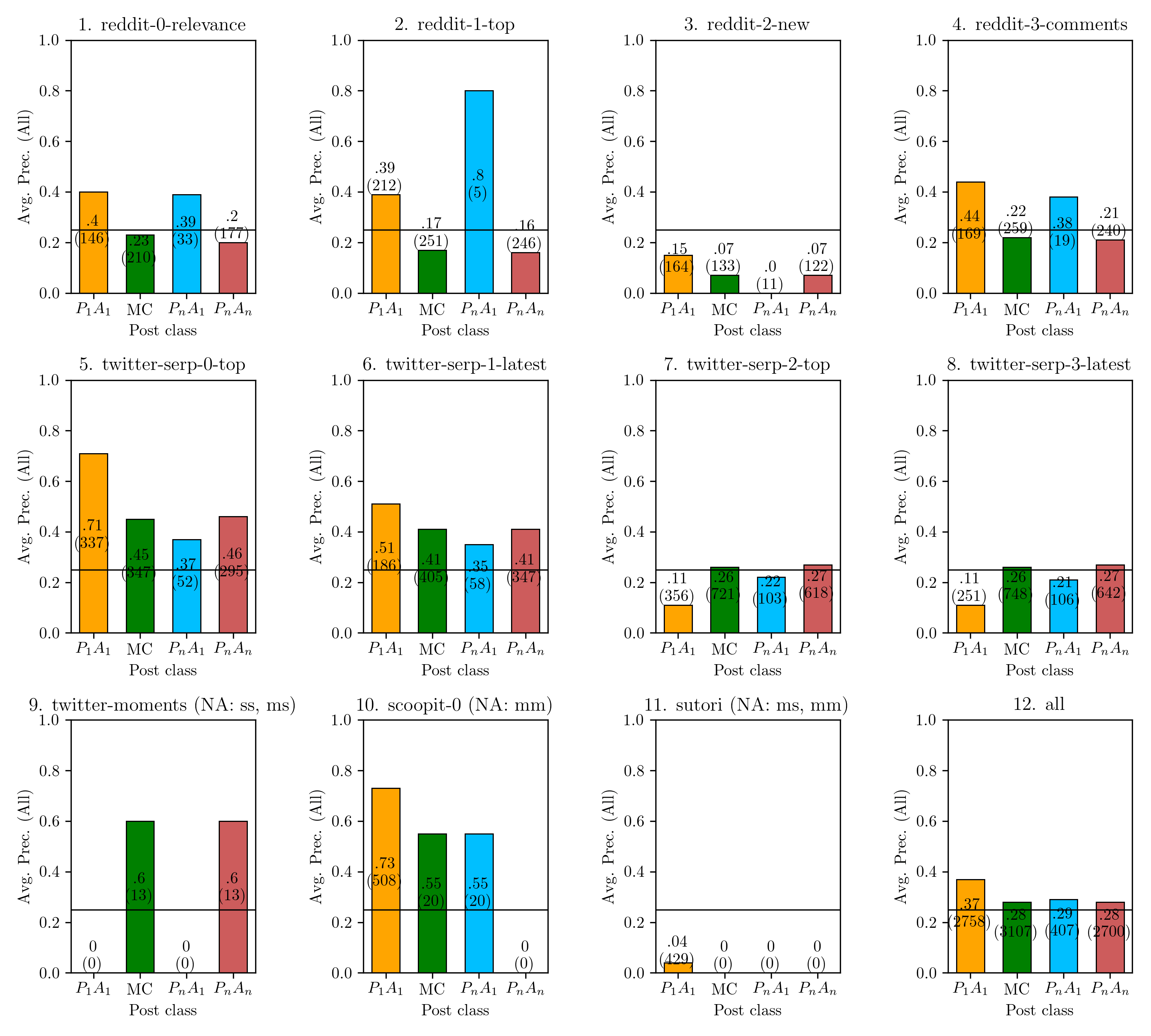}}  
        \caption{Average precision of URIs (HTML and non-HTML) per post class, per social media for \textit{2018 World Cup}}
      \end{figure*}

       \begin{figure*}
        \centering
        \fbox{\includegraphics[width=0.98\textwidth]{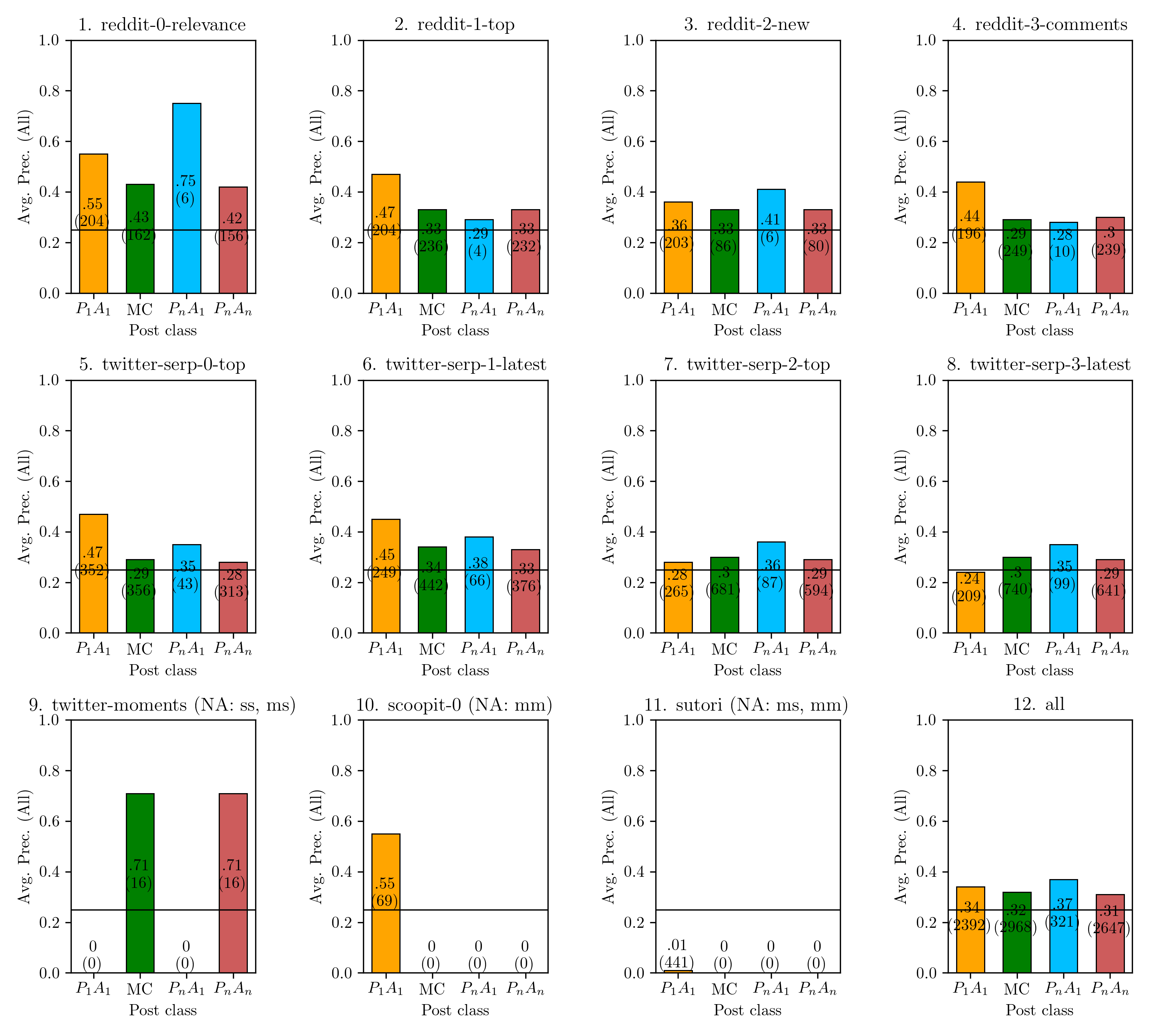}}  
        \caption{Average precision of URIs (HTML and non-HTML) per post class, per social media for \textit{2018 Midterm Elections}}
      \end{figure*}

\clearpage
\begin{figure*}
   \captionsetup{font=Large}
   \centering
   \caption*{APPENDIX 8\\ECDF of URI ages per post class, per social media.}
\end{figure*}
      
      \begin{figure*}
      \centering
      \fbox{\includegraphics[width=0.98\textwidth]{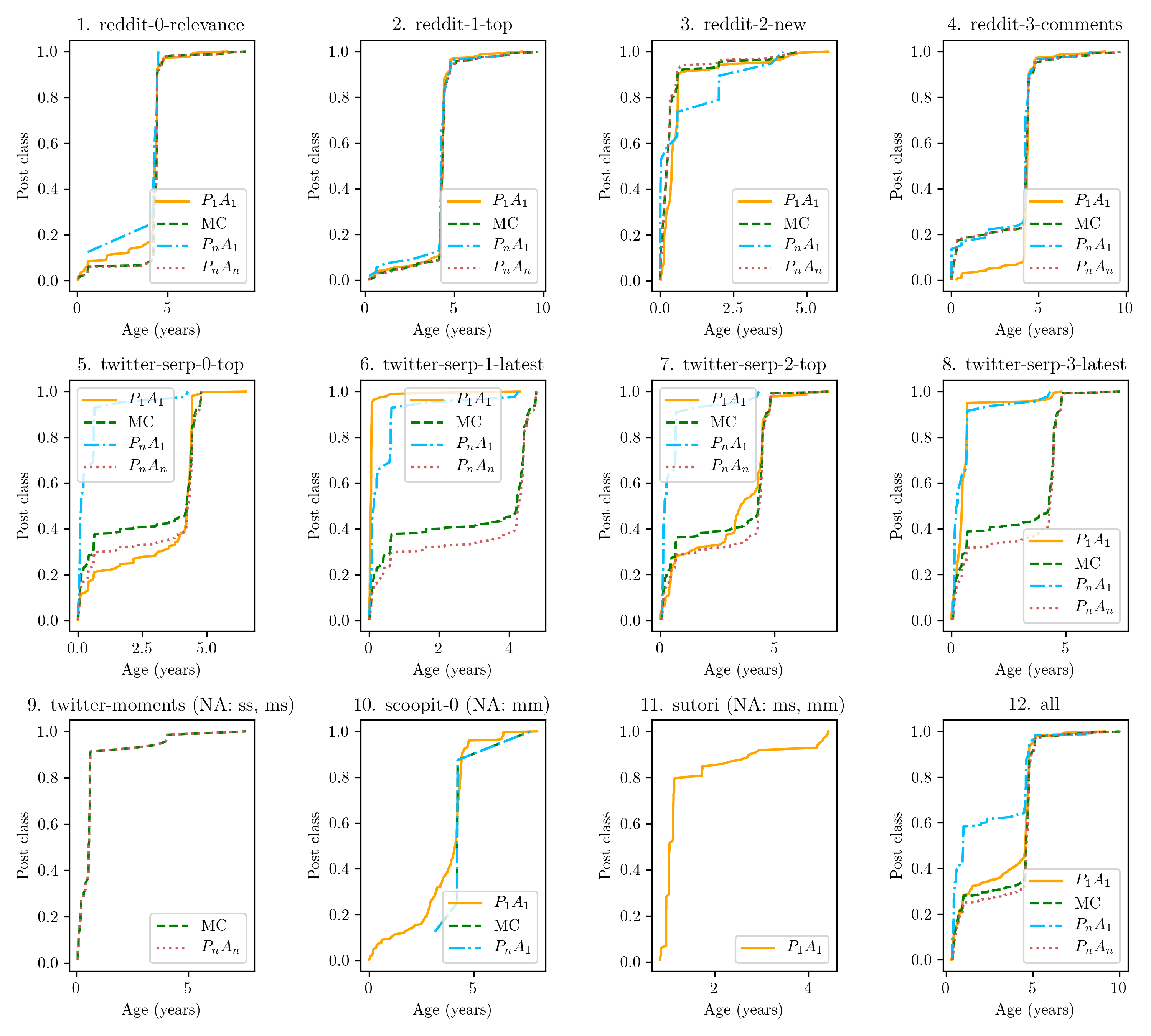}}  
      \caption{Empirical Cumulative Distribution Function (ECDF) of URI ages per post class, per social media for \textit{Ebola Virus Outbreak}. A single sub-figure (e.g., sub-figure 1) reads as follows: 40\% of the Reddit (\textit{relevance} vertical) post class URIs were less than 4 years old. The remaining figures in this appendix are to be read similarly. \textit{twitter-serp-2-top} and \textit{twitter-serp-3-latest} represent collections generated by issuing hashtag (\texttt{\#ebolavirus}) queries.}
      \end{figure*}

      \begin{figure*}
        \centering
        \fbox{\includegraphics[width=0.98\textwidth]{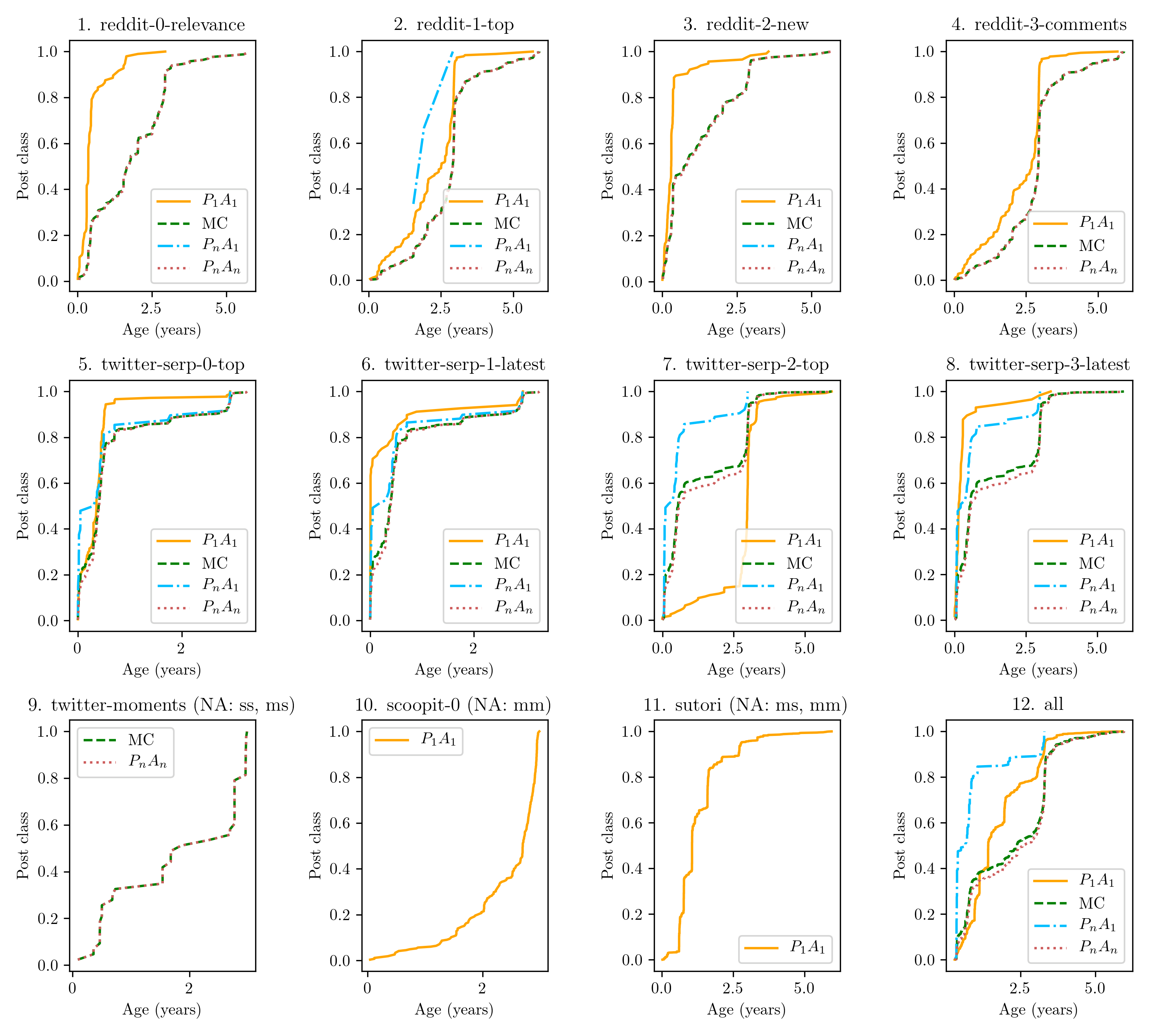}}  
        \caption{ECDF of URI ages per post class, per social media for \textit{Flint Water Crisis}}
      \end{figure*}

      \begin{figure*}
        \centering
        \fbox{\includegraphics[width=0.98\textwidth]{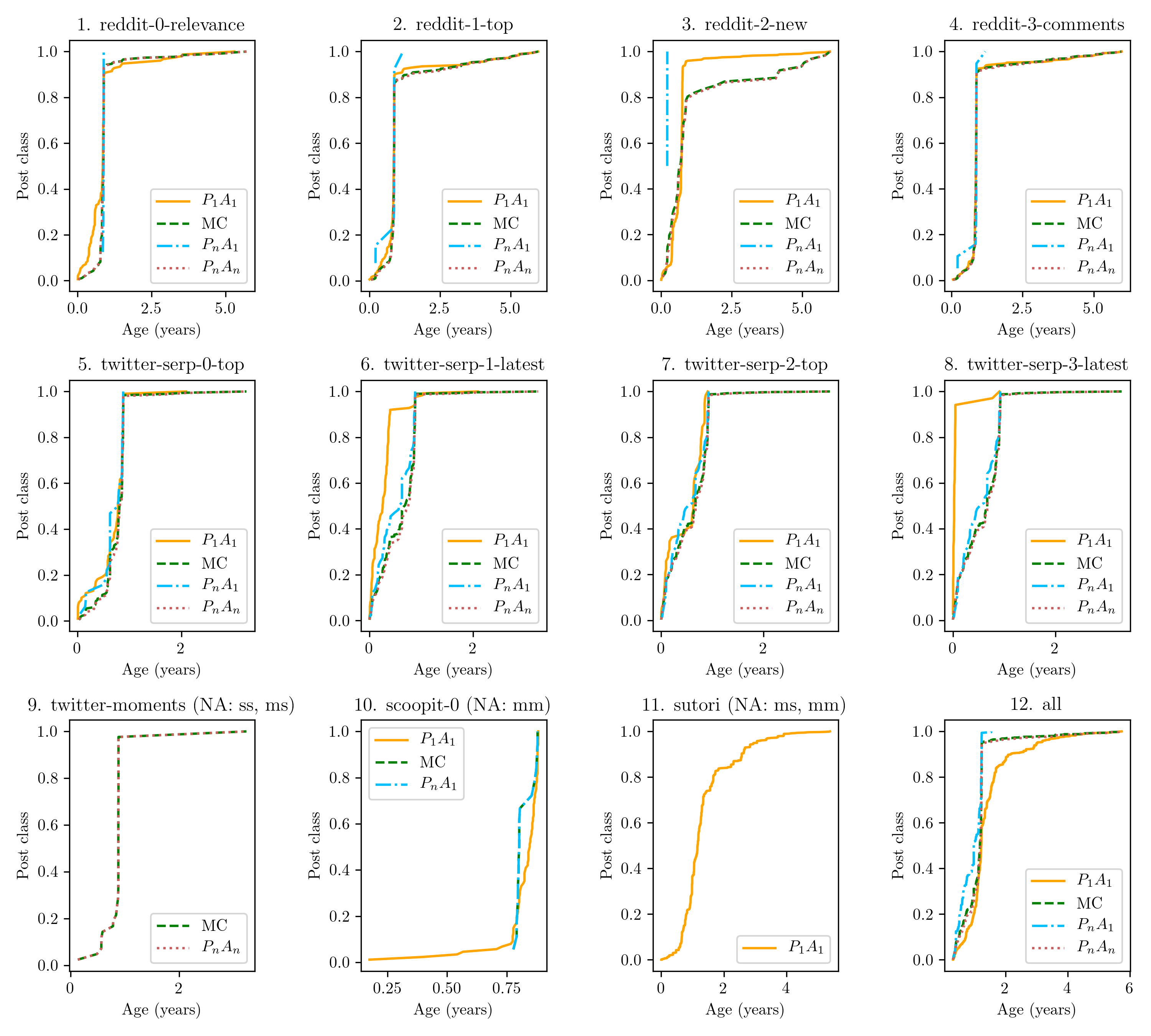}}  
        \caption{ECDF of URI ages per post class, per social media for \textit{MSD Shooting}}
      \end{figure*}

      \begin{figure*}
        \centering
        \fbox{\includegraphics[width=0.98\textwidth]{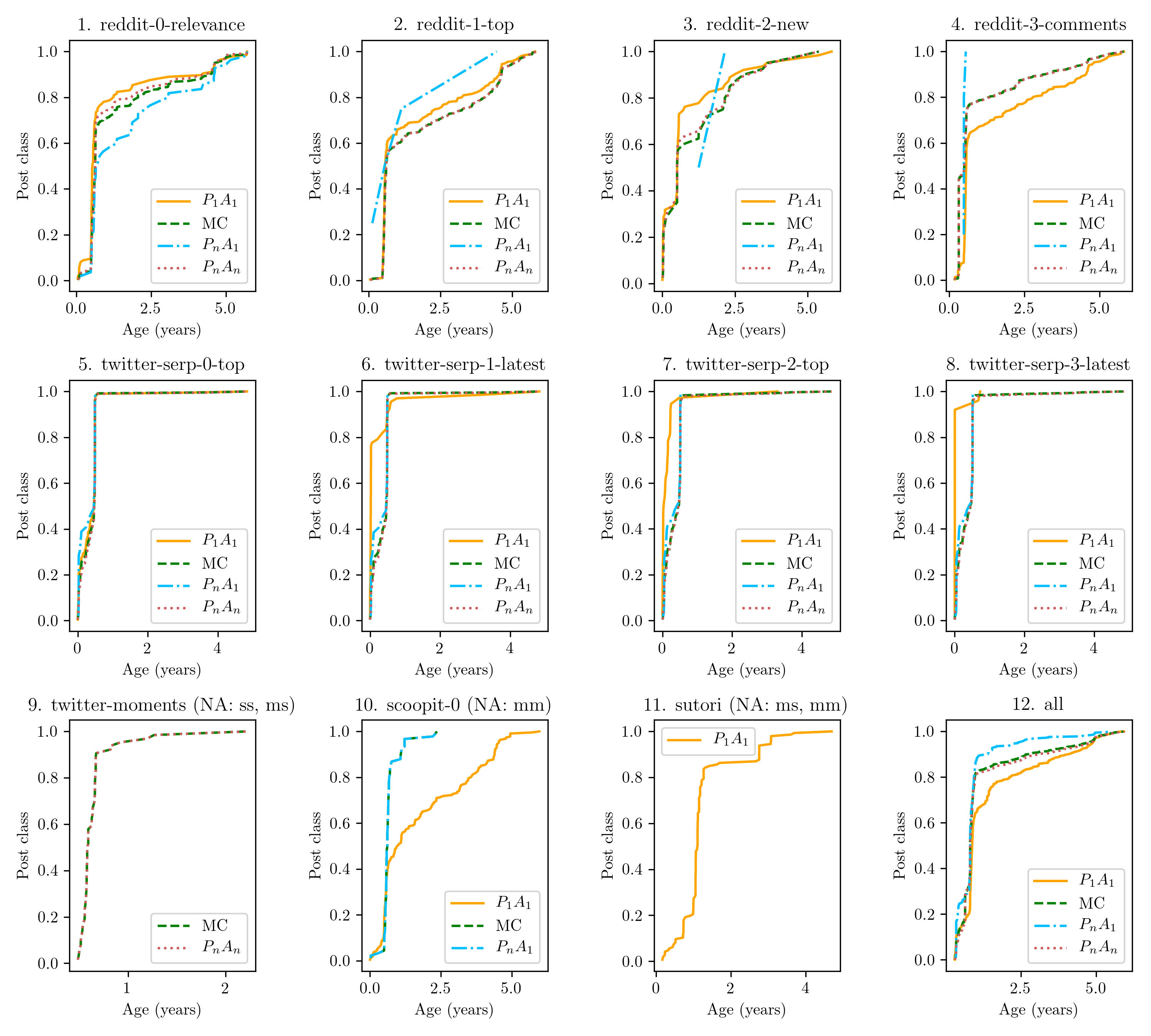}}  
        \caption{ECDF of URI ages per post class, per social media for \textit{2018 World Cup}}
      \end{figure*}

       \begin{figure*}
        \centering
        \fbox{\includegraphics[width=0.98\textwidth]{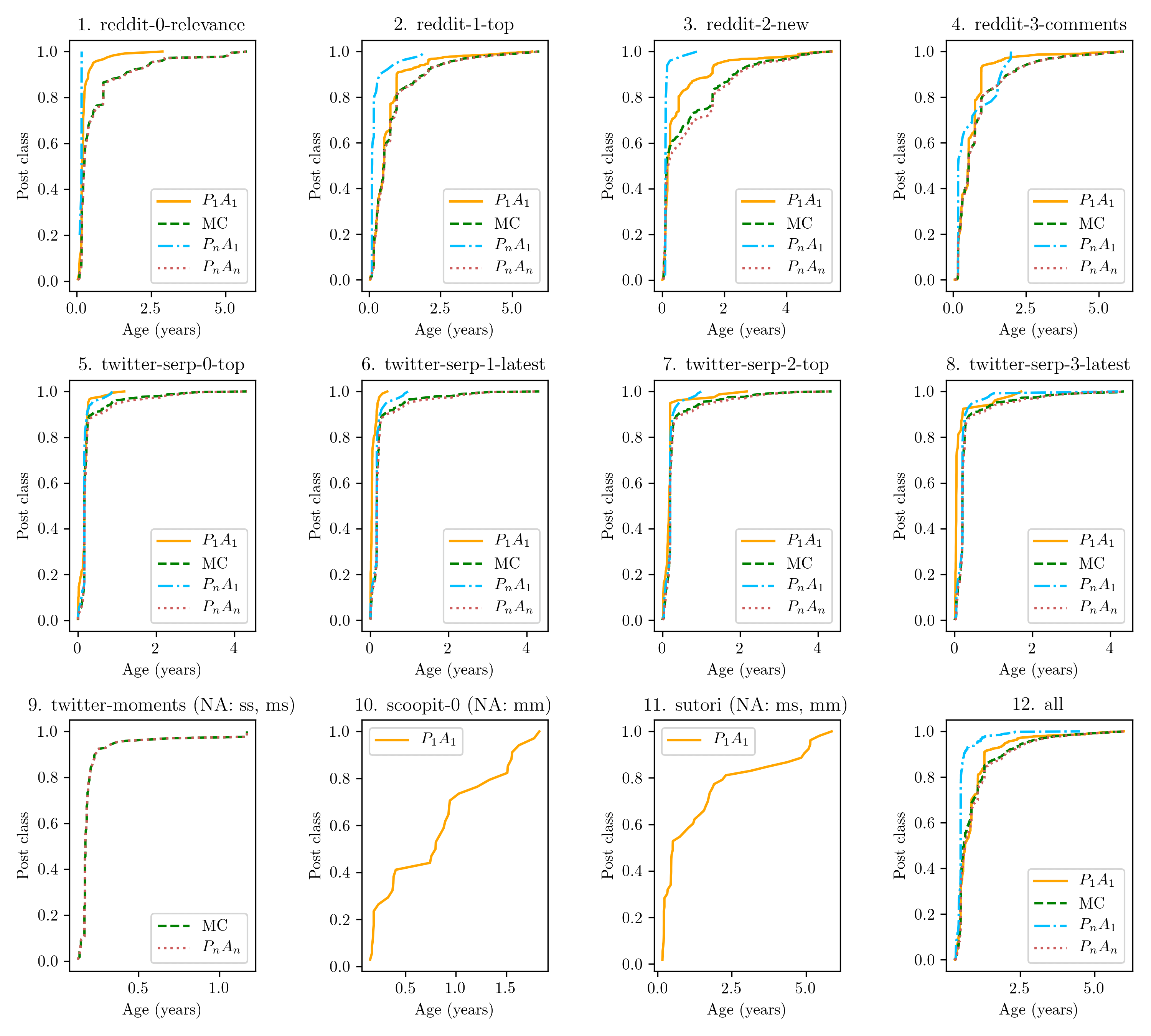}}  
        \caption{ECDF of URI ages per post class, per social media for \textit{2018 Midterm Elections}}
      \end{figure*}

\clearpage
\begin{figure*}
   \captionsetup{font=Large}
   \centering
   \caption*{APPENDIX 9\\Age distribution of URIs per post class, per social media.}
\end{figure*}

      \begin{figure*}
      \centering
      \fbox{\includegraphics[width=0.98\textwidth]{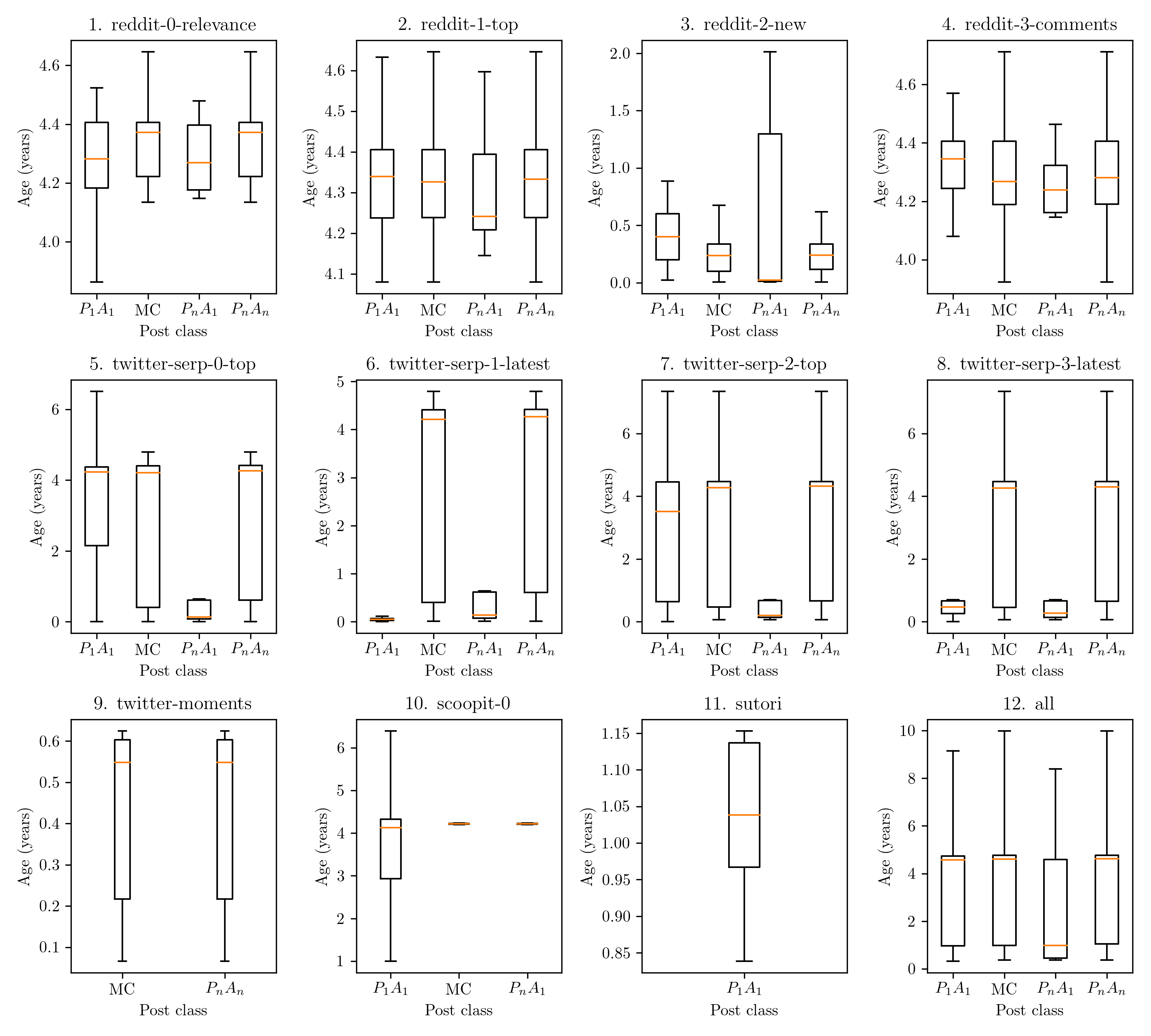}}  
      \caption{Box plot of age distribution of URIs per post class, per social media for \textit{Ebola Virus Outbreak}. A single sub-figure (e.g., sub-figure 1) reads as follows: The minimum, lower-quartile, median, upper-quartile, and maximum ages (in years) of the Reddit (\textit{relevance} vertical) \textbf{P$_1$A$_1$} post class URIs were 3.8, 4.18, 4.28, 4.39, and 4.56, respectively. The remaining figures in this appendix are to be read similarly. \textit{twitter-serp-2-top} and \textit{twitter-serp-3-latest} represent collections generated by issuing hashtag (\texttt{\#ebolavirus}) queries.}
      \end{figure*}

      \begin{figure*}
        \centering
        \fbox{\includegraphics[width=0.98\textwidth]{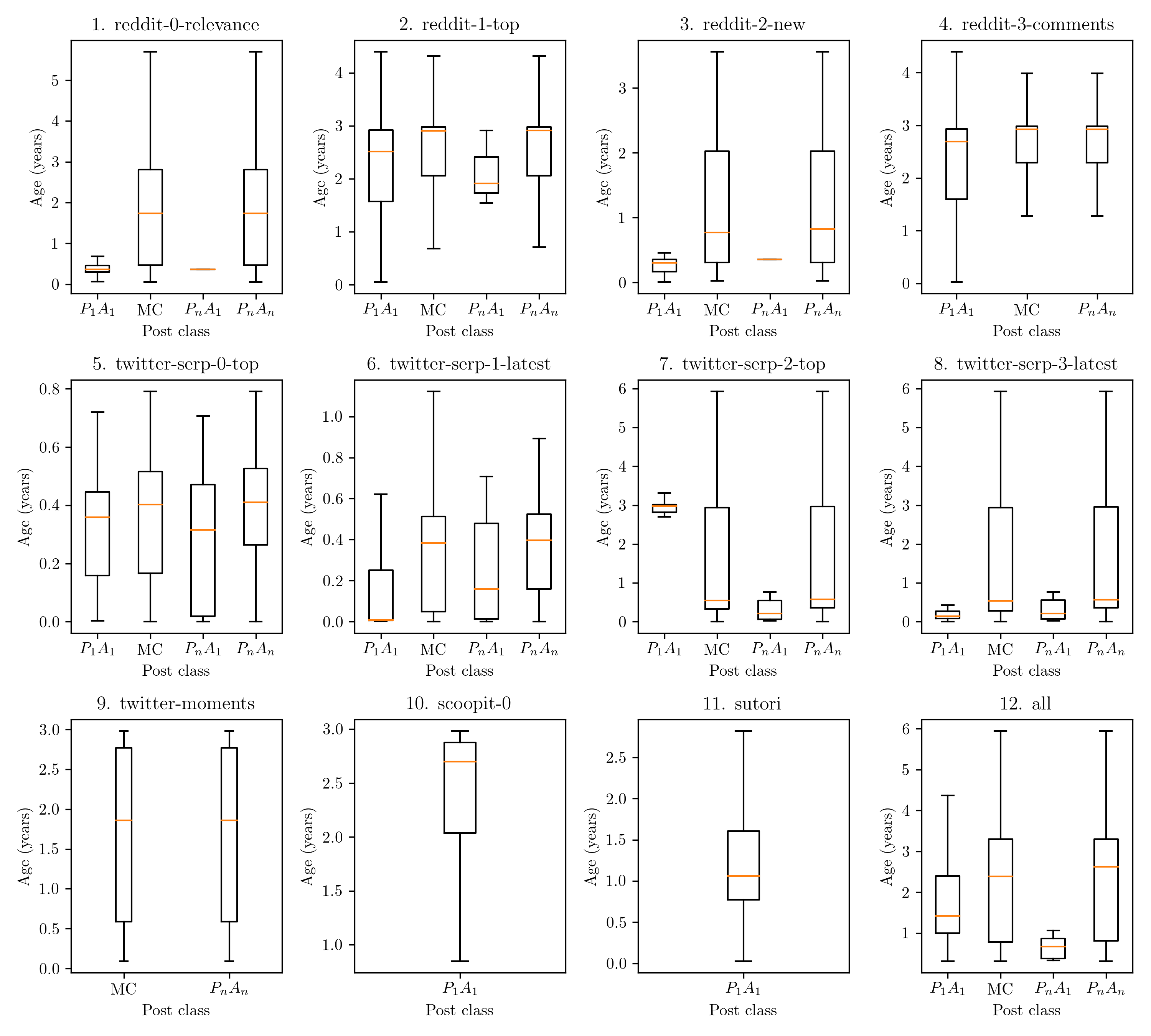}}  
        \caption{Box plot of age distribution of URIs per post class, per social media for \textit{Flint Water Crisis}}
      \end{figure*}

      \begin{figure*}
        \centering
        \fbox{\includegraphics[width=0.98\textwidth]{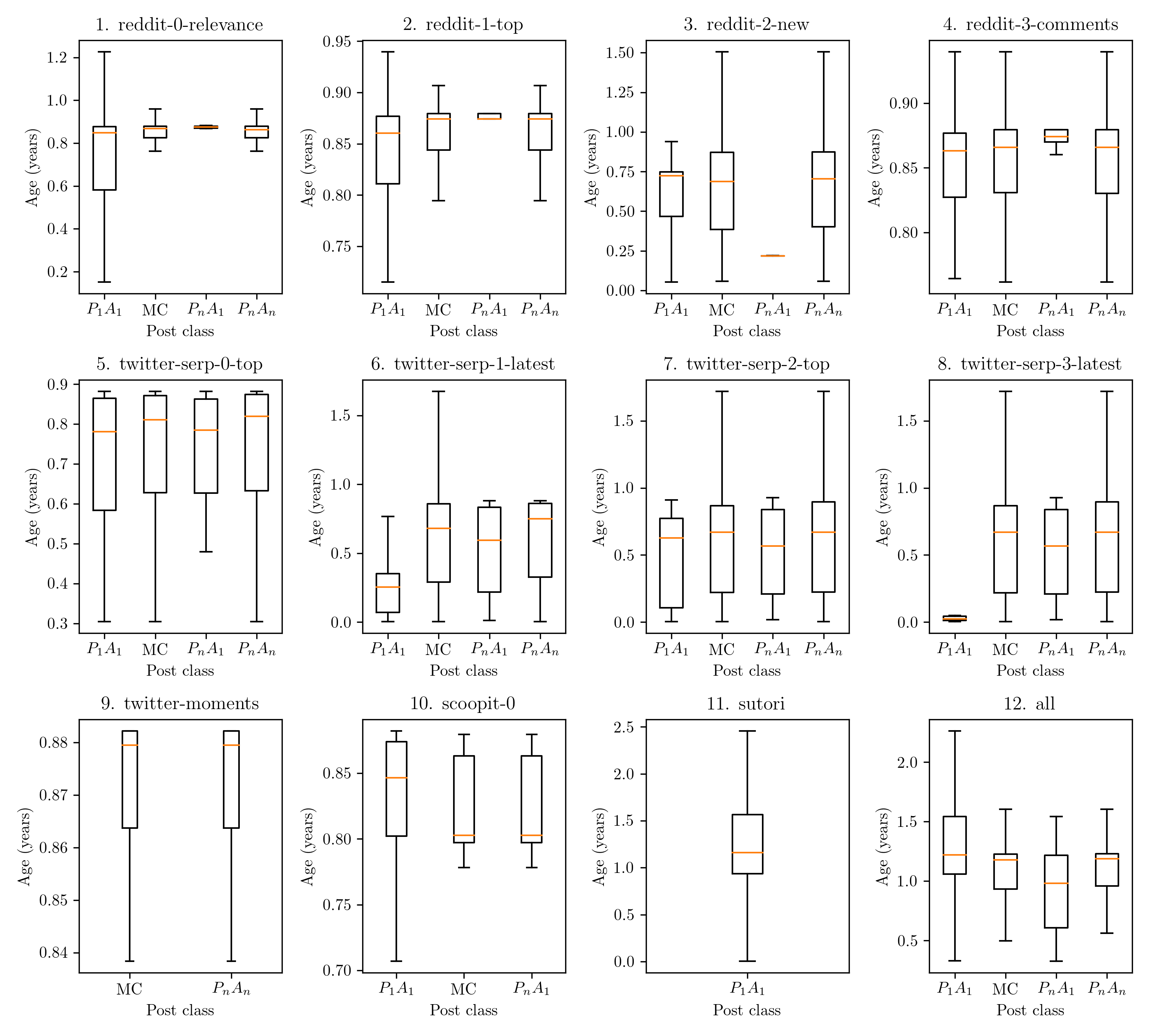}}  
        \caption{Box plot of age distribution of URIs per post class, per social media for \textit{MSD Shooting}}
      \end{figure*}

      \begin{figure*}
        \centering
        \fbox{\includegraphics[width=0.98\textwidth]{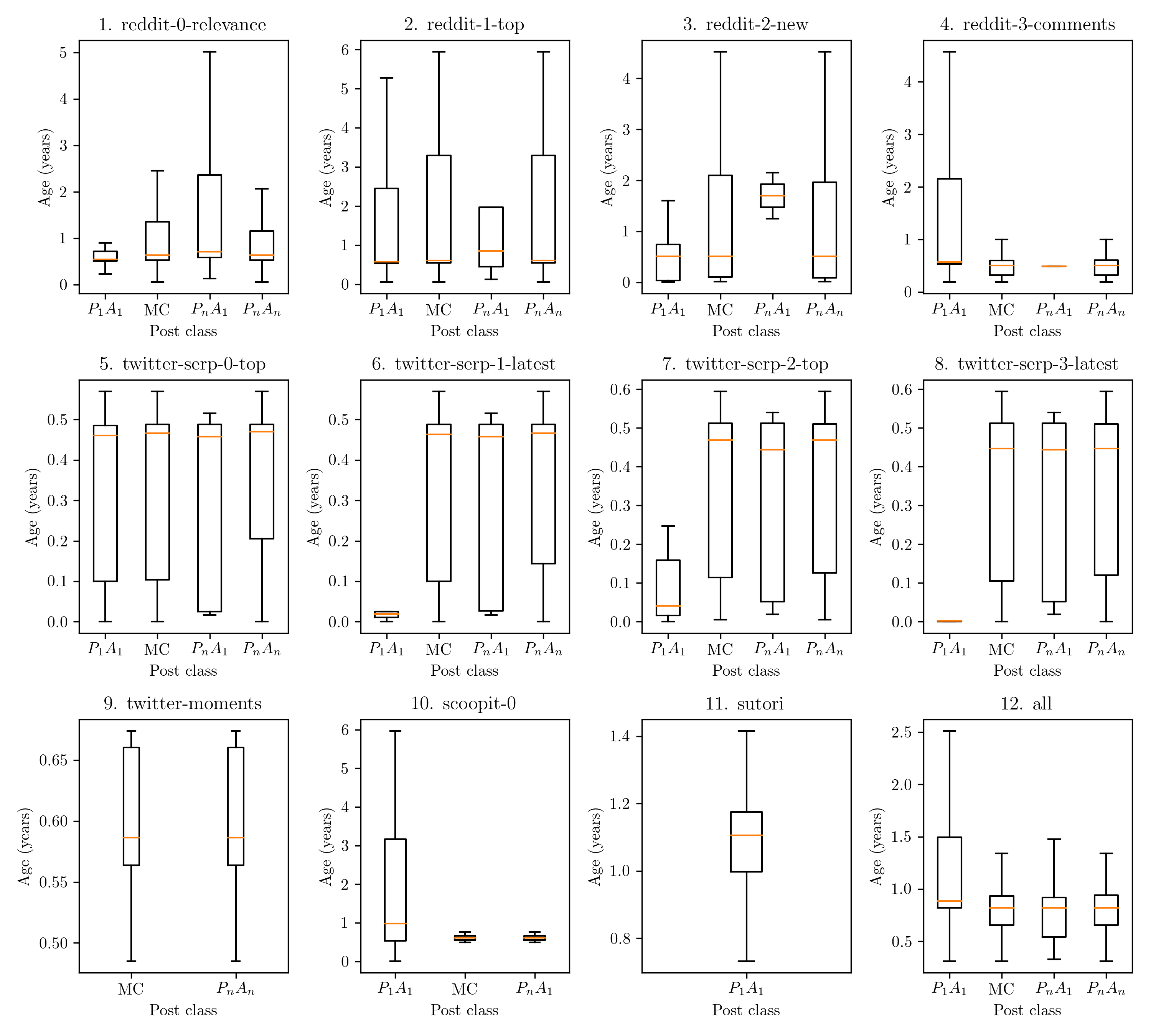}}  
        \caption{Box plot of age distribution of URIs per post class, per social media for \textit{2018 World Cup}}
      \end{figure*}

      \begin{figure*}
        \centering
        \fbox{\includegraphics[width=0.98\textwidth]{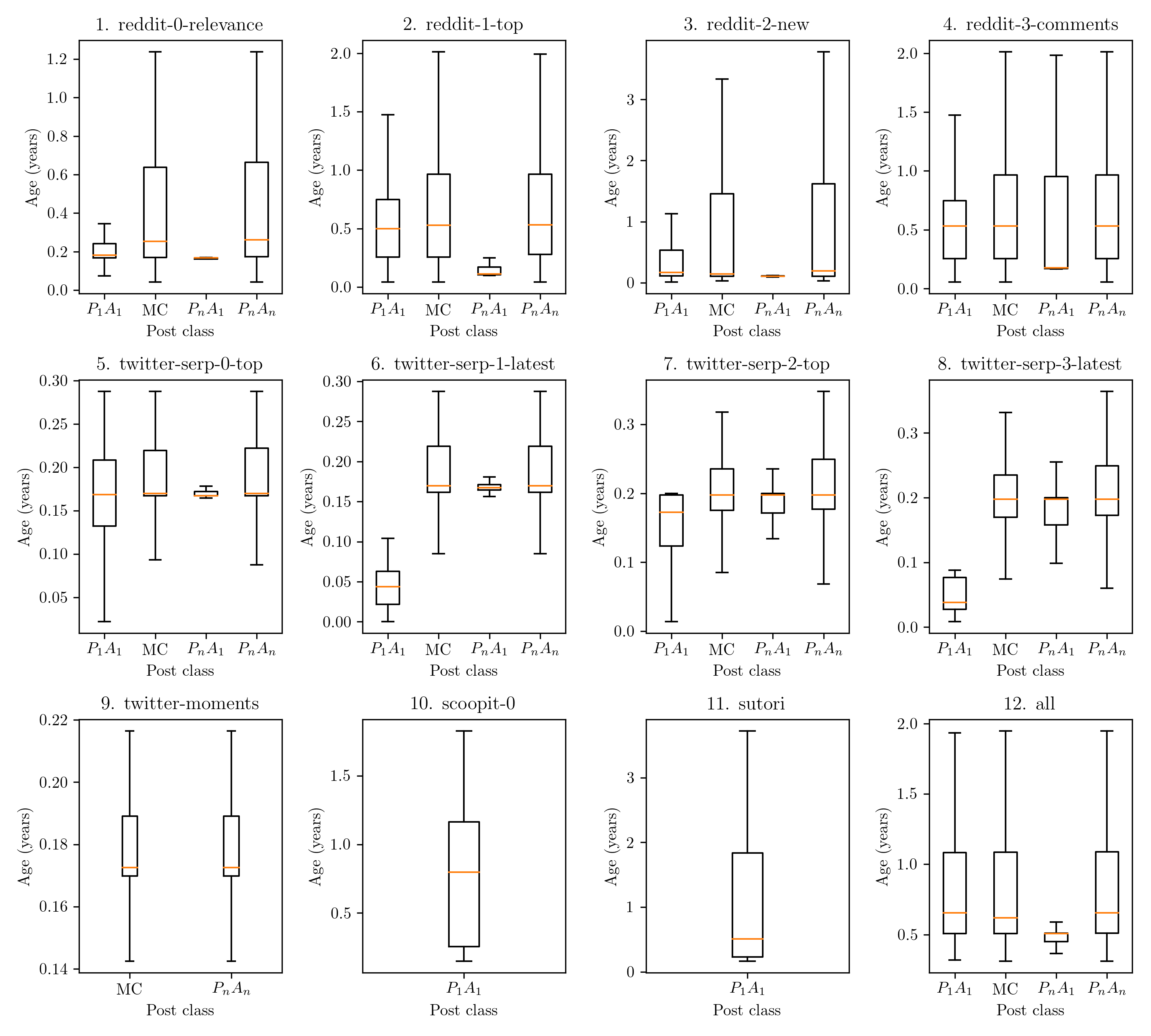}}  
        \caption{Box plot of age distribution of URIs per post class, per social media for \textit{2018 Midterm Elections}}
      \end{figure*}

\clearpage
\begin{figure*}
   \captionsetup{font=Large}
   \centering
   \caption*{APPENDIX 10\\Average precision per count of links in a post, per post class, per social media.}
\end{figure*}
         
      \begin{figure*}
      \centering
      \fbox{\includegraphics[width=0.75\textwidth]{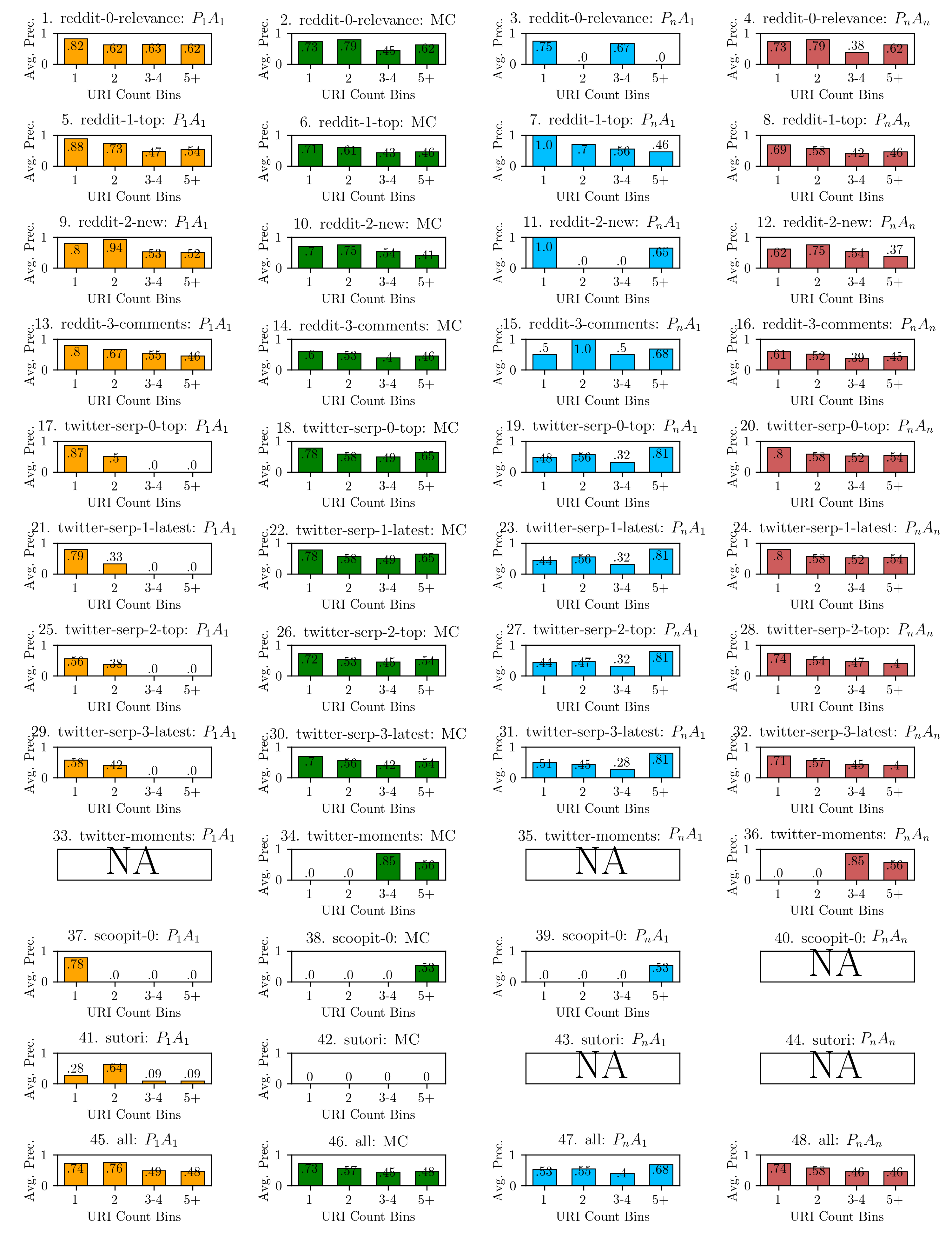}}  
      \caption{Average precision as a function of the number of links in a post, per post class, per social media for \textit{Ebola Virus Outbreak}. A single sub-figure (e.g., sub-figure 1) reads as follows: The average precision of Reddit (\textit{relevance} vertical) posts with 1 URI was 0.82, for posts with 2 URIs, it was, 0.62, etc. The remaining figures in this appendix are to be read similarly. \textit{twitter-serp-2-top} and \textit{twitter-serp-3-latest} represent collections generated by issuing hashtag (\texttt{\#ebolavirus}) queries.}
      \end{figure*}

      \begin{figure*}
        \centering
        \fbox{\includegraphics[width=0.8\textwidth]{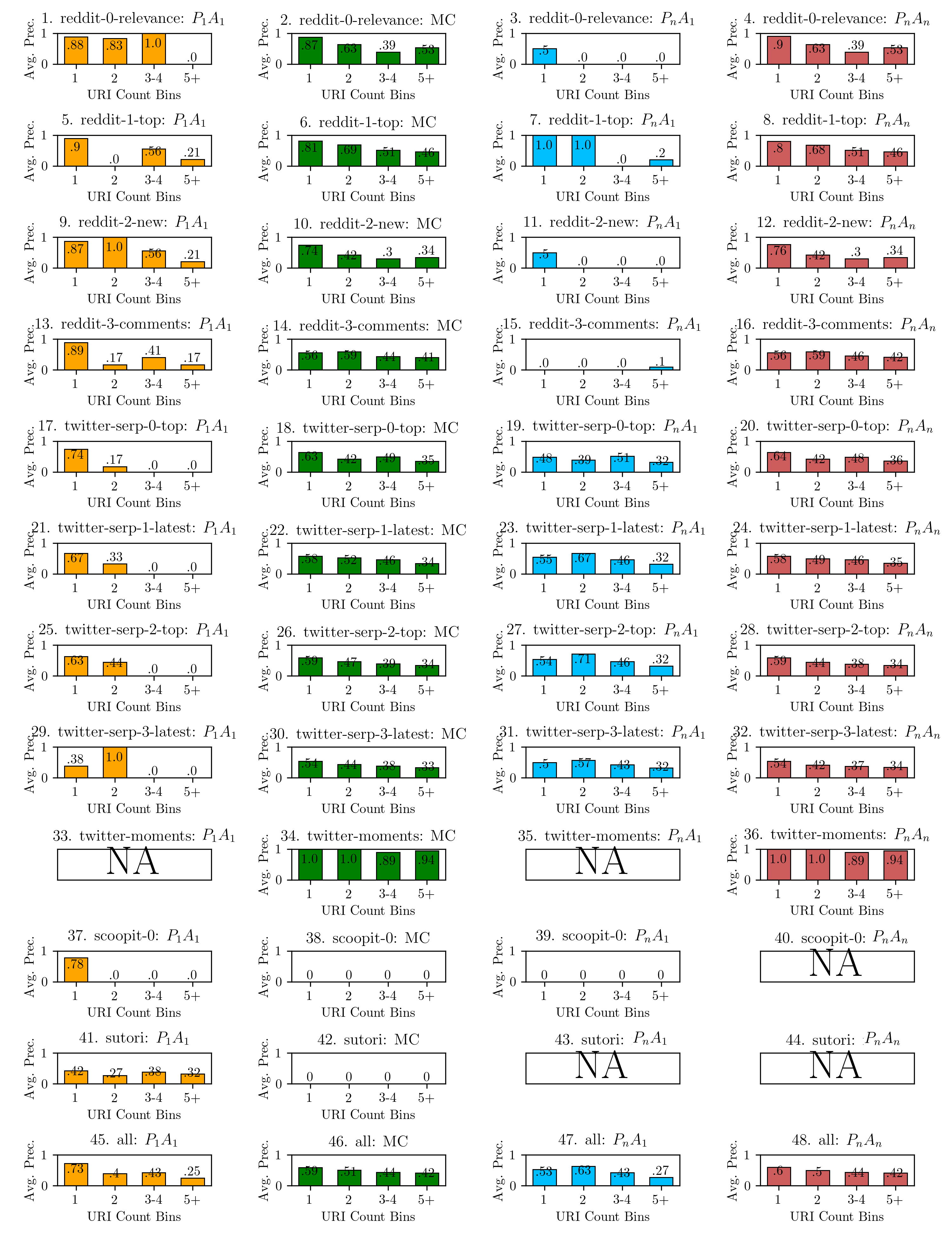}}  
        \caption{Average precision as a function of the number of links in a post, per post class, per social media for \textit{Flint Water Crisis}}
      \end{figure*}

      \begin{figure*}
        \centering
        \fbox{\includegraphics[width=0.8\textwidth]{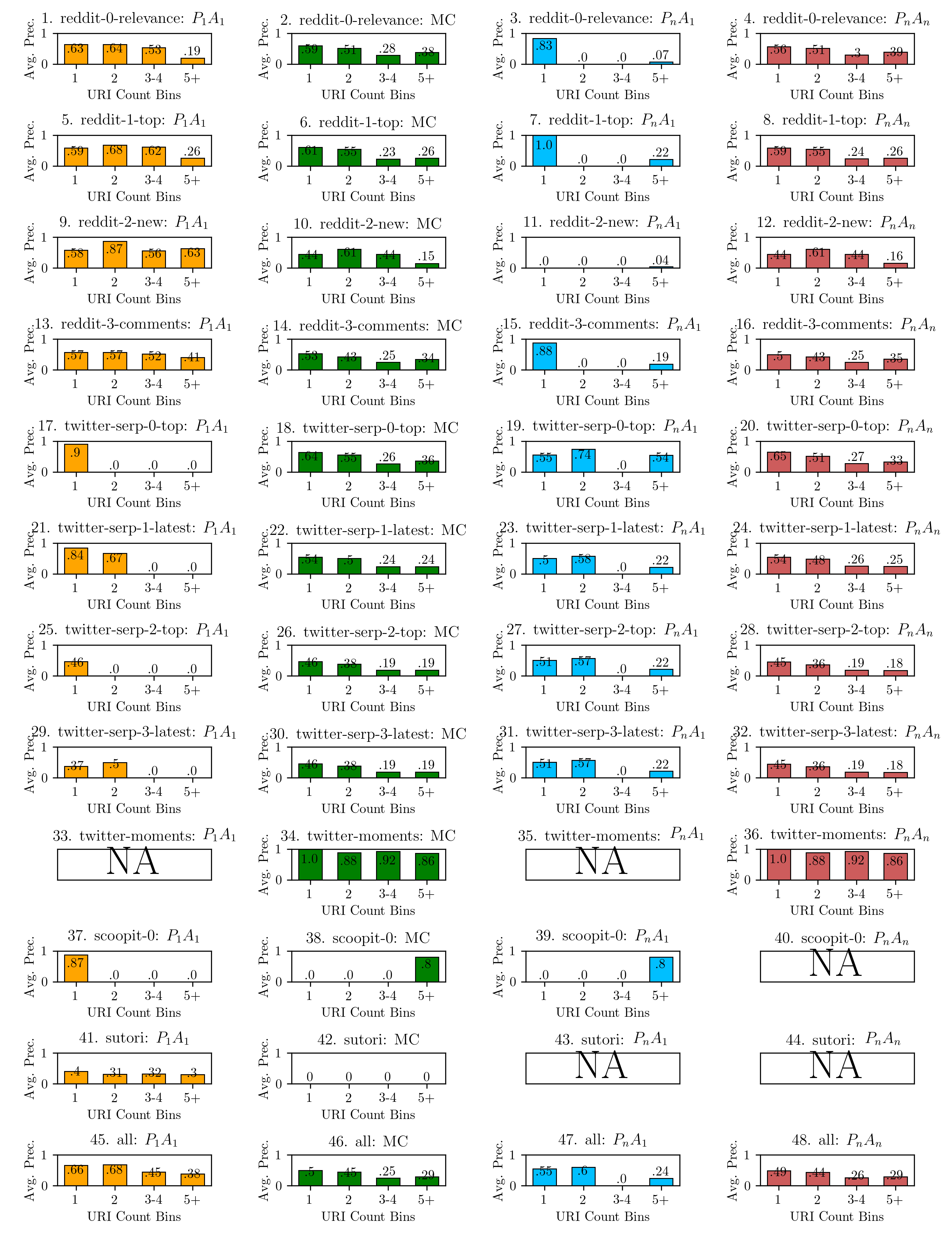}}  
        \caption{Average precision as a function of the number of links in a post, per post class, per social media for \textit{MSD Shooting}}
      \end{figure*}

      \begin{figure*}
        \centering
        \fbox{\includegraphics[width=0.8\textwidth]{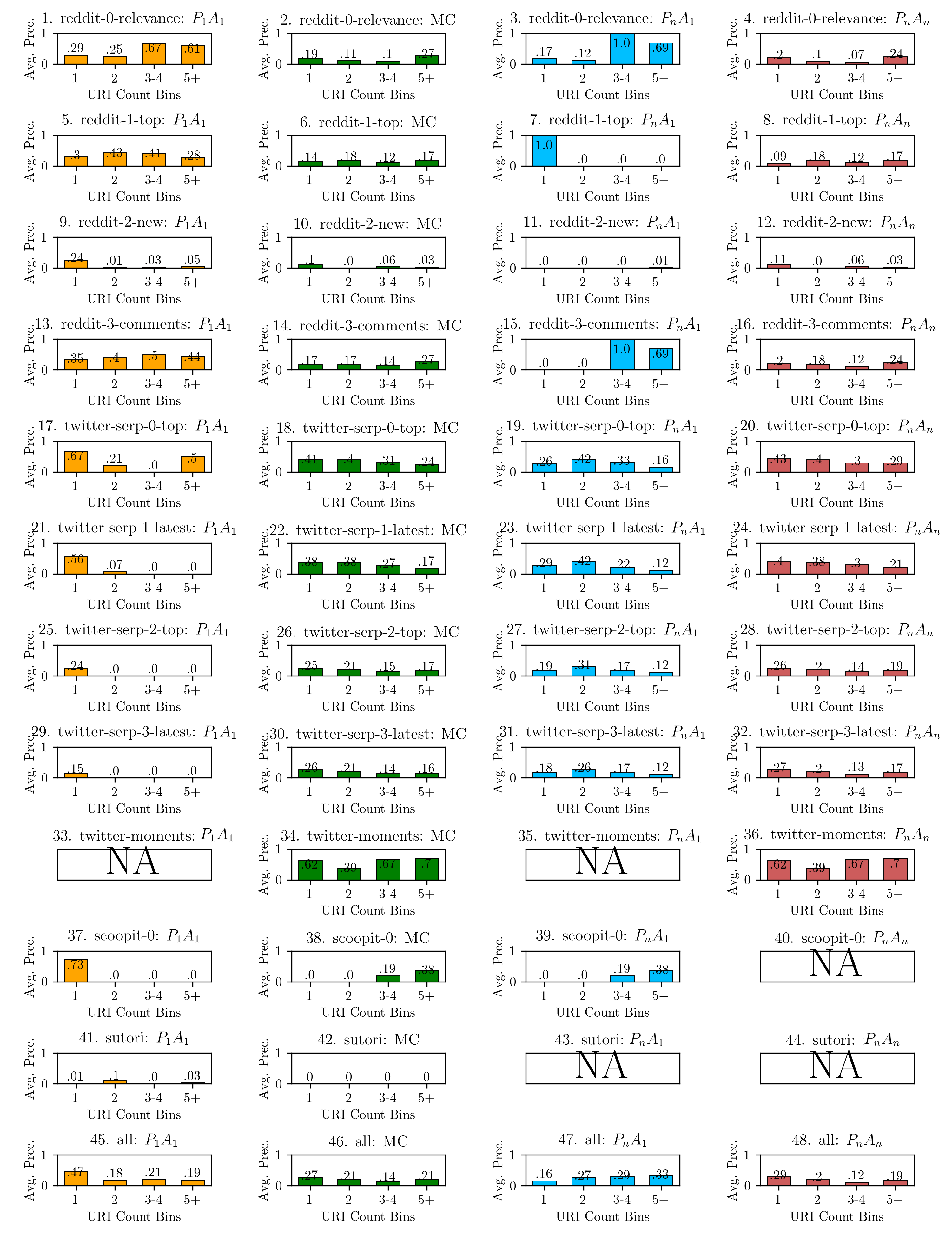}}  
        \caption{Average precision as a function of the number of links in a post, per post class, per social media for \textit{2018 World Cup}}
      \end{figure*}

      \begin{figure*}
        \centering
        \fbox{\includegraphics[width=0.8\textwidth]{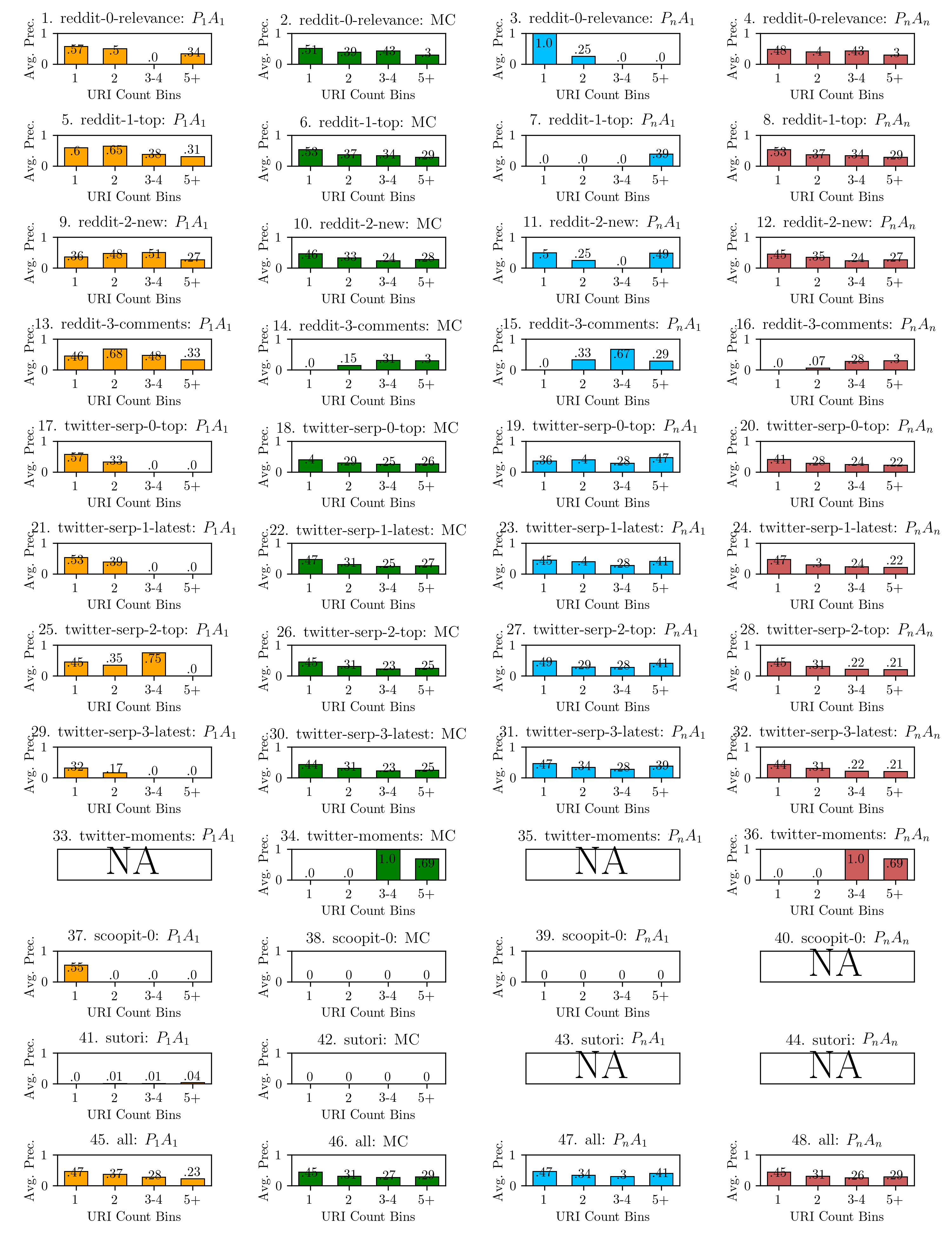}}  
        \caption{Average precision as a function of the number of links in a post, per post class, per social media for \textit{2018 Midterm Elections}}
      \end{figure*}
\end{document}